\documentclass[AMA,STIX1COL]{WileyNJD-v2}

\articletype{ }%

\received{15 December 2023}
\revised{}
\accepted{}

\usepackage{amssymb}
\usepackage{multicol}
\usepackage{amsmath}

\makeatletter
\renewcommand*\env@matrix[1][*\c@MaxMatrixCols c]{%
  \hskip -\arraycolsep
  \let\@ifnextchar\new@ifnextchar
  \array{#1}}
\makeatother

\usepackage{amssymb,mleftright}
\usepackage{amsmath}
\usepackage{setspace}
\usepackage{amsfonts}
\usepackage{amsthm}
\usepackage{bbm}
\usepackage{mathtools}
\usepackage{subfig}
\usepackage{lineno}
\usepackage{hyperref}
\usepackage{float}
\usepackage{array,multirow}
\usepackage{color}
\usepackage{stmaryrd}
\usepackage{tikz}
\usepackage{graphicx}
\usepackage{rotating}

\usepackage{bm}
\usepackage{caption}
\usepackage{booktabs}
\usepackage{lscape}

\definecolor{lightgray}{gray}{0.80}

\newcommand{\mat}[1]{\mathbf{#1}} 

\modulolinenumbers[5]

\begin{document}

\title{The matrix-free macro-element hybridized Discontinuous Galerkin method for steady and unsteady compressible flows}

\author[1]{Vahid Badrkhani}
\author[1]{Marco F.P. ten Eikelder}
\author[1]{Ren\'e R. Hiemstra}
\author[1]{Dominik Schillinger}

\authormark{Badrkhani V,  ten Eikelder M, Hiemstra R AND Schillinger D}

\address[1]{Institute for Mechanics, Computational Mechanics Group, Technical University of Darmstadt, Germany}

\corres{Vahid Badrkhani, Institute for Mechanics, Computational Mechanics Group, Technical University of Darmstadt, Germany. \email{ vahid.badrkhani@tu-darmstadt.de}}
\fundingInfo{German Research Foundation (Deutsche Forschungsgemeinschaft), Grant/Award Numbers: 1249/2-1}

\abstract[Summary]{The macro-element variant of the hybridized discontinuous Galerkin (HDG) method combines advantages of continuous and discontinuous finite element discretization. In this paper, we investigate the performance of the macro-element HDG method for the analysis of compressible flow problems at moderate Reynolds numbers. To efficiently handle the corresponding large systems of equations, we explore several strategies at the solver level. On the one hand, we devise a second-layer static condensation approach that reduces the size of the local system matrix in each macro-element and hence the factorization time of the local solver. On the other hand, we employ a multi-level preconditioner based on the FGMRES solver for the global system that integrates well within a matrix-free implementation. In addition, we integrate a standard diagonally implicit Runge-Kutta scheme for time integration. 
We test the matrix-free macro-element HDG method for compressible flow benchmarks, including Couette flow, flow past a sphere, and the Taylor-Green vortex. 
Our results show that unlike standard HDG, the macro-element HDG method can operate efficiently for moderate polynomial degrees, as the local computational load can be flexibly increased via mesh refinement within a macro-element. Our results also show that due to the balance of local and global operations, the reduction in degrees of freedom, and the reduction of the global problem size and the number of iterations for its solution, the macro-element HDG method can be a competitive option for the analysis of compressible flow problems. 
}

\keywords{Hybridized discontinuous Galerkin method, macro-elements, matrix-free, compressible flows, second-layer static condensation, iterative solvers}

\maketitle

\section{Introduction \label{sec:introduction}}

Discontinuous Galerkin (DG) methods \cite{arnold2002unified} have been widely recognized for their favorable attributes in tackling conservation problems. They possess a robust mathematical foundation, the flexibility to employ arbitrary orders of basis functions on general unstructured meshes, and a natural stability property for convective operators \cite{bassi1997high,cockburn2018discontinuous,hesthaven2007nodal,peraire2008compact}. About a decade ago, the hybridized discontinuous Galerkin (HDG) method was introduced, setting itself apart with distinctive features within the realm of DG methods \cite{cockburn2009unified}. The linear systems arising from the HDG method exhibit equivalence to two different systems: the first globally couples the numerical trace of the solution on element boundaries, leading to a significant reduction in degrees of freedom. The second couples the conserved quantities and their gradients at the element level, allowing for an element-by-element solution. Due to these advantages, numerous research endeavors have extended the application of the HDG method to address a diverse array of initial boundary value problems 
\cite{nguyen2009implicit,cockburn2009superconvergent,nguyen2012hybridizable,peraire2010hybridizable,cockburn2009hybridizable,nguyen2011implicit,
fernandez2017hybridized,vila2021hybridisable,nguyen2011high}. Nevertheless, in the face of extensive computations, relying solely on the hybridization approach may fall short in overcoming limitations related to memory and time-to-solution \cite{pazner2017stage,kronbichler2018performance}. Ongoing research efforts \cite{fabien2019manycore,roca2013scalable,roca2011gpu} highlight these challenges and help provide motivation for the current work.

Compared to the DG method, the classical finite element formulation, also known as the continuous Galerkin (CG) method \cite{hughes2012finite,paipuri2019coupling}, leads to fewer unknowns when the same mesh is used. Nevertheless, unstructured mesh generators frequently produce meshes with high vertex valency, resulting in intricate communication patterns during parallel runs on distributed memory systems, thereby affecting scalability \cite{kirby2012cg,yakovlev2016cg}. The HDG method, although generating a global system with a higher rank than the traditional statically condensed system in CG, demonstrates significantly reduced bandwidth at moderate polynomial degrees.

The macro-element variant of the HDG method, recently motivated by the authors for advection-diffusion problems \cite{badrkhani2023matrix}, investigates a discretization strategy that amalgamates elements from both the CG and HDG approaches, enabling several distinctive features that sets it apart from the CG and the standard HDG method. First and foremost, by incorporating continuous elements within macro-elements, the macro-element HDG method effectively tackles the issue of escalating the number of degrees of freedom in standard HDG methods for a given mesh. Secondly, it provides an additional layer of flexibility in terms of tailoring the macro-element discretization and adjusting the associated local problem size to match the specifications of the available compute system and its parallel architecture. Thirdly, it introduces a direct approach to adaptive local refinement, enabling uniform simplicial subdivision. Fourthly, all local operations are embarrassingly parallel and automatically balanced. This eliminates the necessity for resorting to load balancing procedures external to the numerical method. Consequently, the macro-element HDG method is highly suitable for a matrix-free solution approach.

In this article, we extend the macro-element variant of the HDG method \cite{badrkhani2023matrix} to solving steady and unsteady compressible flow problems given by the Navier--Stokes equations. 
On the one hand, unlike in standard HDG schemes, the time complexity of the local solver in the macro-element HDG method increases significantly as the size of local matrices grows. Therefore, effectively managing the local matrix is pivotal for achieving scalability in the macro-element HDG algorithm. 
On the other hand, the macro-element HDG method can alleviate the accelerated growth of degrees of freedom that stems from the duplication of degrees of freedom along element boundaries. 
Recent research, as documented in \cite{fabien2019manycore,kronbichler2019fast,lermusiaux2007environmental}, has explored alternative techniques to tackle this challenge within the standard HDG method. These techniques pivot towards Schur complement approaches for the augmented system, addressing both local and global unknowns rather than solely focusing on the numerical trace. In our study, we integrate and synthesize some of these techniques to present an efficient, computationally economical and memory-friendly version of the macro-element HDG algorithm. 


The paper is organized as follows. In Section 2, we delineate the differential equations and elucidate the spatial and temporal discretizations employed in this study, leveraging the macro-element HDG method. Section 3 delves into the development of parallel iterative methods designed to solve the nonlinear system of equations resulting from the discretization process. Within this section, we focus on an inexact variant of Newton's method, standard static condensation, and an alternative second-layer static condensation approach. In Section 4, we present the matrix-free implementation utilized in this investigation, alongside a discussion of global solver options and the corresponding choices of preconditioners. Section 5 is devoted to showcasing the outcomes of numerical experiments conducted with our macro-element HDG variant, juxtaposed against the standard HDG method. These experiments encompass several test cases, ranging from three-dimensional steady to unsteady flow scenarios. A comprehensive comparative analysis is undertaken, evaluating the methods in terms of accuracy, iteration counts, computational time, and the number of degrees of freedom in the local / global solver, with a particular focus on the parallel implementation.

\section{The HDG method on macro-elements for compressible flow\label{sec:/The hybridized DG method on macro-element}}

{\color{black}We first present the Navier--Stokes equations for modeling compressible flows. We proceed with a summary of the notation necessary for the description of the macro-element HDG method, following the notation laid out in our earlier work \cite{badrkhani2023matrix}. Next, we briefly describe the macro-element DG discretizaton in space and the implicit Runge-Kutta discretization in time.

\subsection{Governing equations}\label{S30}
The time dependent compressible Navier--Stokes equations are a non-linear conservation law system that can be written as follows:
\begin{subequations}
    \begin{align}
      \partial_{t^*} \rho^* + \nabla^*\cdot(\rho^* \mathbf{v}) =&~ 0\\
      \partial_{t^*} (\rho^*\mathbf{v}) + \nabla^*\cdot(\rho^* \mathbf{v}^* \otimes \mathbf{v}^*) + \nabla^* P^* - \nabla^*\cdot \boldsymbol{\tau}^*  =&~ 0\\
      \partial_{t^*} (\rho^* E^*) + \nabla \cdot ^*\left(\rho^* H^* \mathbf{v}^*\right) - \nabla^*\cdot\left( \boldsymbol{\tau}^*\mathbf{v}^* - \boldsymbol{\phi}^*\right) =&~ 0,
    \end{align}
\end{subequations}
where $\rho^*$ is the density, $\mathbf{v}^*$ the velocity, and $E^*$ the total specific energy, subject to the initial conditions $\rho^*=\rho^*_0$, $\mathbf{v}^* =\mathbf{v}_0^*$, $E^*=E^*_0$. Furthermore, $P^*$ is the pressure, $H^* = E^*+P^*/\rho^*$ the total specific enthalpy, and 
the shear stress and heat flux are respectively given by:
\begin{subequations}\label{C6}
\begin{align}
\mat {\tau}^*=&~\mu^* \left( \nabla \mat{v}^* + (\nabla^*\mat{v}^*)^T + \lambda (\nabla^*\cdot\mat{v}^*) \mat{I}\right),\\
\boldsymbol{\phi}^* =&~ - \kappa^* \nabla^* T^*.
\end{align}
\end{subequations}
Here $\mu$ is the dynamic viscosity, $\lambda = -2/d$, with spatial dimension $d$, $T^*$ is the temperature, and $\kappa^*$ the thermal conductivity. The thermodynamical variables $P^*, \rho^*$ and $T^*$ are related through an equation of state $P^*=P^*(\rho^*,T^*)$. In this work we employ the (calorically) perfect gas equation of state: $P^*=\rho^* R^* T^*$, where $R^*$ is the specific gas constant. The temperature $T^*$ is related to the internal energy via the constitutive relation $e^* = c_v^* T^*$, where $\gamma$ is the ratio of specific heats, and $c_v^* = R^*/(\gamma-1)$ is the specific heat at constant volume. The total specific energy is given by $E^* = e^* + \mat{v}^*\cdot\mat{v}^*/2$.}
{\color{black}
We introduce a rescaling of the system based on the following dimensionless variables:
\begin{align}
  \mathbf{x} =&~ \frac{\mathbf{x}^*}{X_0}, \quad t = \frac{t^* c_\infty}{X_0}, \quad \mathbf{v}=\frac{\mathbf{v}^*}{c_\infty},\quad \rho = \frac{\rho^*}{\rho_\infty},\quad 
  P = \frac{P^*}{\rho_\infty c_\infty^2},\quad 
  e = \frac{e^*}{c_\infty^2},\quad E = \frac{E^*}{c_\infty^2}\nonumber\\T =&~ \frac{T^*}{T_\infty}, \quad \mu = \frac{\mu^*}{\mu_\infty}, \quad \kappa = \frac{\kappa^*}{\kappa_\infty},\quad R = \frac{R^*}{{c_p}_\infty}\quad c_v = \frac{c_v^*}{{c_p}_\infty}\quad c_p = \frac{c_p^*}{{c_p}_\infty}, 
\end{align}
where $X_0$ is a characteristic length, $c_\infty$ the free-stream speed of sound, $\rho_\infty$ the free-stream density, $T_\infty$ the free-stream temperature and $\mu_\infty$ the free-stream dynamic viscosity. 
The non-dimensional system may be written in the compact form \cite{HDGLES,peraire2010hybridizable}:
\begin{subequations}\label{C4}
 \begin{align}
\frac{\partial \mat {u}}{\partial t}+\nabla \cdot(\mat {F}(\mat{u})+\mat {G}(\mat{u},\mat{q}))=&~0,\\
\mat q - \nabla \mat u=&~ 0,
\end{align}   
\end{subequations}
subject to the initial condition $\mat u=\mat{u}_{0}$. Here \(\mat u = {(\rho, \rho \mat v, \rho E)}^T\) denotes the state vector of dimension \({n}_{s}\). It is of dimension \(d+2\), where $d$ is the spatial dimension. The inviscid and viscous fluxes \(\mat {F} = \mat{F}(\mat u)\)  and \(\mat{G}=\mat{G}(\mat u, \mat q)\), respectively, are given by:
\begin{subequations} \begin{align}\label{C5}
 \mat {F}(\mat u)=&~ [\rho \mat{v}, \rho \mat{v}\otimes\mat{v}+P\mat{I},\rho \mat{v} H]^T\\
  \mat{G}(\mat{u},\mat{q})= &~-[0,\mat{\tau}, \mat{\tau}\mat{v} - \boldsymbol{\phi}]^T
  \end{align}
\end{subequations}
The shear stress and heat flux take the form:
\begin{subequations}\label{C6}
\begin{align}
\mat {\tau}=&~\frac{1}{{ Re}_{c_\infty} }\left(\mu \left( \nabla \mat{v} + (\nabla\mat{v})^T + \lambda (\nabla\cdot\mat{v}) \mat{I}\right) \right),\\
\boldsymbol{\phi} =&~ - \frac{1}{{ Re}_{c_\infty} {\rm Pr}} \frac{\kappa}{R}\nabla T.
\end{align}
\end{subequations}
where the acoustic Reynolds number and Prandtl number are given by:
\begin{subequations}
    \begin{align}
        { Re}_{c_\infty} =&~ \dfrac{\rho_\infty c_\infty X_0}{\mu_\infty},\\
        {\rm Pr} =&~ \dfrac{{c_p}_\infty \mu_\infty}{\kappa_\infty}.
    \end{align}
\end{subequations}
Finally, the thermodynamical relations are in non-dimensional form as follows: $\gamma P=\rho T$, $e = c_v T$, $c_v = R/(\gamma-1)$ and $E = e + \mat{v}\cdot\mat{v}/2$.}

\subsection{Finite element mesh and spaces on macro-element}\label{S31}
 We denote by \({\mathcal{T}}_{h}\) a collection of disjoint regular elements \(K\) that partition \(\Omega\), and set \(\partial \mathcal{T}_{h} := \{ \partial K : K \in \mathcal{T}_{h} \}\)  to be the collection of the boundaries of the elements in \({\mathcal{T}}_{h}\).
For an element \(K\) of the collection \({\mathcal{T}}_{h}\), \(e =\partial K  \cap \partial \Omega\) is a boundary face if its \(d-1\) Lebesgue measure is nonzero. For two elements \({K}^{+}\) and
\({K}^{-}\) of \({\mathcal{T}}_{h}\),  \(e =\partial {K}^{+}  \cap \partial {K}^{-}\) is the interior face between  \({K}^{+}\)  and  \({K}^{-}\)  if its  \(d-1\) Lebesgue measure is nonzero. We denote by \({\varepsilon}^{{\rm Int}}\) and \({\varepsilon}^{\partial}\) the set of interior and boundary faces, respectively, and we define
 \({\varepsilon}_{h} :={ \varepsilon}^{{\rm Int}} \cup {\varepsilon}^{\partial}\) 
 as the union of interior and boundary faces.

Let the physical domain be the union of $ N^{{\rm ptch}} $ isogeometric patches, i.e.\ such that $ \overline{\Omega}=\cup^{N^{{\rm ptch}}}_{i=1}\overline\Omega_{i}$ and $ \Omega_{i}\cap\Omega_{j}=\varnothing$ for $ i\neq j $. The patches $ \Omega_{i} $ are consisting of a union of elements in $ \mathcal{T}_{h}  $. 
 We also denote by $ \mathcal{T}_{i} $ a union of elements that satisfies $ \mathcal{T}_{i} \subseteq \mathcal{T}_{h}$. The interface
of subdomain $ \Omega_{i} $ is defined as $\Gamma_{i}=\partial\Omega_{i} $ while the collection of patch interfaces $ \Gamma $ is defined as $ \Gamma=\cup^{N^{{\rm ptch}}}_{i=1} \Gamma_{i}$.  {\color{black}For example, Figure \ref{fig:domain} depicts graphically a sample grid partitioned into non-overlapping subdomains. The domain $\Omega$ is partitioned into $N^{{\rm ptch}}$ patches. 
Each patch consists of $N_{i}^{{\rm Elm}} = N^{{\rm Elm}}/N^{{\rm ptch}}$ elements. In this work, we focus on simplicial meshes obtained by splitting each macro-simplex into a regular number of triangles or tetrahedral elements. 

\begin{figure}
    \centering
    \subfloat{{\includegraphics[width=9cm]{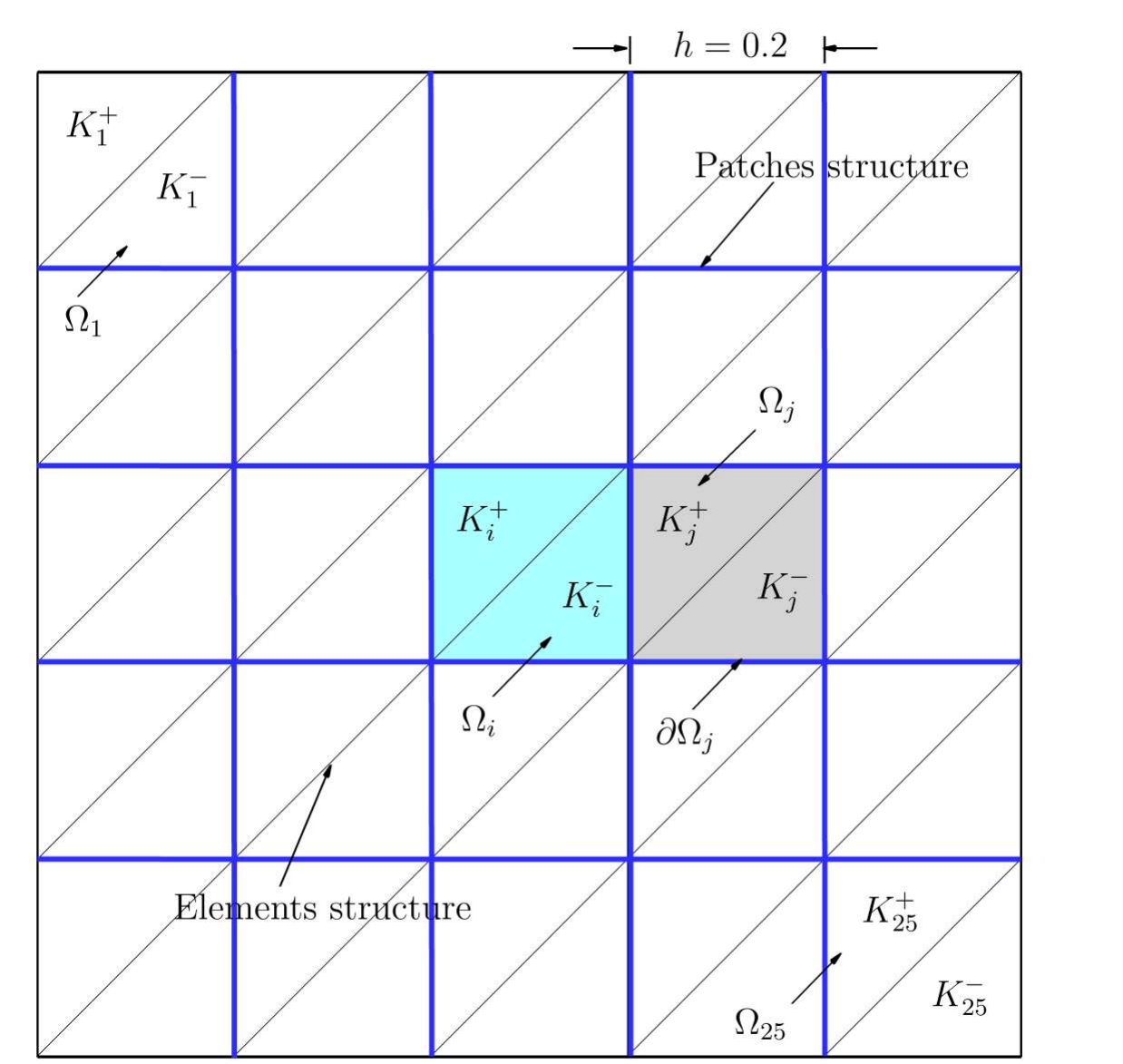} }}%
    \subfloat{{\includegraphics[width=9cm]{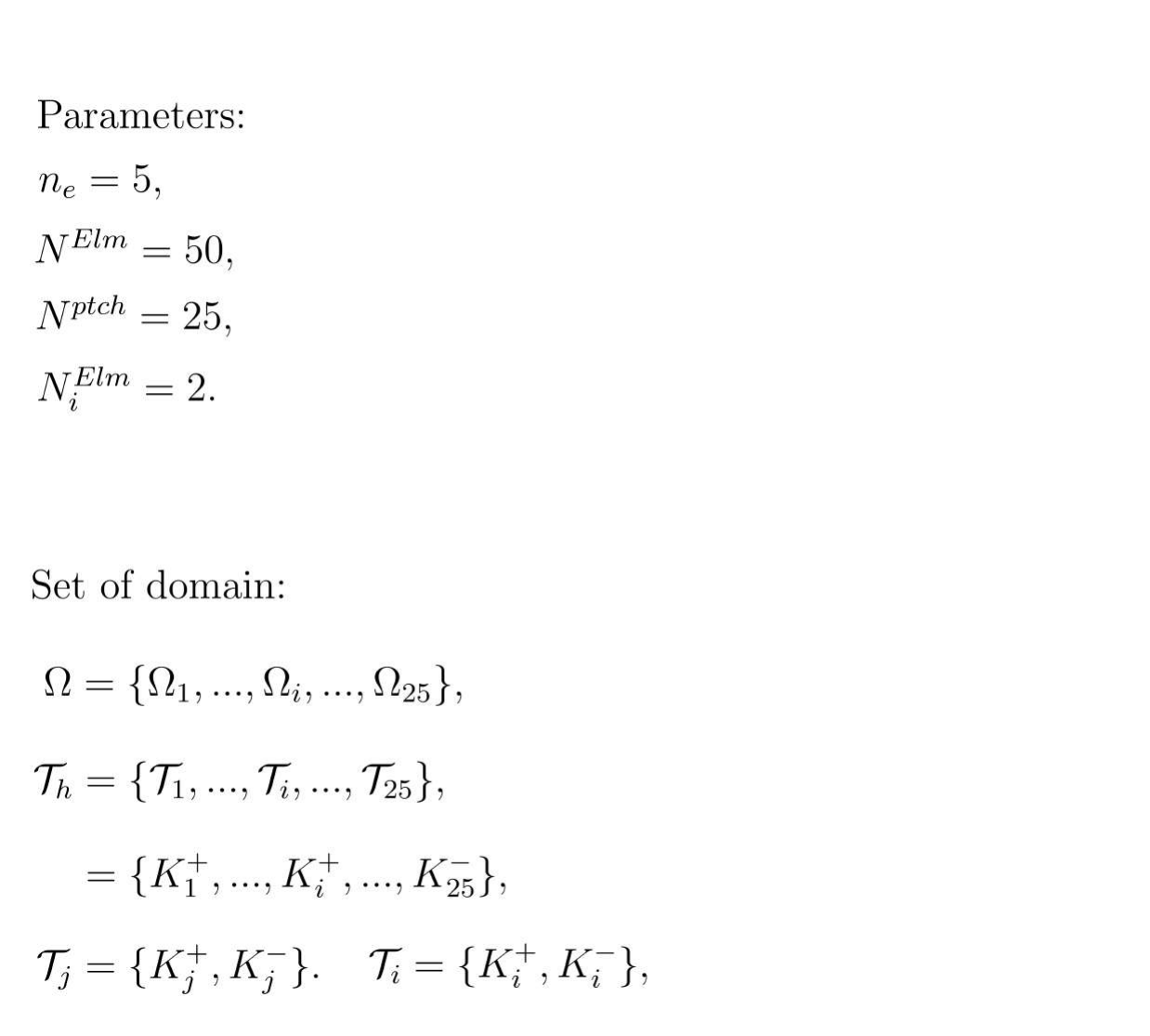} }}%
    \caption{Sample grid partitioned into non-overlapping subdomains.}%
    \label{fig:domain}%
\end{figure}

Let $ \mathcal{P}_{p}(D) $ denote the set of polynomials of degree at most  $ p $ on a domain $ D $ and let $ L^{2}(D) $ be the space of square-integrable functions on $ D $. {\color{black}Our macro-element variant of the HDG method uses patches of standard $C^{0}$ continuous elements that are discontinuous only across patch boundaries. Hence, on macro-elements, we use continuous piece-wise polynomials. We introduce the following finite element spaces:
\begin{subequations}
    \begin{align}
\bm{\mathcal{V}}^{k}_{h} 
 =&~ \lbrace \bm{w} \in [C^{0}(\mathcal{T}_{h})]^{n_{s}} :  
 \bm{w}\mid_{K} \in [\mathcal{P}_{p}(K)]^{n_{s}} \; \;\; \forall K  \in \mathcal{T}_{h} \rbrace,\\
\bm{\mathcal{Q}}^{k}_{h} 
   =&~ \lbrace \bm{r} \in [C^{0}(\mathcal{T}_{h})]^{n_{s}\times d} : 
 \bm{r}\mid_{K} \in [\mathcal{P}_{p}(K)]^{n_{s}\times d} \;\;\; \forall K  \in \mathcal{T}_{h} \rbrace,\\
 \bm{\mathcal{M}}^{k}_{h} =&~ \lbrace \bm{\mu} \in [L^{2}(\varepsilon_{h})]^{n_{s}}  : 
 \bm{\mu}\mid_{e} \in [\mathcal{P}_{p}(e)]^{n_{s}}  \;\;\; \forall e  \in \varepsilon_{h} \rbrace,\\
  \mathcal{M}^{k}_{h} =&~ \lbrace \mu \in L^{2}(\varepsilon_{h})  : 
 \mu \mid_{e} \in \mathcal{P}_{p}(e)  \;\;\; \forall e  \in \varepsilon_{h} \rbrace.
    \end{align}
\end{subequations}}

Next, we define several inner products associated with these finite element spaces. In particular, given $ w,v \in \mathcal{V}^{k}_{h}$, $ \bm{w},\bm{v}\in \bm{\mathcal{V}}^{k}_{h}$ and $ \bm{W},\bm{V}\in \bm{\mathcal{Q}}^{k}_{h} $ we write:
\begin{subequations}\label{C31}
 \begin{align}
(w,v)_{\mathcal{T}_{h}}=&~ \sum_{K\in\mathcal{T}_{h}}(w,v)_{K}= {\color{black}\sum_{K\in\mathcal{T}_{h}}\int_{K} w v_{K}},\\
(\bm{w},\bm{v})_{\mathcal{T}_{h}}=&~ \sum_{i=1}^{n_{s}}(w_{i},v_{i})_{\mathcal{T}_{h}},\\
(\bm{W},\bm{V})_{\mathcal{T}_{h}}=&~\sum_{i=1}^{n_{s}}\sum_{j=1}^{d}(W_{ij},V_{ij})_{\mathcal{T}_{h}}.
 \end{align} 
\end{subequations} 
The corresponding inner product for functions in the trace spaces are given by:
\begin{subequations}\label{C32}
 \begin{align}
\langle\eta,\zeta\rangle_{\partial\mathcal{T}_{h}}=&~ \sum_{K\in\mathcal{T}_{h}}\langle\eta,\zeta\rangle_{\partial K}
= {\color{black}\sum_{K\in\mathcal{T}_{h}}\int_{\partial K} \eta\zeta},\\
\langle\bm{\eta},\bm{\zeta}\rangle_{\partial\mathcal{T}_{h}}=&~ \sum_{i=1}^{n_{s}}\langle\eta_{i},\zeta_{i}\rangle_{\partial\mathcal{T}_{h}},
 \end{align} 
\end{subequations} 
for all $ \eta,\zeta\in \mathcal{M}^{k}_{h} $ and and $ \bm{\eta},\bm{\zeta}\in \bm{\mathcal{M}}^{k}_{h} $.

\subsection{Macro-element discretization}\label{S331}
We seek an approximation $  \left( \mathbf{q}_{h}\left( t\right) ,\mathbf{u}_{h}\left( t\right),\widehat{\mathbf{u}}_{h}\left( t\right)\right)\in \bm{\mathcal{Q}}^{k}_{h} \times \bm{\mathcal{V}}^{k}_{h} \times \bm{\mathcal{M}}^{k}_{h}$ such that {\color{black}
\begin{subequations}\label{C38}
 \begin{align}\label{C38a}
&\left( \mathbf{q}_{h},\bm{r}\right) _{\mathcal{T}_{h}} + 
\left( \mathbf{u}_{h},\nabla\cdot\bm{r}\right) _{\mathcal{T}_{h}} - 
\left\langle   \widehat{\mathbf{u}}_{h} ,\bm{r}\bm{n}\right\rangle  _{\partial \mathcal{T}_{h}}= 0,\\\label{C38b}
&\left( \dfrac{\partial \mathbf{u}_{h}}{\partial t},\bm{w}\right) _{\mathcal{T}_{h}} - 
\left( \mathbf{F}_{h}\left( \mathbf{u}_{h}\right)  + \mathbf{G}_{h}\left( \mathbf{u}_{h},\mathbf{q}_{h}\right) ,\nabla \bm{w}\right) _{\mathcal{T}_{h}} + 
\left\langle \widehat{\mathbf{F}}_{h}+\widehat{\mathbf{G}}_{h},\bm{w}\otimes\bm{n} \right\rangle _{\partial \mathcal{T}_{h}} + 
\sum_{e\in K}\left(\left( \mathbf{A}\cdot \nabla\right) \bm{w}, \tau_{{\rm SUPG}}\mathbf{R} \right) _{\mathcal{T}_{h}}= 0,\\\label{C38c}
&\left\langle \widehat{\mathbf{F}}_{h}+\widehat{\mathbf{G}}_{h},\bm{\mu}\otimes\bm{n}\right\rangle  _{{\partial \mathcal{T}_{h}}\backslash\partial\Omega} 
 + \left\langle   \widehat{\mathbf{B}}_{h}\left( \widehat{\mathbf{u}}_{h},\mathbf{u}_{h},\mathbf{q}_{h}\right) ,\bm{\mu}\otimes\bm{n}\right\rangle  _{\partial\Omega}= 0  , 
\end{align}   
\end{subequations}
for all $ \left( \bm{r},\bm{w},\bm{\mu}\right) \in \bm{\mathcal{Q}}^{k}_{h} \times \bm{\mathcal{V}}^{k}_{h} \times \bm{\mathcal{M}}^{k}_{h}$ and all $ t\in\left( 0,T\right] $.} The boundary trace operator $ \widehat{\mathbf{B}}_{h}\left( \widehat{\mathbf{u}}_{h},\mathbf{u}_{h},\mathbf{q}_{h}\right) $ imposes the boundary conditions along $\partial\Omega  $ exploiting the hybrid variable [11]. 
We take the interior numerical fluxes of the form:
{\color{black}\begin{align}\label{eq: num fluxes}
\left(\widehat{\mathbf{F}}_{h}+\widehat{\mathbf{G}}_{h}\right)\bm{n}= 
\left( \mathbf{F}_{h}\left( \widehat{\mathbf{u}}_{h}\right) + \mathbf{G}_{h}\left( \widehat{\mathbf{u}}_{h},\mathbf{q}_{h}\right) \right) \bm{n} +
 \mathbf{S}\left( \mathbf{u}_{h},\widehat{\mathbf{u}}_{h}\right) \left( \mathbf{u}_{h}-\widehat{\mathbf{u}}_{h}\right)\bm{n}  \;\;\;\; \text{on} \; {\partial \mathcal{T}_{h}},
\end{align}
where $\bm{n}$ denotes the outward unit normal
vector.
The latter member of \eqref{eq: num fluxes} involves the stabilization tensor $\mathbf{S}$ that enhances the stability of the hybridized discontinuous Galerkin method. The inviscid and viscous components of the system are separately stabilized by means of the decomposition \cite{peraire2010hybridizable}:
\begin{subequations}
    \begin{align}
\mathbf{S} =&~ \mathbf{S}^{\rm{inv}} + \mathbf{S}^{\rm{vis}},\\
\mathbf{S}^{\rm{inv}} =&~ \lambda_{{\rm max}}\mathbf{I},\\
\mathbf{S}^{\rm{vis}} =&~ \frac{1}{{Re}}\;\; \text{diag}\left( 0,\Upsilon,1/\left[ (\gamma-1)M_{\infty}^{2}{\rm Pr}\right] \right),
    \end{align}
\end{subequations}
where $ \mathbf{I} $ is the identity matrix, $\Upsilon $ is a ($ n_{s}-2 $)-dimensional vector of ones, and $M_{\infty}$ is the free stream Mach number.
The inviscid stabilization tensor $ \mathbf{S}^{\rm{inv}}$ is a local Lax-Friedrich stabilization in which $ \lambda_{{\rm max}} $ is the maximum absolute eigenvalue of the Jacobian matrix
$\partial\mathbf{F}_h /\partial\widehat{\mathbf{u}}_{h}$.
Finally, the latter term in \eqref{C38b} is a standard residual-based streamline-upwind-Petrov Galerkin (SUPG) stabilization term \cite{brooks1982streamline,donea2003finite} where $\mathbf{R}$ is the local residual of the governing equation \ref{C38b}, $\mathbf{A}$ is the Jacobian of the inviscid flux and $\tau_{{\rm SUPG}}$ is the stabilization matrix. The precise definitions of the $\mathbf{A}$ and $\tau_{{\rm SUPG}}$ are described in \cite{xu2017compressible,shakib1991new,tezduyar2006stabilization,tezduyar2006computation}. This stabilization term is active within the $C^{0}$-macro-elements, and further improves the stability for a wide range of Reynolds and Mach numbers.}

{\color{black}
\subsection{Temporal integration}

We adopt the $s$-stage diagonally implicit Runge-Kutta (DIRK) time-discretization scheme \cite{alexander1977diagonally}. Due to their higher-order accuracy and wide stability range, DIRK methods are widely employed temporal integration schemes for stiff systems. 
We refer the interested reader to comprehensive reviews of higher-order time integration methods suitable for HDG methods \cite{nguyen2012hybridizable,nguyen2011high}. The Butcher’s table associated with the DIRK method can be written as:}
\begin{equation}
\begin{array}{c|cccc}
c_{1} 	&a_{11}		&0			&\dots 		&0		\\
c_{2}    &a_{21}		&a_{22}	&\dots 		&0		\\
\vdots   &\vdots 	&\vdots 	&\ddots 	& \vdots \\
c_{s}    &a_{s1}		&a_{s2}		&\dots 		&a_{ss}\\
\hline
         & b_{1 }	&b_{2 }			&\dots 		&b_{s }	
\end{array} ,
\end{equation}
where we assume the matrix $a_{ij}$ to be non-singular, {\color{black}  $c_{i}$ and $b_{i}$ are numbers that depend on DIRK type \cite{alexander1977diagonally}}. {\color{black}Denoting the time level by $n$, we have $n=s(l-1) + i$, where $s$ is the number of stages, $l$ the current time step, and $i = 1,..., s$ the current stage within the current time step.} Let $d_{ij}$ denote the inverse of $a_{ij}$, and let  $ \left( \mathbf{q}_{h}^{n,i},\mathbf{u}_{h}^{n,i},\widehat{\mathbf{u}}_{h}^{n,i}\right) $ be the intermediate solutions of $ \left( \mathbf{q}_{h}\left( t^{n,i}\right) ,\mathbf{u}_{h}\left( t^{n,i}\right),\widehat{\mathbf{u}}_{h}\left( t^{n,i}\right)\right) $ at the discrete time $t^{n,i}= t_{n} + c_{i}\Delta t^{n}$, where $1\leq i \leq s$. The numerical solution $\mathbf{u}_{h}^{n+1}$ at the time level $n+1$ given by the DIRK method is computed as follows:
{\color{black}\begin{equation}\label{C241}
 \mathbf{u}_{h}^{n+1}=\left(1-\sum_{i=1}^{s} e_j \right)  \mathbf{u}_{h}^{n} + \sum_{j=1}^{s} \mathbf{u}_{h}^{n,j},
\end{equation}}
where  $e_{j}=\sum_{j=1}^{s} b_{i}d_{ij}$. The intermediate solutions are determined as follows: we search for $ \left( \mathbf{q}_{h}^{n,i},\mathbf{u}_{h}^{n,i},\widehat{\mathbf{u}}_{h}^{n,i}\right)\in \bm{\mathcal{Q}}^{k}_{h} \times \bm{\mathcal{V}}^{k}_{h} \times \bm{\mathcal{M}}^{k}_{h} $ such that the following is satisfied:
{\color{black}\begin{subequations}\label{C242}
 \begin{align}
\left( \mathbf{q}_{h}^{n,i},\bm{r}\right) _{\mathcal{T}_{h}} + 
\left( \mathbf{u}_{h}^{n,i},\nabla\cdot\bm{r}\right) _{\mathcal{T}_{h}} - 
\left\langle   \widehat{\mathbf{u}}_{h}^{n,i} ,\bm{r}\bm{n}\right\rangle  _{\partial \mathcal{T}_{h}}&~= 0,\\
\left( \dfrac{\sum_{j=1}^{s}d_{ij}\left(\mathbf{u}_{h}^{n,j}-\mathbf{u}_{h}^{n} \right) }{\Delta t^{n}},\bm{w}\right) _{\mathcal{T}_{h}} - 
\left( \mathbf{F}_{h}^{n,i}  + \mathbf{G}_{h}^{n,i} ,\nabla \bm{w}\right) _{\mathcal{T}_{h}} + 
\left\langle  \widehat{\mathbf{F}}_{h}^{n,i}+\widehat{\mathbf{G}}_{h}^{n,i},\bm{w}\otimes \mathbf{n}\right\rangle _{\partial \mathcal{T}_{h}} + 
\sum_{e\in K}\left(\left( \mathbf{A}^{n,i}\cdot \nabla\right) \bm{w}, \tau_{{\rm SUPG}}^{n,i}\mathbf{R}^{n,i} \right) _{\mathcal{T}_{h}}&~= 0,\\
\left\langle \widehat{\mathbf{F}}_{h}^{n,i}+\widehat{\mathbf{G}}_{h}^{n,i},\bm{\mu}\otimes \mathbf{n} \right\rangle  _{{\partial \mathcal{T}_{h}}\backslash\partial\Omega} 
 + \left\langle   \widehat{\mathbf{B}}_{h}^{n,i}\left( \widehat{\mathbf{u}}_{h}^{n,i},\mathbf{u}_{h}^{n,i},\mathbf{q}_{h}^{n,i}\right) ,\bm{\mu}\otimes \mathbf{n} \right\rangle  _{\partial\Omega}&~= 0,
\end{align}   
\end{subequations}}
for all $ \left( \bm{r},\bm{w},\bm{\mu}\right) \in \bm{\mathcal{Q}}^{k}_{h} \times \bm{\mathcal{V}}^{k}_{h} \times \bm{\mathcal{M}}^{k}_{h}$. Once  $ \mathbf{u}_{h}^{n+1}$ has been determined as above, we search for $ \left( \mathbf{q}_{h}^{n+1},\widehat{\mathbf{u}}_{h}^{n+1}\right)\in \bm{\mathcal{Q}}^{k}_{h}  \times \bm{\mathcal{M}}^{k}_{h} $ such that 
{\color{black}\begin{subequations}\label{C243}
 \begin{align}
\left( \mathbf{q}_{h}^{n+1},\bm{r}\right) _{\mathcal{T}_{h}} + 
\left( \mathbf{u}_{h}^{n+1},\nabla\cdot\bm{r}\right) _{\mathcal{T}_{h}} - 
\left\langle   \widehat{\mathbf{u}}_{h}^{n+1} ,\bm{r}\bm{n}\right\rangle  _{\partial \mathcal{T}_{h}}&~= 0,\\
\left\langle \widehat{\mathbf{F}}_{h}^{n+1}+\widehat{\mathbf{G}}_{h}^{n+1},\bm{\mu}\otimes \mathbf{n} \right\rangle  _{{\partial \mathcal{T}_{h}}\backslash\partial\Omega} 
 + \left\langle   \widehat{\mathbf{B}}_{h}^{n+1}\left( \widehat{\mathbf{u}}_{h}^{n+1},\mathbf{u}_{h}^{n+1},\mathbf{q}_{h}^{n+1}\right) ,\bm{\mu}\otimes \mathbf{n} \right\rangle  _{\partial\Omega}&~= 0.
\end{align}   
\end{subequations}}
for all $ \left( \bm{r},\bm{w},\bm{\mu}\right) \in \bm{\mathcal{Q}}^{k}_{h} \times \bm{\mathcal{V}}^{k}_{h} \times \bm{\mathcal{M}}^{k}_{h}$.

{\color{black}\begin{remark}
The system \eqref{C242} can be advanced in time without solving \eqref{C243}. Hence, in practice we only solve \eqref{C243} at the time steps that we need $ \left( \mathbf{q}_{h}^{n+1},\widehat{\mathbf{u}}_{h}^{n+1}\right) $ for post-processing purposes. Finally, it is worth mentioning that certain specific DIRK schemes, such as the strongly $s$-stable DIRK (2, 2) and DIRK (3, 3) schemes, have the unique property that $c_{s}=1$. As a consequence, \eqref{C243} becomes identical to \eqref{C242} at final stage $i=s$. As a result, these particular DIRK schemes do not require the solution of equation \eqref{C243}.
\end{remark}}

\renewcommand{\arraystretch}{1.3}
\section{Parallel iterative solvers} \label{sec:Solution Method}


{\color{black}We construct parallel iterative methods for the solution of the nonlinear system of equations \eqref{C242}. First, in Section \ref{subsec: newton} we linearize the nonlinear global problem by means of an inexact Newton method. Next, we propose two different options for the solution of the linear system; standard static condensation in Section \ref{subsubsec: static condensation}, and an alternative second-layer static condensation approach in Section \ref{subsubsec: alternative static condensation}.}

\subsection{Nonlinear solver: inexact Newton method}\label{subsec: newton}

{\color{black}At any given (sub) time step $n$, the nonlinear system of equation \eqref{C242} can be written as
\begin{subequations}\label{C31221}
 \begin{align}
{R}_{Q}(\mathbf{q}_{h}^{n},\mathbf{u}_{h}^{n},\widehat{\mathbf{u}}_{h}^{n}) &= 0,\\
{R}_{U}(\mathbf{q}_{h}^{n},\mathbf{u}_{h}^{n},\widehat{\mathbf{u}}_{h}^{n}) &= 0,\\
{R}_{\widehat{U}}(\mathbf{q}_{h}^{n},\mathbf{u}_{h}^{n},\widehat{\mathbf{u}}_{h}^{n}) &=0,
\end{align}   
\end{subequations}
where ${R}_{Q}$, ${R}_{U}$ and ${R}_{\widehat{U}}$ are the discrete nonlinear residuals associated to equations \(\ref{C38a}\),   \(\ref{C38b}\) and  \(\ref{C38c}\), respectively.}

To address the nonlinear system \eqref{C31221}, we use pseudo-transient continuation \cite{bijl2002implicit,jameson1991time}, which is an inexact Newton method. The procedure requires an adaptation algorithm of the pseudo time step size to complete the method. In this study, we employ the successive evolution relaxation (SER) algorithm \cite{mulder1985experiments} with the following parameters:
\begin{equation}\label{C31222}
\Delta\tau^{0}= \tau_{init}, \;\;\Delta\tau^{m+1} = \text{min} \left(\Delta\tau^{m} \dfrac{\Vert {R}_{U}\Vert_{L^{2}}^{m+1}}{\Vert {R}_{U} \Vert_{L^{2}}^{m}} ,\tau_{max} \right) .
\end{equation}
Here, $m$ is the iteration step for the pseudo-transient continuation. In this study, if not otherwise specified, $\tau_{init} = 1.0$ and $ \tau_{max} = 10^{8}$. {\color{black}By linearizing} \eqref{C31221} with respect to the solution  
$ \left( \mathbf{q}_{h}^{m,n} ,\mathbf{u}_{h}^{m,n},\widehat{\mathbf{u}}_{h}^{m,n} \right) $ 
at the Newton step $m=0, 1, ...,$ we obtain the subsequent linear system:
\begin{equation}\label{C312}
\left[ 
 \begin{array}{c | c}
\begin{matrix}  
\mathbf{A}_{qq}^{m,n} & \mathbf{A}_{qu}^{m,n}  \\  
\mathbf{A}_{uq}^{m,n} & \mathbf{A}_{uu}^{m,n} \end{matrix}   &   
\begin{matrix}  \mathbf{A}_{q \widehat{u}}^{m,n} \\ 
\mathbf{A}_{u \widehat{u}}^{m,n} \end{matrix} \\
\hline
\begin{matrix} 
\mathbf{A}_{\widehat{u} q}^{m,n}  & \mathbf{A}_{\widehat{u} u}^{m,n} \end{matrix} & 
\begin{matrix}  \mathbf{A}_{\widehat{u}  \widehat{u}}^{m,n} \end{matrix} 
\end{array}
\right]
 \left[ 
\begin{array}{c}
\begin{matrix}  
\Delta {Q}^{m,n} \\  
\Delta {U}^{m,n} \end{matrix} \\ 
\hline
\begin{matrix} \Delta \widehat{{U}}^{m,n} \end{matrix}
\end{array}
\right]=-
 \left[ 
\begin{array}{c}
\begin{matrix}  {R}_{Q}^{m,n} \\  {R}_{U}^{m,n} \end{matrix} \\ 
\hline
\begin{matrix} {R}_{\widehat{U}}^{m,n} \end{matrix}
\end{array}
\right] \; ,
\end{equation}
where $\Delta {Q}^{m,n}$, $\Delta {U}^{m,n}$ and $\Delta \widehat{{U}}^{m,n}$ are the update of the vector of degrees of freedom of the discrete field solutions $\mathbf{q}_{h}^{m,n}$, $\mathbf{u}_{h}^{m,n}$ and $\widehat{\mathbf{u}}_{h}^{m,n}$, respectively. The next Newton update of these solution fields is defined as 
\begin{align}
\left( \mathbf{q}_{h}^{m+1,n} ,\mathbf{u}_{h}^{m+1,n},\widehat{\mathbf{u}}_{h}^{m+1,n} \right) :=
 \left( \mathbf{q}_{h}^{m,n} ,\mathbf{u}_{h}^{m,n},\widehat{\mathbf{u}}_{h}^{m,n} \right)+
 \left(\Delta \mathbf{q}^{m,n},\Delta \mathbf{u}^{m,n}, \Delta \widehat{\mathbf{u}}^{m,n}\right). 
\end{align}
Newton iterations are repeated until the norm of the full residual vector ${R}_{F} := \left( {R}_{Q},{R}_{U},{R}_{\widehat{U}}\right) $ is smaller than a specified tolerance.

{\color{black}\subsection{Linear solver:  static condensation}\label{subsubsec: static condensation}


The first method that we discuss for the solution of the linear system of equations \eqref{C312} is static condensation. First we discuss the system of equations for each macro element, and subsequently the global system. {\color{black} Eliminating both $\Delta {Q}$ and $\Delta {U}$ in an element-by-element fashion, as mentioned earlier, can be achieved using the first two equations in \eqref{C312}. As a result, we compute for each macro-element $\mathcal{T}_{i}$ the solution updates $\Delta {Q}^{\mathcal{T}_{i}}$ and $\Delta {U}^{\mathcal{T}_{i}}$ as}
\begin{equation}\label{C46}
 \begin{bmatrix} 
 \Delta {Q}^{\mathcal{T}_{i}} \\[7pt] 
 \Delta {U}^{\mathcal{T}_{i}}
\end{bmatrix} = 
\left( \mathbf{A}_{\text{local}}^{\mathcal{T}_{i}}\right) ^{-1}\left(-\begin{bmatrix*}  
 {R}_{Q}^{\mathcal{T}_{i}} \\[7pt]   
 {R}_{U}^{\mathcal{T}_{i}}  
\end{bmatrix*}-
\begin{bmatrix*}  
 \mathbf{A}_{q \widehat{u}}^{\mathcal{T}_{i}} \\[7pt]   
 \mathbf{A}_{u \widehat{u}}^{\mathcal{T}_{i}} 
\end{bmatrix*}
\Delta \widehat{{U}}^{\Gamma_{i}}\right) \\[7pt]
\end{equation}
{\color{black}where the block structured local matrices are given by:}
\begin{equation}\label{AP1}
\mathbf{A}_{\text{local}}^{\mathcal{T}_{i}}
= 
\begin{bmatrix}  
\mathbf{A}_{qq}^{\mathcal{T}_{i}} & \mathbf{A}_{qu}^{\mathcal{T}_{i}}  \\[7pt]   
\mathbf{A}_{uq}^{\mathcal{T}_{i}} & \mathbf{A}_{uu} ^{\mathcal{T}_{i}}.
\end{bmatrix}
\end{equation}

{\color{black}Next, we consider the global system of equations. We note that the matrix  $[\mathbf{A}_{qq}  \mathbf{A}_{qu}; \mathbf{A}_{uq}  \mathbf{A}_{uu}]$ has a block-diagonal structure. This permits expressing $\Delta {U}$ and $\Delta {Q}$ in terms of $\Delta \widehat{{U}}$. By eliminating $\Delta {Q}$ and $\Delta {U}$ from  \eqref{C312}, we obtain the globally coupled reduced system of linear equations:}
\begin{equation}\label{C44}
\widehat{\mathbf{A}}\;
\Delta \widehat{{U}}
=
\widehat{\mathbf{b}},
\end{equation}
{\color{black}which has to be solved in every Newton iteration. 
The macro-element contributions to the global reduced system are given by:}
\begin{subequations}\label{C47}
\begin{align}
&\widehat{\mathbf{A}}^{\mathcal{T}_{i}} = 
\mathbf{A}_{\widehat{u}  \widehat{u}}^{\mathcal{T}_{i}} -
\begin{bmatrix*}  
\mathbf{A}_{\widehat{u}q}^{\mathcal{T}_{i}} & \mathbf{A}_{\widehat{u}u}^{\mathcal{T}_{i}}
\end{bmatrix*} 
\left( \mathbf{A}_{\text{local}}^{\mathcal{T}_{i}}\right) ^{-1}
\begin{bmatrix*}  
\mathbf{A}_{q \widehat{u}}^{\mathcal{T}_{i}} \\[7pt]
 \mathbf{A}_{u \widehat{u}}^{\mathcal{T}_{i}}
\end{bmatrix*},\\[7pt]
&\widehat{\mathbf{b}}^{\mathcal{T}_{i}}= 
-{R}_{\widehat{U}}^{\mathcal{T}_{i}} +
\begin{bmatrix*}  
\mathbf{A}_{\widehat{u}q}^{\mathcal{T}_{i}} & \mathbf{A}_{\widehat{u}u}^{\mathcal{T}_{i}}
\end{bmatrix*} 
\left( \mathbf{A}_{\text{local}}^{\mathcal{T}_{i}}\right) ^{-1}
\begin{bmatrix*}  
 {R}_{Q}^{\mathcal{T}_{i}} \\[7pt] 
 {R}_{U}^{\mathcal{T}_{i}}
\end{bmatrix*}.
\end{align}
\end{subequations}
{\color{black}As a result of the single-valued trace quantities $\widehat{\mathbf{u}}_{h}$, the final matrix system of the HDG method is smaller than that of many other DG methods \cite{peraire2010hybridizable, cockburn2016static, nguyen2015class}. Moreover, the matrix $\widehat{\mathbf{A}}$ has a small bandwidth since solely the degrees of freedom between neighboring faces that share the same macro-element are connected \cite{peraire2010hybridizable}.

\begin{remark}
We note that the local vector updates $\Delta {Q}^{\mathcal{T}_{i}}$ and 
$\Delta {U}^{\mathcal{T}_{i}}$, and the global reduced system (\ref{C44}) need to be stored for each macro-element. 
\end{remark}}

{\color{black}\subsection{Linear solver: second-layer static condensation}\label{subsubsec: alternative static condensation}

We propose an an alternative to the static condensation strategy described in Section \ref{subsubsec: static condensation}, which we refer to as second-layer static condensation in the following. To this end, we start by exploiting the structure of the local block matrices, which read:}
\begin{equation}\label{AP1}
\mathbf{A}_{\text{local}}^{\mathcal{T}_{i}}
= 
\begin{bmatrix}  
\mathbf{A}_{qq}^{\mathcal{T}_{i}} & \mathbf{A}_{qu}^{\mathcal{T}_{i}}  \\[7pt]   
\mathbf{A}_{uq}^{\mathcal{T}_{i}} & \mathbf{A}_{uu} ^{\mathcal{T}_{i}}  
\end{bmatrix}=
\left[ 
 \begin{array}{c | c}
\begin{matrix}  
\mathbf{A}_{q_{x}q_{x}}^{\mathcal{T}_{i}} &0								   			&0											 \\[7pt]   
0								 		 &\mathbf{A}_{q_{y}q_{y}}^{\mathcal{T}_{i}}	&0  											 \\[7pt]
0								 		 &0											&\mathbf{A}_{q_{z}q_{z}}^{\mathcal{T}_{i}}	\\[7pt]
\end{matrix}   &   
\begin{matrix} 
\mathbf{A}_{q_{x}u}^{\mathcal{T}_{i}}  \\[7pt]
\mathbf{A}_{q_{y}u}^{\mathcal{T}_{i}} \\[7pt]
\mathbf{A}_{q_{y}u}^{\mathcal{T}_{i}}
\end{matrix} \\[7pt]
\hline\\ [-1.3ex]
\begin{matrix} 
\mathbf{A}_{uq_{x}}^{\mathcal{T}_{i}} 	 &\mathbf{A}_{uq_{y}}^{\mathcal{T}_{i}}	    &\mathbf{A}_{uq_{z}}^{\mathcal{T}_{i}}
\end{matrix} & 
\begin{matrix}  
\mathbf{A}_{uu} ^{\mathcal{T}_{i}}
\end{matrix} 
\end{array}
\right],
\end{equation}
{\color{black}The form of \eqref{AP1} permits an efficient storage and inversion strategy. Namely the inverse of the local matrix is given by:}
\begin{equation}\label{AP12}
\left( \mathbf{A}_{\text{local}}^{\mathcal{T}_{i}}\right) ^{-1}
= 
\begin{bmatrix}  
\left( \mathbf{A}_{qq}^{\mathcal{T}_{i}}\right) ^{-1} + \left( \mathbf{A}_{qq}^{\mathcal{T}_{i}}\right) ^{-1} \mathbf{A}_{qu}^{\mathcal{T}_{i}} \left(\mathbf{S}^{\mathcal{T}_{i}}\right)^{-1} \mathbf{A}_{uq}^{\mathcal{T}_{i}} \left(\mathbf{A}_{qq}^{\mathcal{T}_{i}}\right)^{-1} & - \left(\mathbf{A}_{qq}^{\mathcal{T}_{i}}\right)^{-1} \mathbf{A}_{qu}^{\mathcal{T}_{i}} \left(\mathbf{S}^{\mathcal{T}_{i}}\right)^{-1} \\[7pt]   
- \left(\mathbf{S}^{\mathcal{T}_{i}}\right)^{-1} \mathbf{A}_{uq}^{\mathcal{T}_{i}} \left(\mathbf{A}_{qq}^{\mathcal{T}_{i}}\right)^{-1} &\left(\mathbf{S}^{\mathcal{T}_{i}}\right)^{-1}
\end{bmatrix},
\end{equation}
{\color{black}where 
$\mathbf{S}^{\mathcal{T}_{i}} = \mathbf{A}_{uu}^{\mathcal{T}_{i}}-\mathbf{A}_{uq}^{\mathcal{T}_{i}}\left(\mathbf{A}_{qq}^{\mathcal{T}_{i}}\right)^{-1}\mathbf{A}_{qu}^{\mathcal{T}_{i}}$ is the Schur complement of $\mathbf{A}_{\text{local}}^{\mathcal{T}_{i}}$. It is needless store the complete dense inverse $\left( \mathbf{A}_{\text{local}}^{\mathcal{T}_{i}}\right) ^{-1}$. We observe that the inverse contains the inverse of the Schur complement $\mathbf{S}^{\mathcal{T}_{i}} $ as well as the inverse of $\mathbf{A}_{qq}^{\mathcal{T}_{i}}$. We compute and store the Schur complement $\left(\mathbf{S}^{\mathcal{T}_{i}}\right)^{-1}$ on each macro-element, which is the same size as $\mathbf{A}_{uu}^{\mathcal{T}_{i}}$. }
{\color{black}The block matrix $\mathbf{A}_{qq}^{\mathcal{T}_{i}}$ consists of the components $\mathbf{A}_{q_{k}q_{k}}^{\mathcal{T}_{i}} = \text{diag}\left(J^{\mathcal{T}_i}\right) \mathbf{M}$, where $k$ is the coordinate direction, $\mathbf{M}$ is the mass matrix over a reference macro-element, and $J^{\mathcal{T}_i}$ is the determinant of the Jacobian of the geometrical map. Its inverse is given by: 
\[\left(\mathbf{A}_{q_{k}q_{k}}^{\mathcal{T}_{i}}\right) ^{-1} = \text{diag}\left(1/J^{\mathcal{T}_i}\right) \mathbf{M}^{-1},\]
where $\mathbf{M}^{-1}$ is the inverse mass matrix on a reference macro-element. The inverse mass matrix may be precomputed and stored}. {\color{black} Substituting $\mathbf{A}_{\text{local}}^{\mathcal{T}_{i}} $ from of \eqref{AP12} into equation \eqref{C47}, the contributions to the global system take the form}:
\begin{subequations}\label{ASC2}
\begin{align}\label{ASC2a}
\widehat{\mathbf{A}}^{\mathcal{T}_{i}} &= 
\left(\mathbf{A}_{\widehat{u}u}^{\mathcal{T}_{i}}-\mathbf{A}_{\widehat{u}q}^{\mathcal{T}_{i}}\left(\mathbf{A}_{qq}^{\mathcal{T}_{i}}\right) ^{-1}\mathbf{A}_{qu}^{\mathcal{T}_{i}}\right)\left(\mathbf{S}^{\mathcal{T}_{i}}\right) ^{-1}\left(-\mathbf{A}_{u\widehat{u}}^{\mathcal{T}_{i}}+\mathbf{A}_{uq}^{\mathcal{T}_{i}}\left(\mathbf{A}_{qq}^{\mathcal{T}_{i}}\right)^{-1}\mathbf{A}_{q\widehat{u}}^{\mathcal{T}_{i}}\right)\nonumber\\ &+ \left(\mathbf{A}_{\widehat{u}\widehat{u}}^{\mathcal{T}_{i}}-\mathbf{A}_{\widehat{u}q}^{\mathcal{T}_{i}}
\left(\mathbf{A}_{qq}^{\mathcal{T}_{i}}\right)^{-1}\mathbf{A}_{q\widehat{u}}^{\mathcal{T}_{i}}\right),\\
\widehat{\mathbf{b}}^{\mathcal{T}_{i}} &= 
\left(-\mathbf{A}_{\widehat{u}u}^{\mathcal{T}_{i}}+\mathbf{A}_{\widehat{u}q}^{\mathcal{T}_{i}}\left(\mathbf{A}_{qq}^{\mathcal{T}_{i}}\right) ^{-1}\mathbf{A}_{qu}^{\mathcal{T}_{i}}\right)\left(\mathbf{S}^{\mathcal{T}_{i}}\right) ^{-1}
\left(-{R}_{U}^{\mathcal{T}_{i}}+\mathbf{A}_{uq}^{\mathcal{T}_{i}}\left(\mathbf{A}_{qq}^{\mathcal{T}_{i}}\right)^{-1}{R}_{Q}^{\mathcal{T}_{i}}\right)\nonumber \\ 
&+ \left(-{R}_{\widehat{U}}^{\mathcal{T}_{i}}+\mathbf{A}_{\widehat{u}q}^{\mathcal{T}_{i}}\left(\mathbf{A}_{qq}^{\mathcal{T}_{i}}\right)^{-1}{R}_{Q}^{\mathcal{T}_{i}}\right).
\end{align}
\end{subequations}
{\color{black}Finally, the local solution updates $\Delta {Q}^{\mathcal{T}_{i}}$ and $\Delta {U}^{\mathcal{T}_{i}}$ are given by:}
\begin{subequations}\label{Csc4}
\begin{align} \label{Csc4a}
  \Delta {U}^{\mathcal{T}_{i}} &=\left(\mathbf{S}^{\mathcal{T}_{i}}\right)^{-1}\left[\left( \mathbf{A}_{uq}^{\mathcal{T}_{i}}\left(\mathbf{A}_{qq}^{\mathcal{T}_{i}}\right) ^{-1}{R}_{Q}^{\mathcal{T}_{i}}-{R}_{U}^{\mathcal{T}_{i}}\right) +\left(-\mathbf{A}_{u\widehat{u}}^{\mathcal{T}_{i}}\:  \Delta \widehat{{U}}^{\Gamma_{i}}+\mathbf{A}_{uq}^{\mathcal{T}_{i}}\left(\mathbf{A}_{qq}^{\mathcal{T}_{i}}\right) ^{-1}\mathbf{A}_{q\widehat{u}}^{\mathcal{T}_{i}} \:  \Delta \widehat{{U}}^{\Gamma_{i}}\right) \right],\\\label{Csc4b}
     \Delta {Q}^{\mathcal{T}_{i}} &= \left(\mathbf{A}_{qq}^{\mathcal{T}_{i}}\right)^{-1}\left(-{R}_{Q}^{\mathcal{T}_{i}}
 -\mathbf{A}_{qu}^{\mathcal{T}_{i}} \:   \Delta {U}^{\mathcal{T}_{i}}- \mathbf{A}_{q \widehat{u}}^{\mathcal{T}_{i}} \:   \Delta \widehat{{U}}^{\Gamma_{i}}\right).
\end{align} 
\end{subequations}

{\color{black}\begin{remark}
We do not explicitly compute the inverse of the Schur complement $\mathbf{S}^{-1}$, but instead compute an appropriate factorization that we store, and then apply the inverse to a vector $\mathbf{a}$ by solving the system $\mathbf{S}\mathbf{x} = \mathbf{a}$ via back-substitution.
\end{remark}}
\section{Implementation aspects}

In the following, we provide details of a matrix-free implementation and a brief description of the preconditioning approach, which we will use in the computational study thereafter.

 \subsection{Matrix-free implementation}
{\color{black}We apply a straightforward matrix-free parallel implementation of the macro-element HDG method, for the both linear solvers presented in Section \ref{subsubsec: static condensation} and Section \ref{subsubsec: alternative static condensation}. We provide the details for the second-layer static condensation of Section \ref{subsubsec: alternative static condensation} in Algorithm 1, and note that the static condensation of Section \ref{subsubsec: alternative static condensation} follows similarly. We provide a few core details of the algorithm for the sake of clarity.} In Lines 2-5 the vector contributions of each macro-element local to process $n$ are assembled in a global vector. Due to the discontinuous nature of macro-elements, this procedure requires 
only data from the macro-element local to each process, $\mathcal{T}_{h}^{n}$, and hence implies no communication between processes. 
{\color{black}The global linear system solve in Line 6 is carried out via a matrix-free iterative procedure such as GMRES, relying on efficient matrix-vector products. As the globally coupled system is potentially very large, the matrix-vector products are carried out in a matrix-free fashion.} {\color{black}Finally, $\Delta {Q}^{\mathcal{T}{i}}$ and $\Delta {U}^{\mathcal{T}{i}}$ are obtained from $\Delta \widehat{{U}}^{\Gamma_{i}}$ in Lines 7-10, where $\Delta {U}^{\mathcal{T}{i}}$ is derived from equation \eqref{Csc4a}, and $\Delta {Q}^{\mathcal{T}{i}}$ is derived from equation \eqref{Csc4b}}. We note that the procedures local to each macro-element can, but do not have to be implemented in a matrix-free fashion, as the corresponding matrices are comparatively small.
 \begin{algorithm}
\caption{Solution procedure on each macro-element associated with one parallel process}
 \begin{algorithmic}[1]
\State $ n \gets$ Current process

\For {$\mathcal{T}_{i} \in \mathcal{T}_{h}^{n}$}
\State $\widehat{\mathbf{b}}^{\mathcal{T}_{i}}   \gets$ 
Vector contribution from ${Q}^{\mathcal{T}_{i}}$, ${U}^{\mathcal{T}_{i}}$, $\widehat{{U}}^{\Gamma_{i}}$
\State $\widehat{\mathbf{b}}  \gets$ 
Assemble $\widehat{\mathbf{b}}^{\mathcal{T}_{i}}$ for all $\mathcal{T}_{i} \in \mathcal{T}_{h}^{n}$
\EndFor

\State $   \Delta \widehat{{U}} \gets$ Matrix-free iterative solve $\widehat{\mathbf{A}}\;   \Delta \widehat{{U}}=\widehat{\mathbf{b}}$
\For {$\mathcal{T}_{i} \in \mathcal{T}_{h}^{n}$}
\State $   \Delta {Q}^{\mathcal{T}_{i}} \gets \left(\mathbf{A}_{qq}^{\mathcal{T}_{i}}\right)^{-1}\left(-{R}_{Q}^{\mathcal{T}_{i}}
 -\mathbf{A}_{qu}^{\mathcal{T}_{i}} \:   \Delta {U}^{\mathcal{T}_{i}}- \mathbf{A}_{q \widehat{u}}^{\mathcal{T}_{i}} \:   \Delta \widehat{{U}}^{\Gamma_{i}}\right).$
\State $    \Delta {U}^{\mathcal{T}_{i}} \gets \left(\mathbf{S}^{\mathcal{T}_{i}}\right)^{-1}\left[\left( \mathbf{A}_{uq}^{\mathcal{T}_{i}}\left(\mathbf{A}_{qq}^{\mathcal{T}_{i}}\right) ^{-1}{R}_{Q}^{\mathcal{T}_{i}}-{R}_{U}^{\mathcal{T}_{i}}\right) +\left(-\mathbf{A}_{u\widehat{u}}^{\mathcal{T}_{i}}\:  \Delta \widehat{{U}}^{\Gamma_{i}}+\mathbf{A}_{uq}^{\mathcal{T}_{i}}\left(\mathbf{A}_{qq}^{\mathcal{T}_{i}}\right) ^{-1}\mathbf{A}_{q\widehat{u}}^{\mathcal{T}_{i}} \:  \Delta \widehat{{U}}^{\Gamma_{i}}\right) \right].$ 

\EndFor
\end{algorithmic}
 \end{algorithm}

Algorithm 2 outlines an efficient matrix-free matrix-vector product, {\color{black} equation \eqref{ASC2a}}. The vector $\Delta \widehat{{U}}$ contains all degrees of freedom on the macro-element interfaces $\varepsilon$. The degrees of freedom on each interior interface $e\in \varepsilon$ will be operated on by $\widehat{\mathbf{A}}^{\mathcal{T}_{+}}$ and $\widehat{\mathbf{A}}^{\mathcal{T}_{-}}$. {\color{black}We adopt the  notation $\mathcal{T}_{\pm}$ to denote the macro-elements on the left and right side of an interface.} 
{\color{black}Lines 6-8 constitute the iterations of the matrix-free algorithm that are performed for each macro-element, $\mathcal{T}_{i}$,  in parallel, in the following four steps:}
\begin{enumerate}
	\item $\mathbf{y}_{1}=\left(-\mathbf{A}_{u\widehat{u}}^{\mathcal{T}_{i}}\: \Delta \widehat{{U}}^{\Gamma_{i}}+\mathbf{A}_{uq}^{\mathcal{T}_{i}}\left(\mathbf{A}_{qq}^{\mathcal{T}_{i}}\right)^{-1}\mathbf{A}_{q\widehat{u}}^{\mathcal{T}_{i}}\: \Delta \widehat{{U}}^{\Gamma_{i}}\right),$
	\item $\mathbf{y}_{2}= \left(\mathbf{S}^{\mathcal{T}_{i}}\right) ^{-1} \mathbf{y}_{1},$ 
	\item $ \mathbf{y}_{3}= \left(\mathbf{A}_{\widehat{u}u}^{\mathcal{T}_{i}}-\mathbf{A}_{\widehat{u}q}^{\mathcal{T}_{i}}\left(\mathbf{A}_{qq}^{\mathcal{T}_{i}}\right) ^{-1}\mathbf{A}_{qu}^{\mathcal{T}_{i}}\right) \mathbf{y}_{2}.$
	\item $ \mathbf{y}_{4}=   \mathbf{y}_{3}+\left(\mathbf{A}_{\widehat{u}\widehat{u}}^{\mathcal{T}_{i}}\: \Delta \widehat{{U}}^{\Gamma_{i}}-\mathbf{A}_{\widehat{u}q}^{\mathcal{T}_{i}}
\left(\mathbf{A}_{qq}^{\mathcal{T}_{i}}\right)^{-1}\mathbf{A}_{q\widehat{u}}^{\mathcal{T}_{i}}\: \Delta \widehat{{U}}^{\Gamma_{i}}\right) $
\end{enumerate}

{\color{black}Steps 1, 3, and 4 act on the global system, whereas step 2 operates on the local system. Step 4} involves a reduction that relies on the data associated with the mesh skeleton faces and data from macro-elements, and thus requires communication among processors.

\begin{algorithm}
\caption{Distributed matrix-free procedure for the matrix-vector product $\widehat{\mathbf{A}}\; \Delta \widehat{{U}}$}
 \begin{algorithmic}[1]
\State $ \mathbf{y} \gets 0$
\State $ n \gets$ Current processor

\For {$\mathcal{T}_{i} \in \mathcal{T}_{h}^{n}$}
\State $\varepsilon \gets$ 
Extract global DOF indices on $ \partial \mathcal{T}_{i}$
\State $ \Delta \widehat{{U}}^{\Gamma_{i}} \gets  \Delta \widehat{{U}}\left[ \varepsilon \right] $ 
\State $ \mathbf{y}\left[ \varepsilon \right] \gets \mathbf{y}\left[ \varepsilon \right]$
\State $+\left(\mathbf{A}_{\widehat{u}u}^{\mathcal{T}_{i}}-\mathbf{A}_{\widehat{u}q}^{\mathcal{T}_{i}}\left(\mathbf{A}_{qq}^{\mathcal{T}_{i}}\right) ^{-1}\mathbf{A}_{qu}^{\mathcal{T}_{i}}\right)\left(\mathbf{S}^{\mathcal{T}_{i}}\right) ^{-1}
\left(-\mathbf{A}_{u\widehat{u}}^{\mathcal{T}_{i}}\: \Delta \widehat{{U}}^{\Gamma_{i}}+\mathbf{A}_{uq}^{\mathcal{T}_{i}}\left(\mathbf{A}_{qq}^{\mathcal{T}_{i}}\right)^{-1}\mathbf{A}_{q\widehat{u}}^{\mathcal{T}_{i}}\: \Delta \widehat{{U}}^{\Gamma_{i}}\right)$
\State $+ \left(\mathbf{A}_{\widehat{u}\widehat{u}}^{\mathcal{T}_{i}}\: \Delta \widehat{{U}}^{\Gamma_{i}}-\mathbf{A}_{\widehat{u}q}^{\mathcal{T}_{i}}
\left(\mathbf{A}_{qq}^{\mathcal{T}_{i}}\right)^{-1}\mathbf{A}_{q\widehat{u}}^{\mathcal{T}_{i}}\: \Delta \widehat{{U}}^{\Gamma_{i}}\right)$

\EndFor

\State $ \Delta \widehat{{U}}  \gets \mathbf{y}$
\end{algorithmic}
 \end{algorithm}

Algorithms 1 and 2 together constitute the complete algorithmic procedure for performing a macro-element HDG solve, storing only the reference-to-physical macro-element transformation data, the inverse Schur complements $\left(\mathbf{S}^{\mathcal{T}_{i}}\right)^{-1}$, and the right-hand-side vector $\widehat{\mathbf{b}}$. 
{\color{black}We refer readers interested in further details on the matrix-free implementation for the macro-element HDG method to our earlier paper \cite{badrkhani2023matrix} which specifically focuses on this aspect.}

 \subsection{Preconditioning methods}
In this work, we consider two options for solving the global linear system \eqref{C44}.
The first and straightforward one solves the linear system in parallel using the restarted GMRES method \cite{saad1986gmres} with iterative classical Gram-Schmidt (ICGS) orthogonalization. In order to accelerate convergence, a left preconditioner is used by the inverse of the block matrix $\mathbf{A}_{\widehat{u}\widehat{u}}$. We emphasize that the matrix $\left( \mathbf{A}_{\widehat{u}\widehat{u}}\right) ^{-1}$ is never actually computed. Since all blocks $\mathbf{A}_{\widehat{u}\widehat{u}}^{\mathcal{T}_{i}}$ are positive definite and symmetric, we compute and store its Cholesky factorization in-place. 
 
In the second option, we use an iterative solver strategy, where we use a flexible implementation of the GMRES (FGMRES) method for our spatial discretization scheme. We note that within matrix-free approaches such as the one followed here, multilevel algorithms have emerged as a promising choice to precondition a flexible GMRES solver, see for example the reference \cite{saad1993flexible}. This methodology has demonstrated success when applied for compressible flow problems \cite{wang2007implicit,diosady2009preconditioning, fidkowski2005p,badrkhani2017development} and incompressible flow problems \cite{botti2017h, franciolini2020p,vakilipour2019developing}. In this regard, we use the FGMRES iteration, where GMRES itself is employed as a preconditioner, and a GMRES approach, where the inverse of the global matrix $\left( \mathbf{A}_{\widehat{u}\widehat{u}}\right) ^{-1}$ is employed as a preconditioner. We note that we determine this inverse in the same way as above via a matrix-free Cholesky factorization. We note that preconditioned iterative solvers become more effective as the time step size of the discretization increases, which implies that the condition number increases.
  
 \subsection{Computational setup}  
  
We implement our methods in Julia\footnote{The Julia Programming Language, \href{https://julialang.org/}{https://julialang.org/}}, by means of element formation routines. In the context of the local solver, the associated tasks involve the assembly of local matrices, the computation of local LU factorizations of $\mathbf{S}^{\mathcal{T}_{i}} $, and the extraction of local values from the global solution. We use FGMRES / GMRES iterative methods for the global solver provided by the open-source library PETSc\footnote{Portable, Extensible Toolkit for Scientific Computation, \href{https://petsc.org/}{https://petsc.org/}}. {\color{black}Motivated by the very small size of the matrices on each element, we employ the dense linear algebra from the LAPACK package\footnote{Linear Algebra PACKage, \href{https://netlib.org/lapack/}{https://netlib.org/lapack/}} for the standard HDG method. The matrices for the macro-element HDG method are larger and therefore we utilize sparse linear algebra provided by the UMFPACK library \cite{davis2004algorithm}.}
  
Our implementation has been adapted to the parallel computing environment Lichtenberg II (Phase 1), provided by the High-Performance Computing Center at the Technical University of Darmstadt. It was compiled using GCC (version 9.2.0), Portable Hardware Locality (version 2.7.1), and OpenMPI (version 4.1.2). Our computational experiments were conducted on this cluster system, utilizing multiple compute nodes. Each compute node features two Intel Xeon Platinum 9242 processors, each equipped with 48 cores running at a base clock frequency of 2.3 GHz. Additionally, each compute node provides a main memory capacity of up to 384 GB. For more detailed information about the available compute system, please refer to the Lichtenberg webpage\footnote{High performance computing at TU Darmstadt, \href{https://www.hrz.tu-darmstadt.de/hlr/hochleistungsrechnen/index.en.jsp}{https://www.hrz.tu-darmstadt.de/hlr/hochleistungsrechnen/index.en.jsp}}.

\color{black}

\section{Numerical results \label{sec:Numerical results}}

In the following, we will demonstrate the computational advantages of our macro-element HDG approach for compressible flow problems, in particular in comparison with the standard HDG method. Our numerical experiments encompass several three-dimensional test cases that include steady and unsteady flow scenarios. {\color{black} Here we consider the DIRK(3,3) scheme for the discretization in time.} We conduct a comprehensive comparative analysis in terms of accuracy, number of iterations {\color{black}  (all time steps)}, computing times {\color{black}  (wall time for all time steps)}, and required number of degrees of freedom in the local / global solver, with a particular focus on the parallel implementation.

\begin{remark}
We will use the parameter $m$ to denote the number of $C^0$-continuous elements along each macro-element edge. Therefore, the parameter $m$ defines the size of the mesh of $C^0$-continuous elements in each tetrahedral macro-element. For $m=1$, we obtain the standard HDG method. 
\end{remark}

\subsection{Compressible Couette flow}

To demonstrate our method for a simple example and to illustrate its optimal convergence, we consider the problem of steady-state compressible Couette flow with a source term \citep{nguyen2012hybridizable, schutz2012hybridized} on a three-dimensional cubic domain $\Omega=\left(x_1,x_2,x_3\right) = \left(0,1\right) ^{3}$. The analytical expression of the axis-aligned solution is:
\begin{subequations}\label{CF0}
 \begin{align}
 &v_{1}=x_{2} \text{log}\left( 1+x_{2}\right) ,\\
&v_{2}=0  ,\\
&v_{3}=0  ,\\
&T=T_{\infty}\left(  \alpha_{c}+x_{2}\left(\beta_{c}-\alpha_{c}\right) +\frac{\gamma-1}{2\gamma\rho_{\infty}} \; Pr\; x_{2} \left( 1-x_{2}\right)\right) ,
\end{align}   
\end{subequations}
where we assume the following parameters: $\alpha_{c}=0.8$, $\beta_{c}=0.85$, $\gamma=1.4$, $T_{\infty}=1/(1-\gamma)M_{\infty}^{2}$ 
and $\rm Pr = 0.71$. {\color{black}
In addition, the Mach number is taken as $M_{\infty}=v_{\infty}/c_{\infty}=0.15$, where $c_{\infty}$ is the speed of sound corresponding to the temperature $T_{\infty}$ and $v_{\infty}$ is infinity velocity.}
The viscosity is assumed constant and the source term, which is determined from the exact solution, is given by
\begin{equation}\label{sourse}
\mathbf{S} = \frac{-1}{Re}\left\lbrace 0,\frac{2+x_{2}}{(1+x_{2})^{2}}, 0,0, \text{log}^{2}(1+x_{2})+
\frac{x_{2}\text{log}(1+x_{2})}{1+x_{2}}+\frac{y(3+2x_{2})\text{log}(1+x_{2})-2x_{2}-1}{(1+x_{2})^{2}}\right\rbrace ^{t}.
\end{equation}

\begin{figure}
    \centering
    \subfloat[\centering 3D structured mesh.]{{\includegraphics[width=0.53\textwidth]{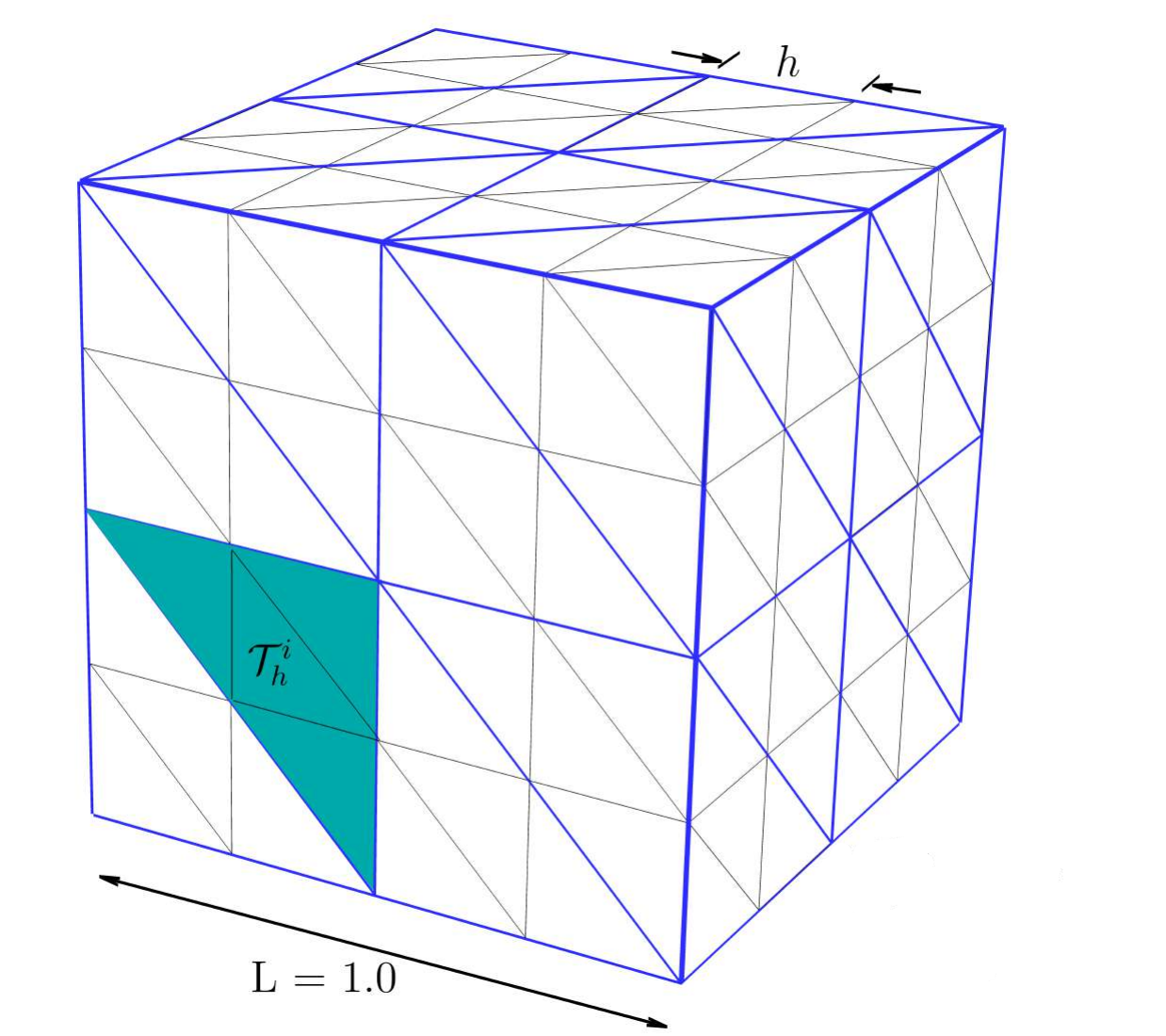} }}%
    \subfloat[\centering Exact solution in $x_1-x_2$ plane.]{{\includegraphics[width=0.47\textwidth]{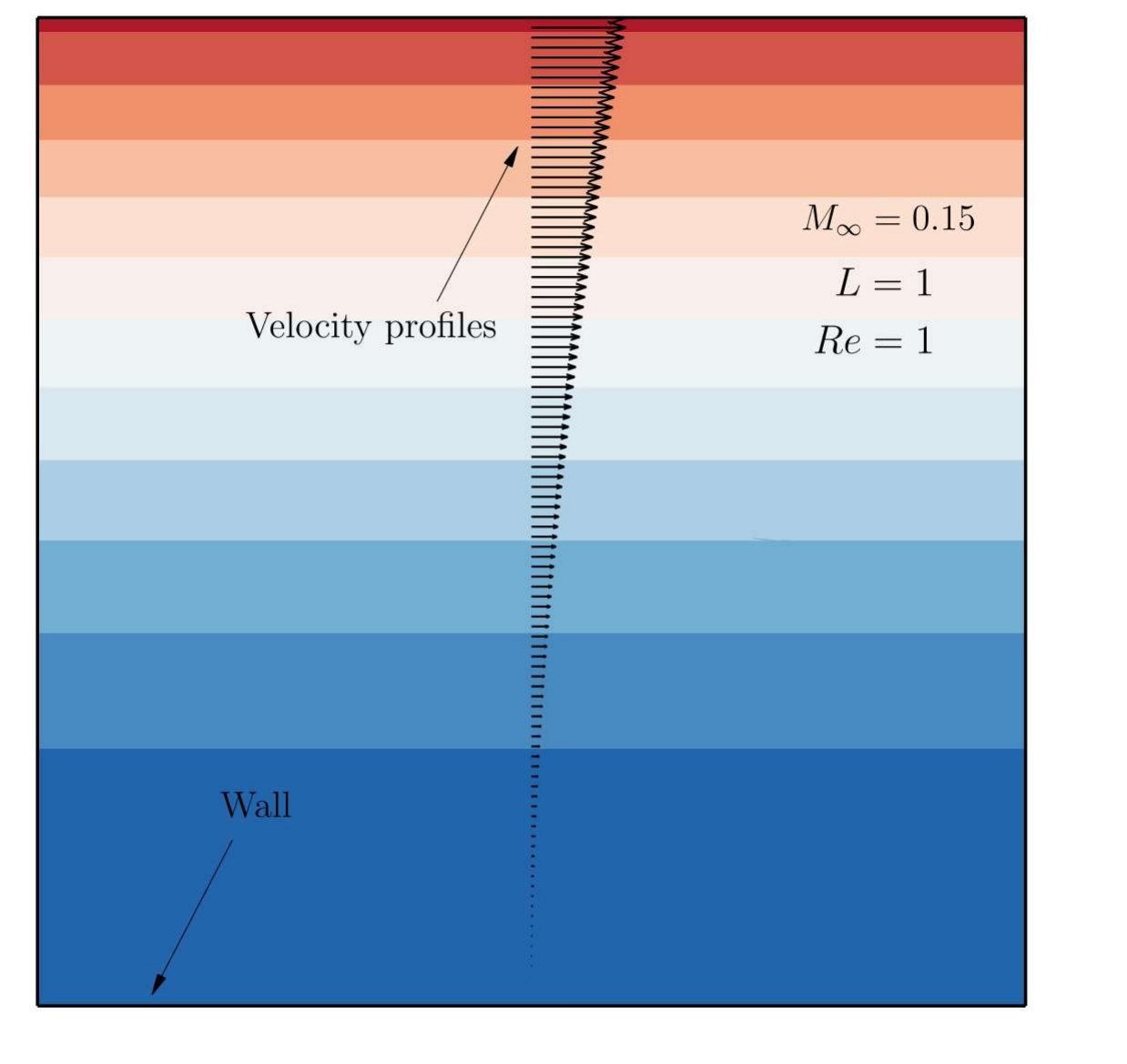} }}%
    \caption{Compressible Couette Flow on a unit cube.}%
    \label{fig:Couette test case}%
\end{figure}

While the exact solution is independent of the Reynolds number {\color{black}$Re= \rho_{\infty}v_{\infty}L/\mu_{\infty}=Re_{c_{\infty}}M_{\infty}$}, we set it to 1 in order to replicate the case presented in \citep{nguyen2012hybridizable, schutz2012hybridized}, taking a characteristic length of $L=1$. 


Figure \ref{fig:Couette test case}(b) plots the exact solution \ref{CF0} in the $x_1-x_2$ that corresponds to the source term \ref{sourse}. We consider polynomial orders $p=1$ to $p=5$ and consider meshes as shown in Figure \ref{fig:Couette test case}(a), where we choose the number $m$ of elements along a macro-element edge in each direction to be two. In Figures \ref{fig:Couette converge}(a) and (b), we plot the error in the $L^{2}$ norm versus the element size under uniform refinement of the macro-elements for velocity and energy, respectively. We observe that we achieve the optimal convergence rate $p+1$ in all cases. 

 \begin{figure}
    \centering
    \subfloat[\centering Velocity, $\Vert {v_{1}}_{h}-v_{1}\Vert_{L^{2}\left(\Omega \right)}$ ]{{\includegraphics[width=0.5\textwidth]{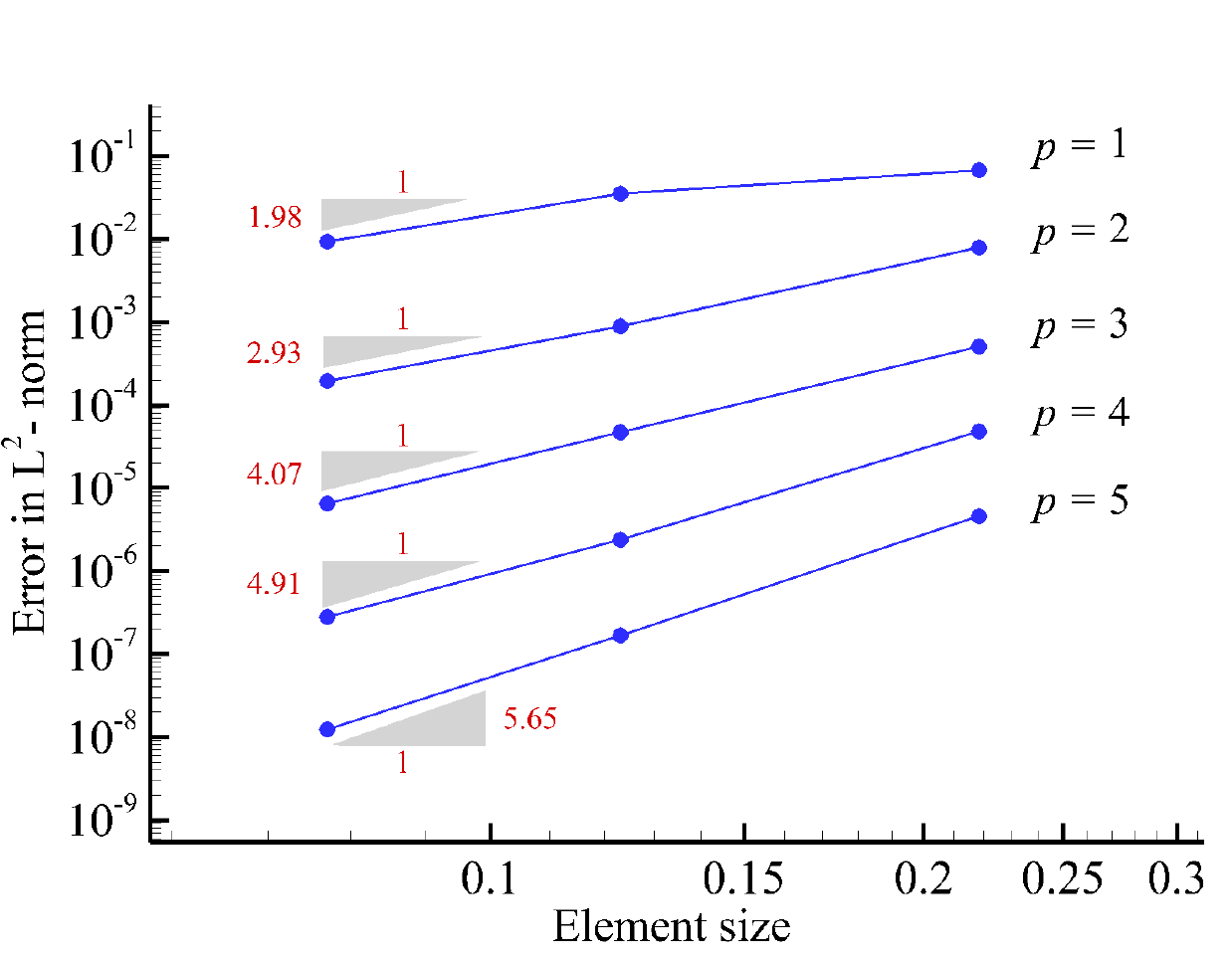} }}%
    \subfloat[\centering Energy, $\Vert\rho E_{h}-\rho E\Vert_{L^{2}\left(\Omega \right)}$  ]{{\includegraphics[width=0.5\textwidth]{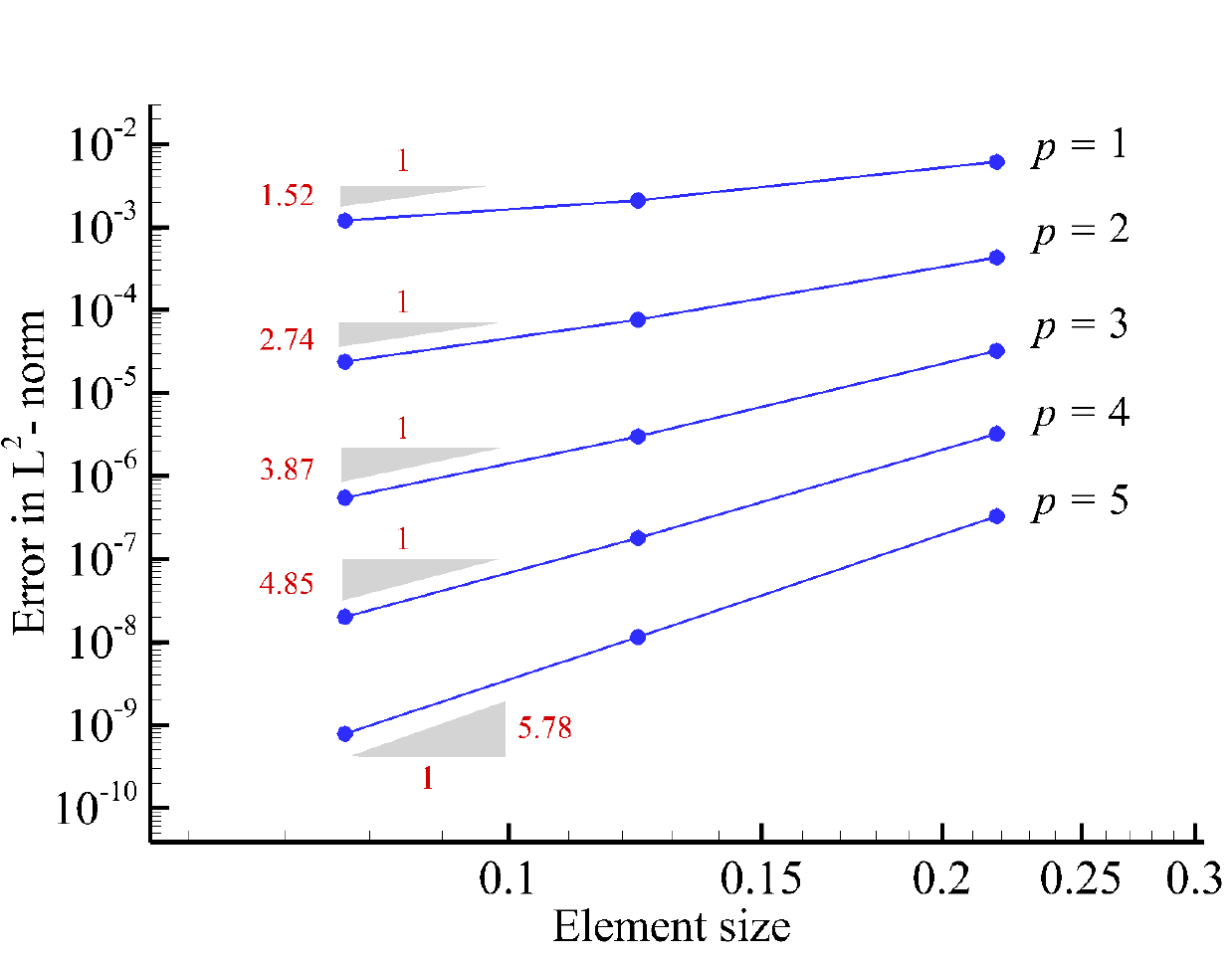} }}%
    \caption{Couette flow: convergence of the macro-element HDG method with $m = 2$.}%
    \label{fig:Couette converge}%
\end{figure}   
    
\subsubsection{Computational performance for a fixed global mesh size}
To assess the performance of the local and global solvers with respect to a change in the number of elements $m$ per macro-element, we consider a discretization of the cube with 768 elements with polynomial degree $p=3$ that we apply to the current problem. We compute in parallel on 12 processors. We compare the standard HDG method ($m=1$) versus the macro-element HDG method with eight elements per macro-element ($m=2$) and 64 elements per macro-element ($m=4$). Each HDG variant uses the same overall mesh of $N^{Elm} = 768$ elements to discretize the cube, only changing the number of macro-elements, $N^{\textit{Mcr}}$. In Table \ref{Tab: DOfs size of precon}, we report the number of degrees of freedom for the global problem for $m=1$ (standard HDG method), $m=2$ and $m=4$ (macro-element HDG method). In Table \ref{Tab:Time of precon}, we report the time and number of iterations of the FGMRES method, where we use a drop tolerance of 10e-12. In addition, we compare for each HDG variant a GMRES linear solver approach versus a FGMRES linear solver approach, where both variants use the $\left( \mathbf{A}_{\widehat{u}\widehat{u}}\right) ^{-1}$ preconditioner for the GMRES solver as described above.
 
We see that in both variants of the linear solver considered here, the computing time required for the local solver increases when $m$ is increased. This increase is expected as the number of degrees of freedom in the local solver increases. We observe that the FGMRES approach is more efficient than the GMRES approach for all HDG variants. The decrease in the local and global computing time is even larger for the macro-element HDG methods than for the standard HDG method. We can also see that compared to the standard HDG method ($m=1$), the global solver operations can be significantly reduced for the macro-element HDG variants with increasing $m$. We note that this is the opposite when the GMRES linear solver is employed. Based on this observation, we will focus on the FGMRES solver in the remainder of this study.

\begin{table}
\caption{Couette flow: number of local and global unknowns for $p=3$ and varying $m$ on a fixed global mesh.} 
\centering
\begin{tabular}{l|ccc|ccc|ccc}\toprule
& \multicolumn{3}{c}{$\text{dof}^{local}$ (total)} & \multicolumn{3}{|c}{$\text{dof}^{global}$} & \multicolumn{3}{|c}{$\text{dof}^{local}_{i}$(per macro-element)}
\\\cmidrule(lr){2-4}\cmidrule(lr){5-7}\cmidrule(lr){8-10}
          					& $m = 1$      & $m = 2$ 		& $m = 4$   		& $m = 1$    & $m = 2$ 	& $m = 4$	& $m = 1$    & $m = 2$ 	& $m = 4$  \\\midrule           
 $N^{Elm}=768$  			&307,200	&161,280		&109,200		&81,600	&30,240	&13,650	&400		&1,680	&9,100 \\\bottomrule
\end{tabular}
\label{Tab: DOfs size of precon}%
\end{table}

\begin{table}
\caption{Couette flow: we compare the time for the local solver and the local part of the matrix-free global solver (step 2) vs.\ the time for the remaining parts of the global solver (residual drop ${{10}^{-12}}$). We use standard HDG ($m=1$) vs.\ macro-element HDG $\left(m=2,4\right)$ (all with $p=3$, GMRES preconditioner $\left( \mathbf{A}_{\widehat{u}\widehat{u}}\right) ^{-1}$, proc's = 12).}
\centering
\begin{tabular}{l|ccc|ccc|ccc}\toprule
& \multicolumn{3}{c}{Time local op's [s]} & \multicolumn{3}{|c}{Time global op's [s]}  & \multicolumn{3}{|c}{\# Newton iterations}
\\\cmidrule(lr){2-4}\cmidrule(lr){5-7}\cmidrule(lr){8-10}
          		& $m = 1$     & $m = 2$ 	& $m = 4$   		& $m = 1$    & $m = 2$ 	& $m = 4$	  & $m = 1$    & $m = 2$ 	& $m = 4$ \\\midrule           
 GMRES  	&11.1		&15.4		&120.1		&27.3		&15.1		&58.9		 &199		&160		&160	\\[5pt]
 FGMRES  	&8.6	 	&5.2 		&26.6			&20.6 	&4.5 		&10.1		 &56		&50		&49\\\bottomrule				
\end{tabular}
\label{Tab:Time of precon}%
\end{table}
 
\subsubsection{Computational performance for a fixed number of local unknowns per macro-element}
  
In the next step, we test the HDG variants for a fixed number of local unknowns that we choose as $\text{dof}^{local}_{i}=3,300$. We focus on investigating the performance of one and two level static condensation in conjunction with FGMRES iterations. 
To this end, instead of keeping the number of processors fixed, we switch to a variable number of processors, but keep the ratio of the number of macro-elements to the number of processors fixed at one (macro-elements / proc’s = 1). The macro-element HDG method combines patches of eight ($m=2$), 64 ($m=4$) or 512 ($m=8$) $C^{0}$-continuous tetrahedral elements into one macro-element. To always obtain the same number of local unknowns $\text{dof}^{local}_{i}$, we adjust the polynomial degree $p$ in such a way that the product of $m$ and $p$ is always $mp = 8$. Consequently, with an increase in the macro-element size $m$, the polynomial order $p$ decreases. Table \ref{Tab:DOfsCoue} reports the number of degrees of freedom for two of the meshes with 12 and 44 macro-elements that we generated by uniformly refining the initial mesh given in Figure \ref{fig:Couette test case}. 

Table \ref{Tab:PerfCoue} reports the computing time for the local operations and the time for the global operations with increasing number of elements per macro-element, parametrized by $m$. As we use the same preconditioner, the number of Newton iterations in the static condensation and alternative second-layer static condensation approaches are practically the same. We note that the large number of iterations is not uncommon in the present case, as we solve a steady-state problem with a matrix-free method and a preconditioner that can only use the locally available part of the matrix \cite{franciolini2020efficient, franciolini2020p}.
We observe that as $m$ increases, the number of iterations decreases, despite the constant number of degrees of freedom in the local / global solver. This phenomenon can be attributed to the change in structure of the global matrices, leading to improved conditioning. Due to the reduced number of iterations, the time required for both local and global operations is reduced. When looking at the results of Table \ref{Tab:PerfCoue}, it is important to keep in mind that a change in $m$ also involves a change in $p$. The variation in polynomial order, however, strongly influences the accuracy of the approximate solution. Table \ref{Tab:ErroCoue} presents the approximation error corresponding to different values of $p$ for the number of degrees of freedom reported in Table \ref{Tab:DOfsCoue}. We observe that with increasing the polynomial degree, we obtain a significantly better accuracy.


\begin{table}
\caption{Couette flow: number of local and global unknowns at a constant number of local unknowns per macro-element. We report values for two different meshes with 12 and 44 macro-elements, which hold for the pairs $(m,p)=(2,4)$, $(m,p)=(4,2)$ and $(m,p)=(8,1)$.} 
\centering
\begin{tabular}{l|ccc}\toprule  
					&$\text{dof}^{local}$   	&$\text{dof}^{global}$  	&$\text{dof}^{local}_{i}$\\\midrule           
 $N^{\textit{Mcr}}=12$  			&39,600	&6,750	&3,300		 \\[5pt]
 $N^{\textit{Mcr}}=44$  			&145,200	&25,200	&3,300		\\\bottomrule
\end{tabular}
\label{Tab:DOfsCoue}%
\end{table}

\begin{table}
\caption{Couette flow: we compare the time for the local solver and the local part of the matrix-free global solver (step 2) vs.\ the time for the remaining parts of the global solver (residual drop ${{10}^{-12}}$, number of macro-elements $N^{\textit{Mcr}}$ / proc's = 1).}
\centering
\begin{tabular}{l|ccc|ccc|ccc}\toprule
& \multicolumn{3}{c|}{Time local op's [s]} & \multicolumn{3}{c|}{Time global op's [s]} & \multicolumn{3}{c}{\# iterations}
\\\cmidrule(lr){2-4}\cmidrule(lr){5-7}\cmidrule(lr){8-10}
$(m,p)$          			      		& $(2,4)$ 		& $(4,2)$   		& $(8,1)$    
          						& $(2,4)$		& $(4,2)$    		& $(8,1)$ 	
          			 			& $(2,4)$  	 	& $(4,2)$ 		& $(8,1)$ \\\midrule   \multicolumn{10}{c}{Static Condensation }\\\midrule         
 $N^{\textit{Mcr}}=12$  			&2.3				&1.9			&1.6		
 									&1.7				&1.4			&1.2	
 									&4,481			&3,452		&2,984\\[5pt]
 					
 $N^{\textit{Mcr}}=44$  			&4.4				&3.4			&3.3		
 									&4.3				&3.2			&3.4	
 									&10,314		&8.137		&7,710\\\midrule	 \multicolumn{10}{c}{Second-layer Static Condensation }\\\midrule

 $N^{\textit{Mcr}}=12$  			&0.6				&0.5			&0.4		
 									&1.3				&0.9			&0.7	
 									&4,481			&3,452		&2,984\\[5pt]
 					
 $N^{\textit{Mcr}}=44$  			&1.1				&0.9			&0.8		
 									&3.2			&2.2			&1.9	
 									&10,314		&8.137		&7,710\\\bottomrule	 								
\end{tabular}
\label{Tab:PerfCoue}%
\end{table}

\begin{table}
\caption{Couette flow: mass, velocity and energy error in the $L^{2}$-norm for different $(m,p)$ and number of macro-elements $N^{\textit{Mcr}}$. Subscript $h$ denotes the numerical solution and no subscript denotes the exact solution.} 
\centering
\begin{tabular}{l|ccc|ccc|ccc}\toprule  
& \multicolumn{3}{c|}{$\Vert\rho_{h}-\rho\Vert_{L^{2}\left(\Omega \right)}$} 
& \multicolumn{3}{c|}{$\Vert {v_{1}}_{h}-v_{1}\Vert_{L^{2}\left(\Omega \right)}$} 
& \multicolumn{3}{c}{$\Vert\rho E_{h}-\rho E\Vert_{L^{2}\left(\Omega \right)}$}
\\\cmidrule(lr){2-4}\cmidrule(lr){5-7}\cmidrule(lr){8-10}    	
          $(m,p)$ 					& $(2,4)$		&$(4,2)$		&$(8,1)$  
          					&$(2,4)$  		&$(4,2)$		&$(8,1)$     
          					&$(2,4)$  		&$(4,2)$		&$(8,1)$\\\midrule           
 $N^{\textit{Mcr}}=12$  			& 1.72e-6		&6.71e-5		&6.41e-4	 
 					& 7.05e-6		&2.21e-4		&1.67e-3
 					& 1.13e-4		&4.30e-3		&4.15e-2 \\[5pt]
 $N^{\textit{Mcr}}=44$  			& 1.37e-7		&1.52e-5		&2.22e-4	
 					& 8.90e-7		&6.40e-5 		&9.15e-4
 					& 8.94e-6		&9.43e-4 		&1.40e-2 \\\bottomrule
\end{tabular}
\label{Tab:ErroCoue}%
\end{table}

\subsection{Laminar compressible flow past a sphere}

In the next step, we consider the test case of laminar compressible flow past a sphere at Mach number $M_{\infty}=0.1$ and Reynolds number $Re = 100$. At these conditions, we expect a steady state solution with a large toroidal vortex formed just aft of the sphere. As we would like to arrive at the steady-state solution in a computational efficient manner, we would like to choose an implicit time integration scheme with a large time step, resulting in a large \textit{CFL} number (in our case in the order of $10^{5}$). The computational mesh, shown in Figure \ref{fig:sphere mesh}, consists of 139,272 tetrahedral elements at moderate polynomial order $p=3$. We compare the standard HDG method, where all 139,272 elements are discontinuous, and the macro-element HDG method with $m=2$ that uses the same elements grouped into 17,409 discontinuous macro-elements. 

Figure \ref{fig:sphere test case} illustrates the solution characteristics of the problem achieved with the macro-element HDG method in terms of the streamlines and the Mach number. We expect that as both methods use the same mesh at the same polynomial degree, they also achieve practically the same accuracy.
This assumption is supported by Table \ref{Tab:sphere}, where we report the values obtained with our matrix-free implementation of the two HDG variants for the length of the ring vortex, $x_{s}$, and the angle $\theta_s$ at which separation occurs. We observe that both HDG variants produce values that agree well with each other as well as commonly accepted values reported in the literature \citep{johnson1999flow,taneda1956experimental}. 

\begin{table}[ht]
\centering
\caption{Predicted length of the ring vortex and separation angle for flow past sphere at $M_{\infty}=0.1$ and $Re = 100$.}
\begin{tabular}[t]{lccc}
\hline
Author(s) &Method &$x_{s}$ &$\theta_{s}$\\
\hline
Taneda \cite{taneda1956experimental}							&Experimental					&0.89			&127.6\\
Johnson \& Patel \citep{johnson1999flow}		&Finite volume method					&0.88			&126.6\\
Our matrix-free implementation			&HDG							&0.87			&128.1\\
Our matrix-free implementation			&Macro-element HDG	&0.86			&127.8\\
\hline
\end{tabular}
\label{Tab:sphere}%
\end{table}%


\begin{figure}
    \centering
    \subfloat[\centering Element mesh]{{\includegraphics[width=0.28\textwidth]{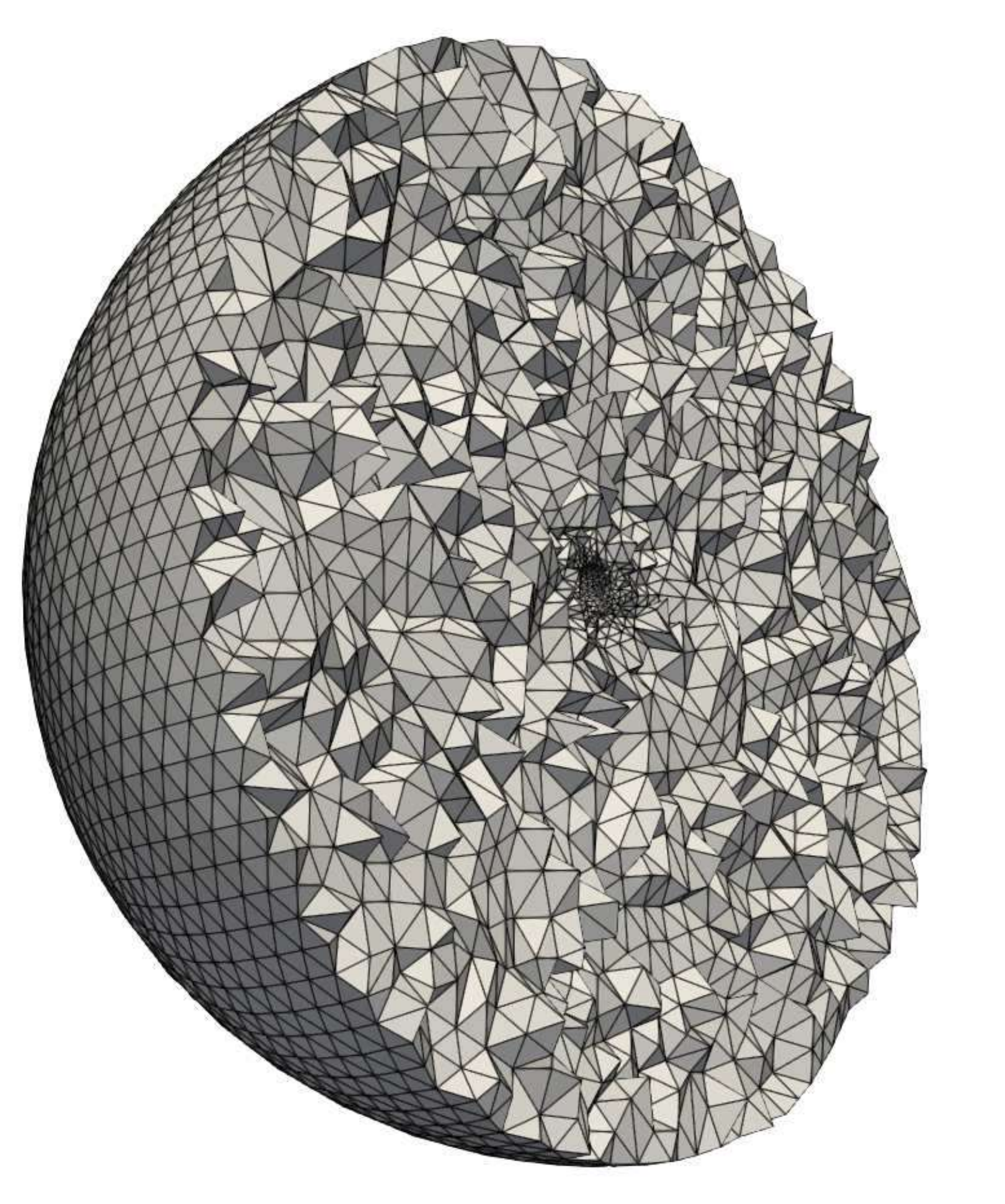} }}%
    \hspace{1cm}
    \subfloat[\centering Macro-element mesh  ]{{\includegraphics[width=0.30\textwidth]{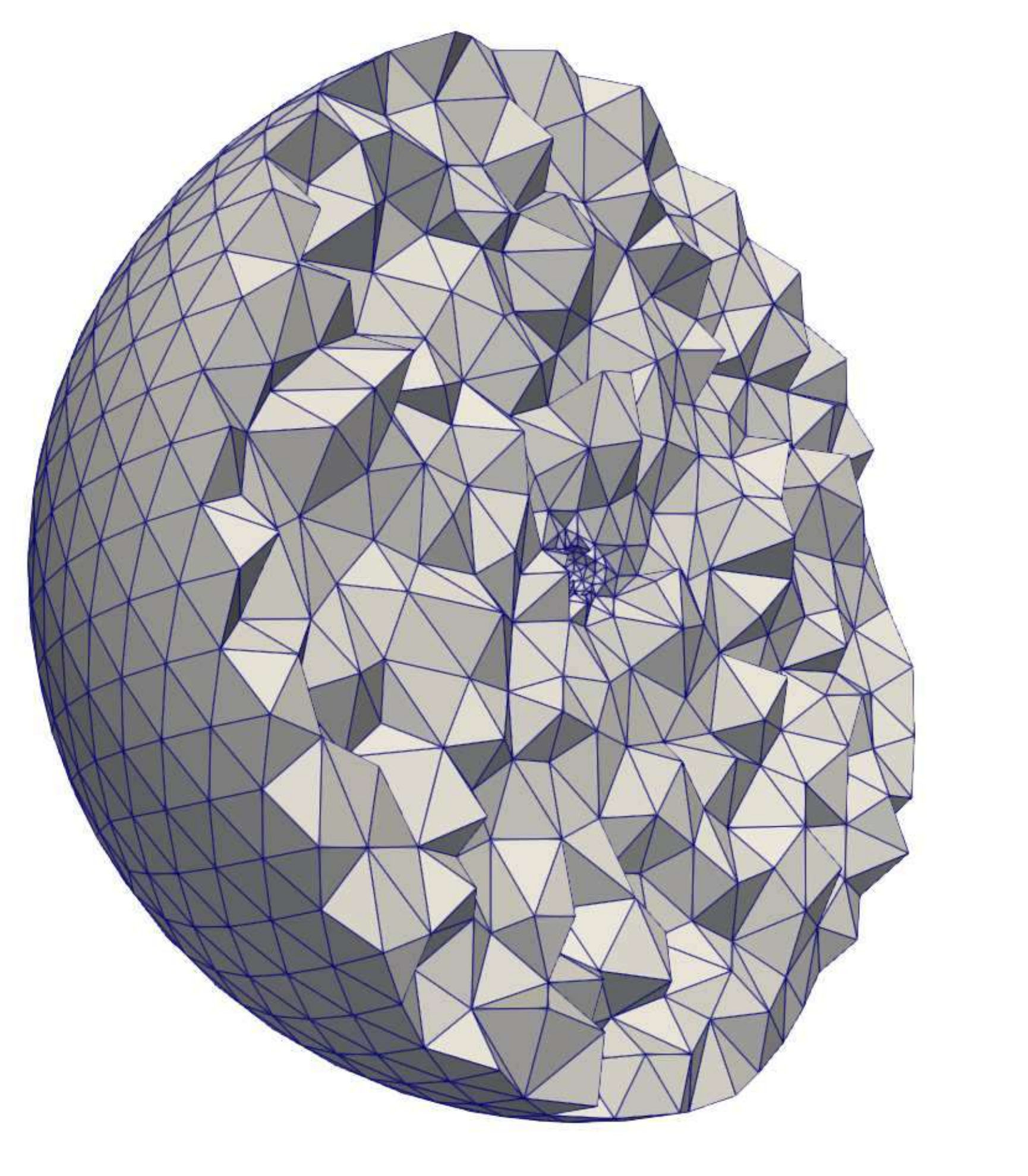} }}%
    \caption{Flow past a sphere: cut halfway through the unstructured mesh, adaptively refined around the sphere in the center.}%
    \label{fig:sphere mesh}%
\end{figure}

\begin{figure}
    \centering
    \subfloat[\centering 3D streamlines, toroidal vortex aft of the sphere]{{\includegraphics[width=0.5\textwidth]{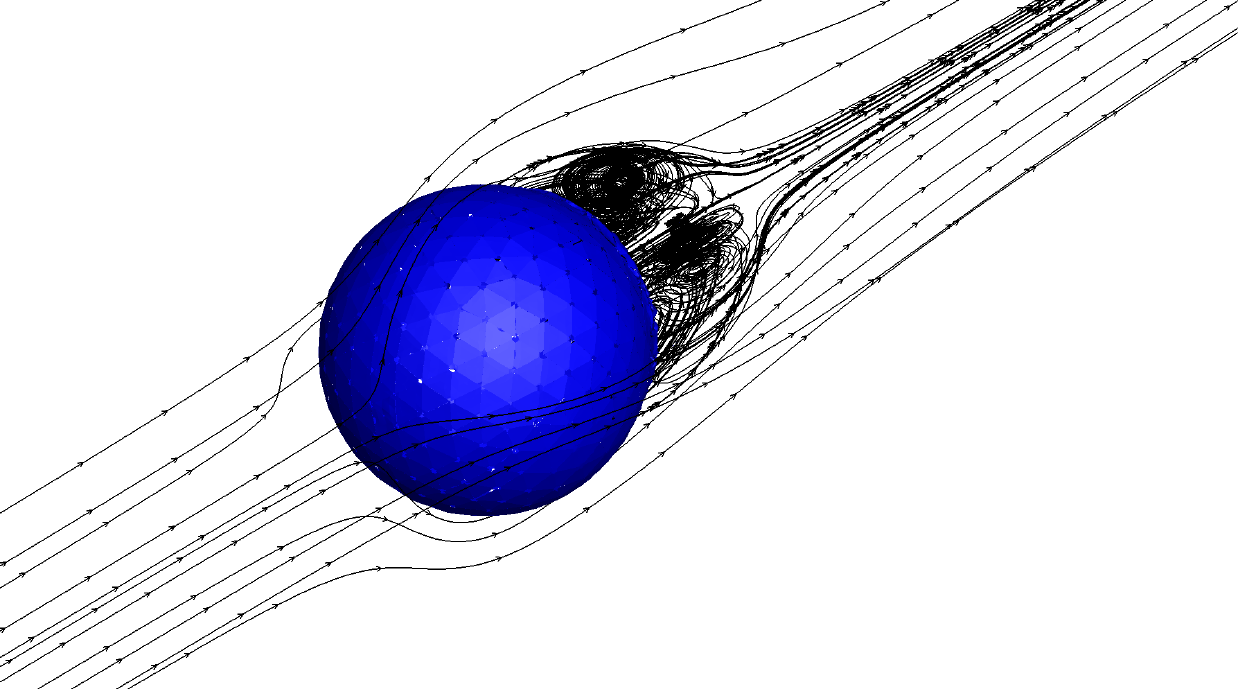} }}
    \subfloat[\centering Mach number contours]{{\includegraphics[width=0.5\textwidth]{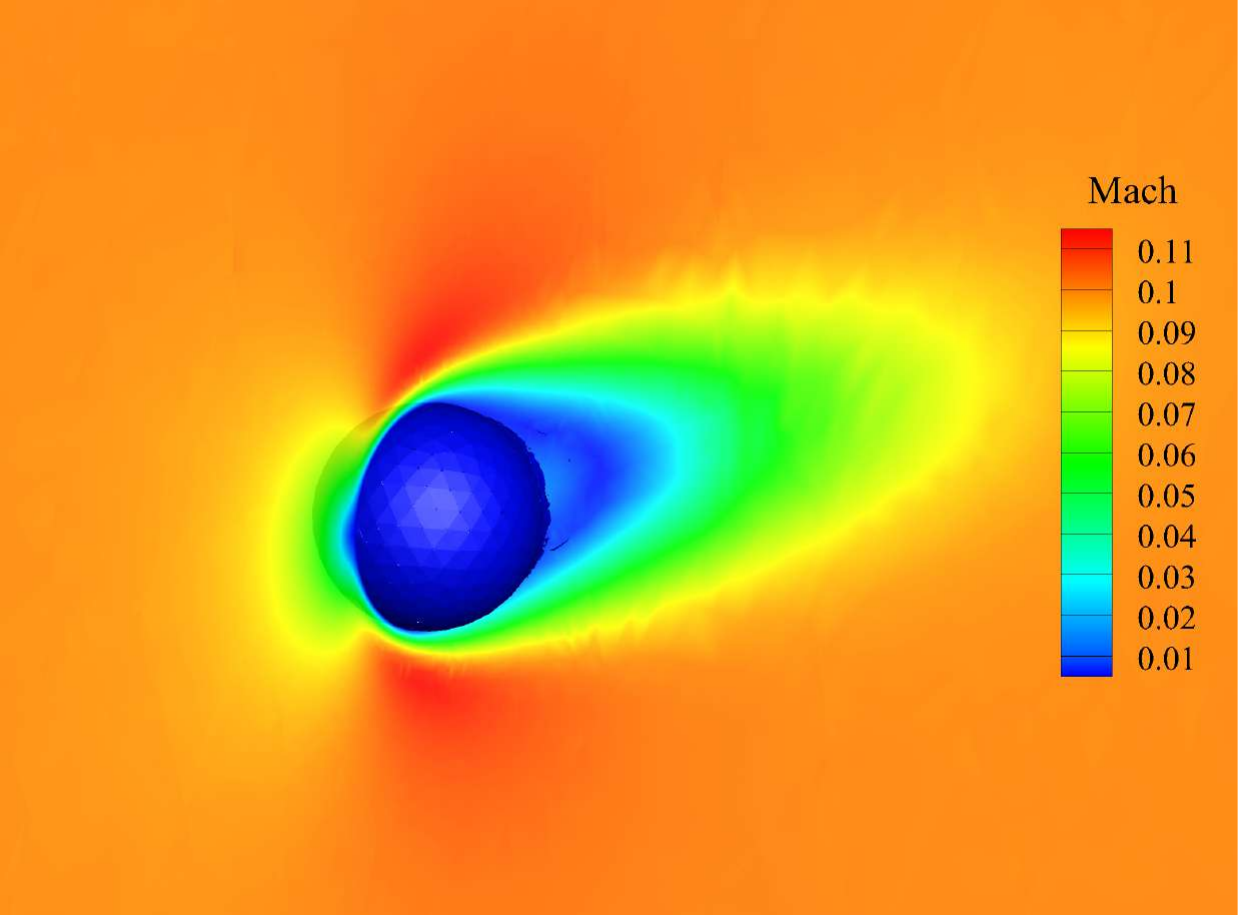} }}%
    \caption{Flow past a sphere at $Re = 100$ and $M_{\infty}=0.1$, computed on the mesh shown in Figure \ref{fig:sphere mesh} at $p=3$.}%
    \label{fig:sphere test case}%
\end{figure}

Table \ref{Tab:laminar sphere} reports the number of degrees of freedom for the mesh shown in Figure \ref{fig:sphere test case} and polynomial degree $p=3$. Both HDG variants use the same mesh, where the standard HDG method assumes fully discontinuous elements ($m=1$) and the macro-element HDG method combines groups of eight tetrahedral elements ($m=2$) into one $C^{0}$ continuous macro-element. We observe that the macro-element HDG variant, with increasing $m$, significantly reduces the total amount of degrees of freedom in both the local and global problems. 

We observe that the macro-element HDG method exhibits a notable speed advantage, primarily reducing the time for the global solver, while the time for the parallelized local solver remains at the same level. 
In the macro-element HDG method, computation time thus shifts from the global solver to the local solver, enhancing parallelization efficiency. Therefore, the time ratio between the local and global solver is thus reduced from approximately 16 in the standard HDG method to approximately four in the macro-element HDG method at $m=2$. This example demonstrates that at moderate polynomial degrees such as $p=3$, where the standard HDG method is typically not efficient, subdivision of one tetrahedral hybridized DG element into a macro-element with just eight $C^0$ continuous elements already leads to a practical advantage in terms of computational efficiency. 

\begin{table}
\caption{Flow past a sphere: we compare the time for the local solver and the local part of the matrix-free global solver (step 2) vs.\ the time for the remaining parts of the global solver for $p = 3$ and $Proc's = 2,048$ (on the mesh shown in Figure \ref{fig:sphere mesh}).}
\centering
\begin{tabular}{cc|cc|cc|cc}\toprule
 \multicolumn{2}{c|}{Time local op's [min]} & \multicolumn{2}{c|}{Time global op's [min]} & \multicolumn{2}{c|}{$\text{dof}^{local}$ } & \multicolumn{2}{c}{$\text{dof}^{global}$ }
\\\cmidrule(r){1-2}\cmidrule(lr){3-4}\cmidrule(lr){5-6}\cmidrule(lr){7-8}
          			 $m = 1$ 		& $m = 2$   	    
          			& $m = 1$		     & $m = 2$    	 	
          			& $m = 1$  	 	& $m = 2$ 	
          			 & $m = 1$  	 	& $m = 2$ 	 
          			\\\midrule         
  		    5.38											&6.34					    					&83.6										   &26.6				
 					&  	55,708,800	        &  29,247,120         &  14,123,600          &  5,012,000		    \\\bottomrule	 						
\end{tabular}
\label{Tab:laminar sphere}%
\end{table}

\subsection{Compressible Taylor-Green Vortex}


As the final benchmark, we consider the Taylor-Green vortex flow \cite{taylor1937mechanism,brachet1991direct,fehn2018efficiency} on a cube $\left[ -\pi L,\pi L\right]^3$. We prescribe periodic boundary conditions in all coordinate directions, assume $M_{0}=0.1$, and use the following initial conditions:
\begin{equation}\label{TGV0}
 \begin{split}
 & v_{1}\left( \mathbf{x}, t=0\right)  = V_{0} \sin\left(\dfrac{x_{1}}{L}\right)\cos\left(\dfrac{x_{2}}{L}\right)\cos\left(\dfrac{x_{3}}{L}\right) ,\\
&  v_{2}\left( \mathbf{x}, t=0\right) =-V_{0} \cos\left(\dfrac{x_{1}}{L}\right)\sin\left(\dfrac{x_{2}}{L}\right)\cos\left(\dfrac{x_{3}}{L}\right),\\
&  v_{3}\left( \mathbf{x}, t=0\right) =0,\\
& P\left( \mathbf{x}, t=0\right) = P_{0}  + \dfrac{\rho_{0} V_{0}^{2}}{16} \left(\cos\left(\dfrac{2x_{1}}{L}\right)+\cos\left(\dfrac{2x_{2}}{L}\right) \right) \left( \cos\left(\dfrac{2x_{3}}{L}\right)+2\right),
\end{split}   
\end{equation}
where $L = 1$, $\rho_{0} = 1$ and $P_{0} =1/\gamma$. The flow is initialized to be isothermal, that is, $P / \rho=P_{0} / \rho_{0}$. The flow is computed at two Reynolds numbers, which are $Re=100$ and $Re=400$. The unsteady simulation is performed for a duration of $15 t_{c}$, where 
$t_{c}=L/ V_{0}$ is the characteristic convective time, {\color{black}$V_{0}=M_{0} c_{0}$, and $c_{0}$ is the speed of sound corresponding to  $P_{0} $ and $\rho_{0}$, ${c_{0}}^2 = \gamma P_{0}/ \rho_{0}$. }

\begin{figure}
    \centering
    \subfloat[\centering Mesh for $Re=100$]{{\includegraphics[width=0.5\textwidth]{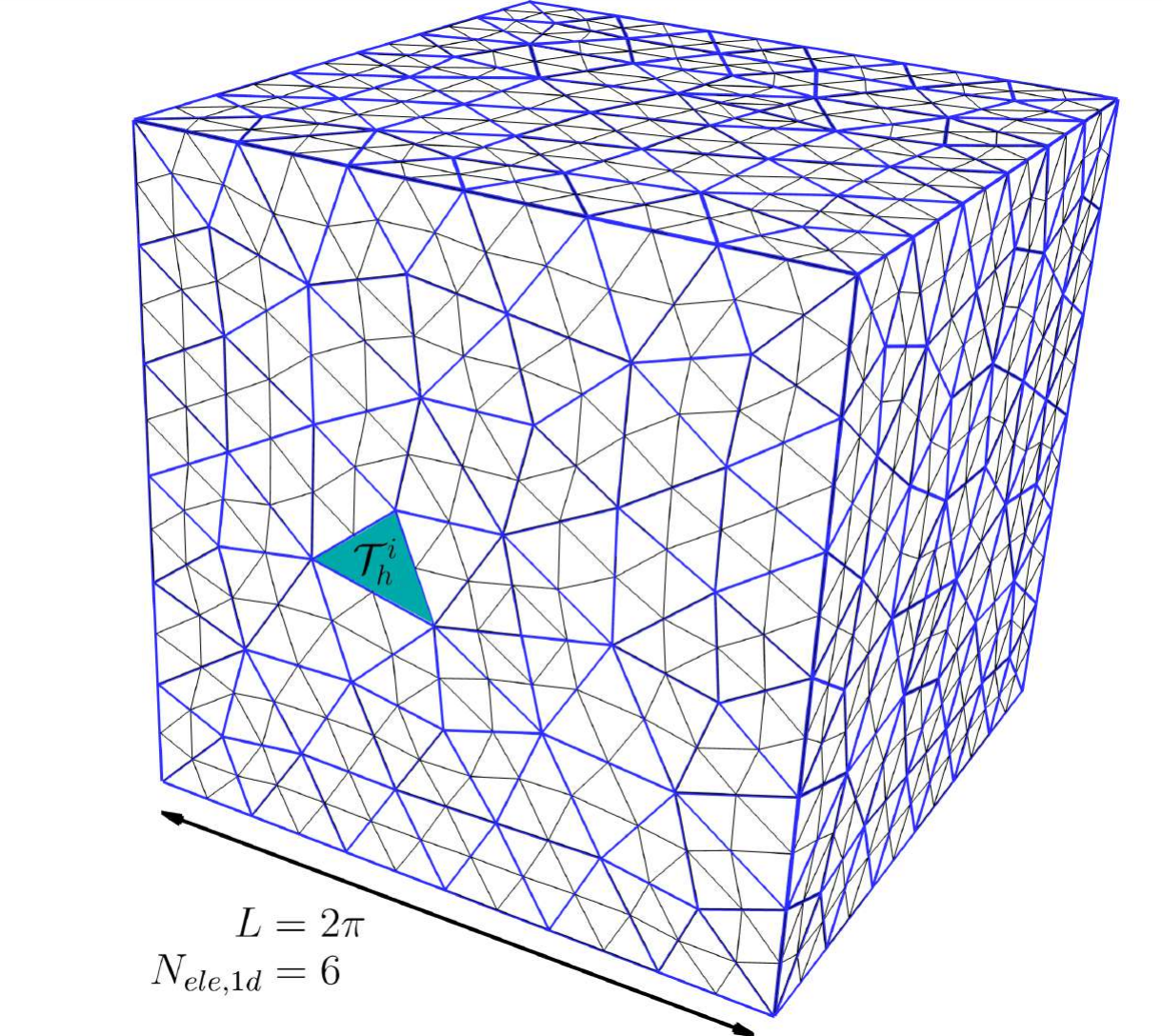} }}%
    \subfloat[\centering Mesh for $Re=400$]{{\includegraphics[width=0.5\textwidth]{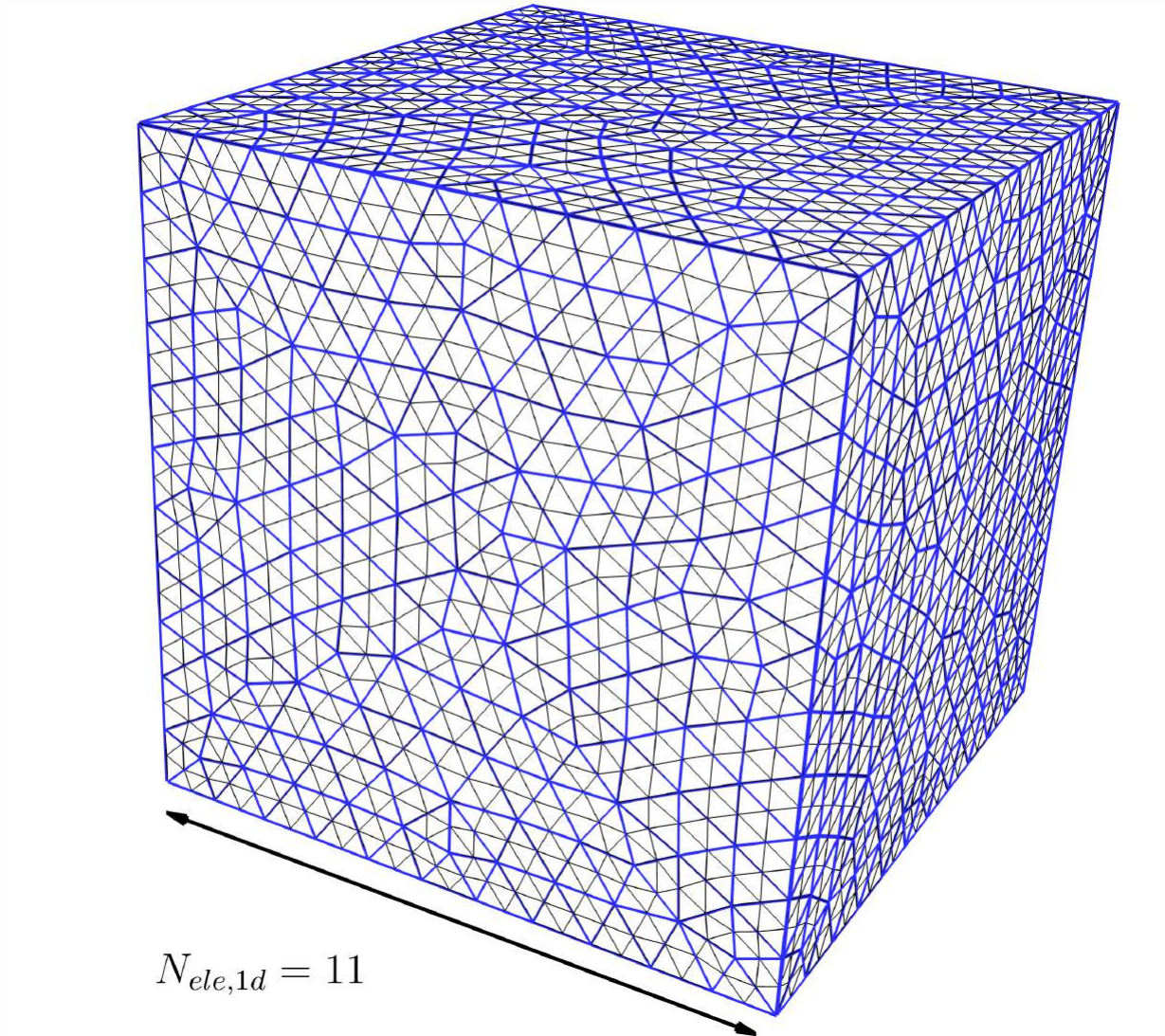} }}%
    \caption{Taylor Green vortex: unstructured meshes of the cube for $m=2$.}%
    \label{fig:Couette test case}%
\end{figure}

\subsubsection{Macro-element HDG discretization}
We employ our macro-element HDG scheme with $mp=8$, that is, the polynomial degrees chosen are $p=4$ for $m=2$ and $p=2$ for $m=4$. The corresponding effective resolutions of the cube are denoted as $N_{\text{eff}} = \left(N_{\text{ele,1d}} \cdot (mp+1)\right)^{3}$, where $N_{\text{ele,1d}}$ denotes the number of macro-elements in one spatial direction, see Figure \ref{fig:Couette test case} for an illustration. We adjust the effective resolution of the mesh according to the Reynolds number. In our case, we employ $N_{\text{eff}} = 54^{3}$ for $Re = 100$, and $N_{\text{eff}} = 99^{3}$ for $400$. Figure \ref{fig:Couette test case} illustrates the two macro-element HDG meshes for $m=2$, where the edges of the discontinuous macro-elements are plotted in black and the edges of the $C^0$-continuous elements within each macro-element are plotted in black. In Table \ref{Tab:DOfsCoue TwoRe}, we report the number of degrees of freedom for both local and global problems, along with the corresponding number of processes, for the two Reynolds numbers under consideration. 

We set the time step such that a \textit{CFL} number of the order of one is maintained, which represents the relation between the convective speed, the resolution length and the time scale. Specifically, we use a time step of 
\begin{equation}
\Delta t = \textit{CFL} \dfrac{h}{V_{0} (p+1)} \, ,
\end{equation} 
where 
$h$ is the characteristic element length and $p$ is the polynomial of order. For this test case, our solution approach is used with absolute solver tolerance
of $10^{-12}$ and relative solver tolerance of $10^{-6}$. 

\begin{table}
\caption{Taylor Green vortex: number of local and global unknowns for $mp=8$ at two different mesh resolutions for $Re = 100$ and $Re=400$.} 
\centering
\begin{tabular}{l|ccc|cc}\toprule  
 & \multicolumn{3}{c|}{Mesh} & \multicolumn{2}{c}{$\text{\# degrees of freedom}$} \\
\cmidrule(lr){2-4} \cmidrule(lr){5-6} 
	&$N_{ele,1d}$ & $N_{\text{eff}}$  &$N^{\textit{Mcr}}$ 					& $\text{dof}^{local}$   		& $\text{dof}^{global}$  	       		\\\midrule   
	$Re=100$ 	& 6	& $54^3$ &1,592						&5,253,600		&716,400							 \\[5pt]
$Re=400$  	& 11	& $99^3$ &  10,143		            	&33,471,900		&4,564,350				 \\\bottomrule        
\end{tabular}
\label{Tab:DOfsCoue TwoRe}%
\end{table} 
 
\begin{figure}
    \centering
   {\includegraphics[width=0.80\textwidth]{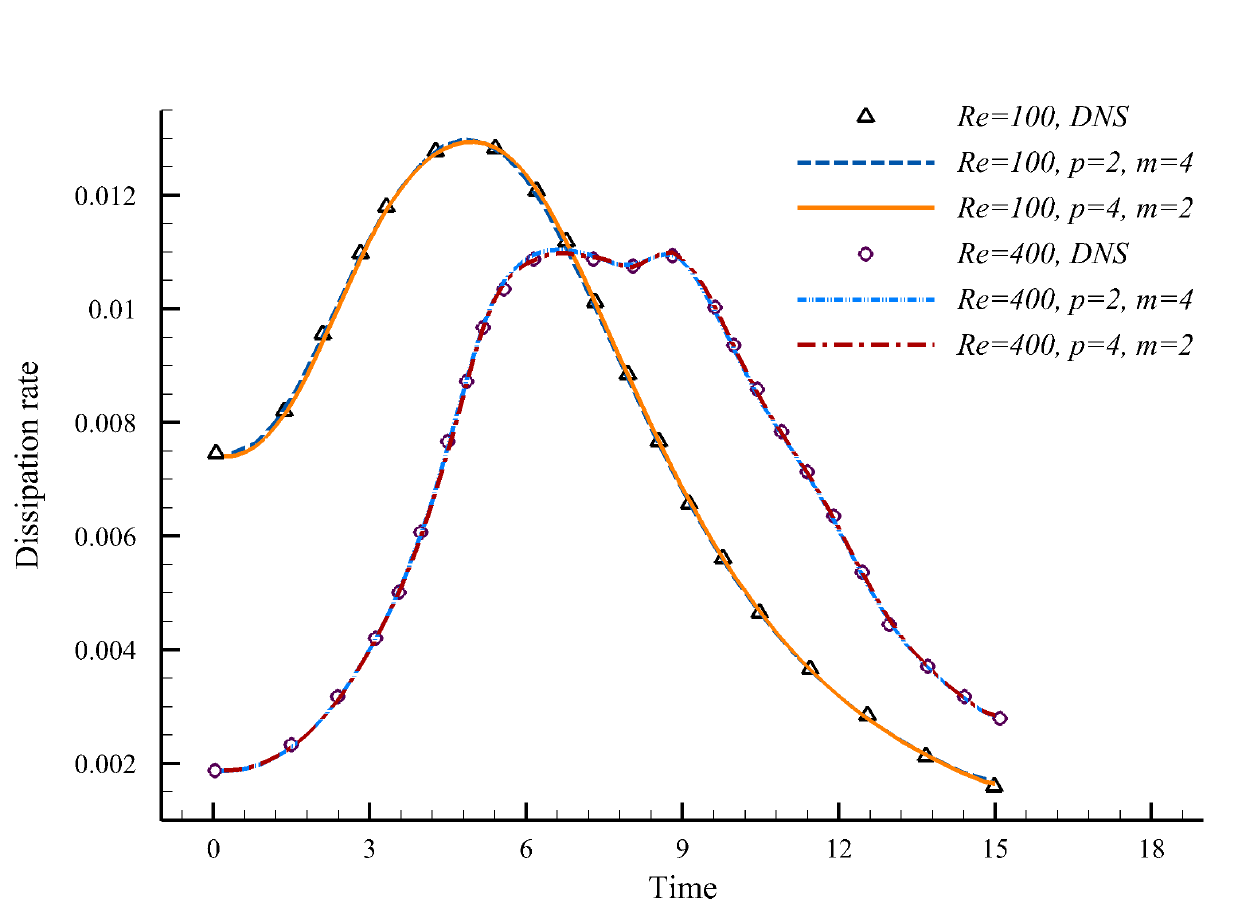} }
    \caption{Taylor–Green vortex: time evolution of kinetic energy dissipation rates, computed for $Re = 100$ and $Re= 400$ on the two different meshes shown in Figure \ref{fig:Couette test case} with $(m,p)=(2,4)$ and $(m,p)=(4,2)$.}
    \label{fig: TGV Re}%
\end{figure}

 \begin{figure}
    \centering
    \subfloat[\centering $t/t_{c} = 0.0$ ]{{\includegraphics[width=0.4\textwidth]{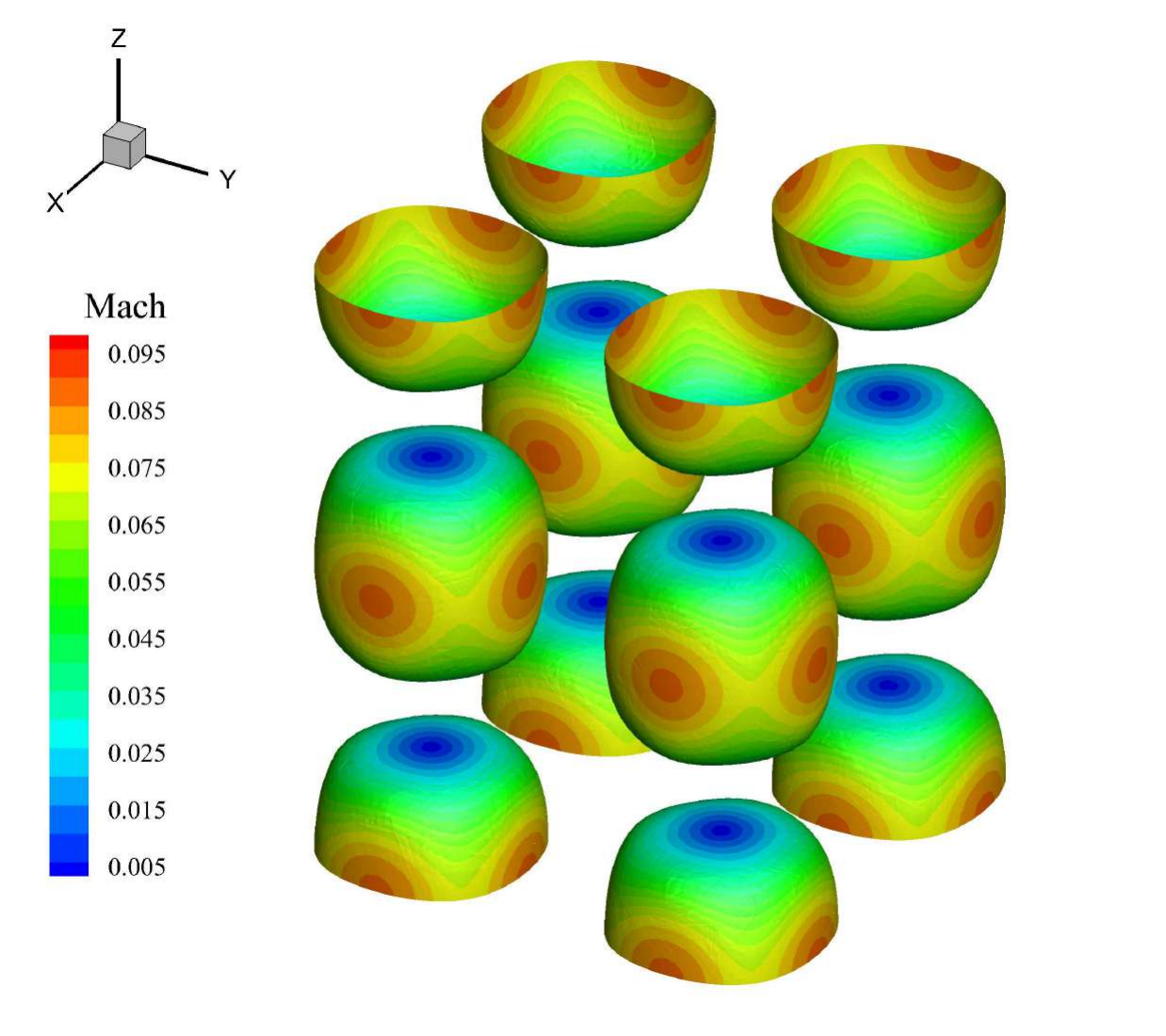} }}%
    \subfloat[\centering $t/t_{c} = 3.0$  ]{{\includegraphics[width=0.4\textwidth]{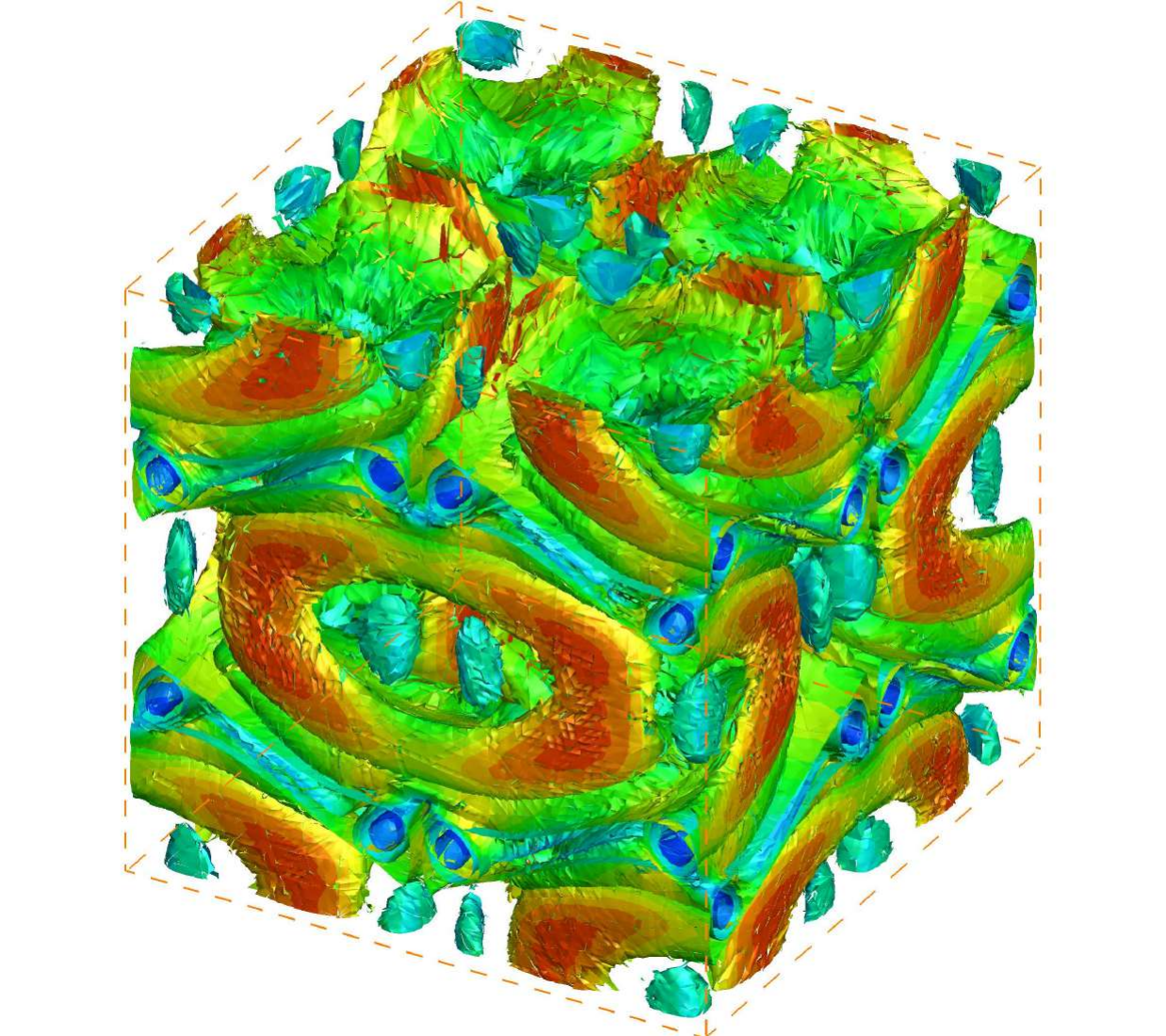} }}\\
    \subfloat[\centering $t/t_{c} = 6.0$ ]{{\includegraphics[width=0.4\textwidth]{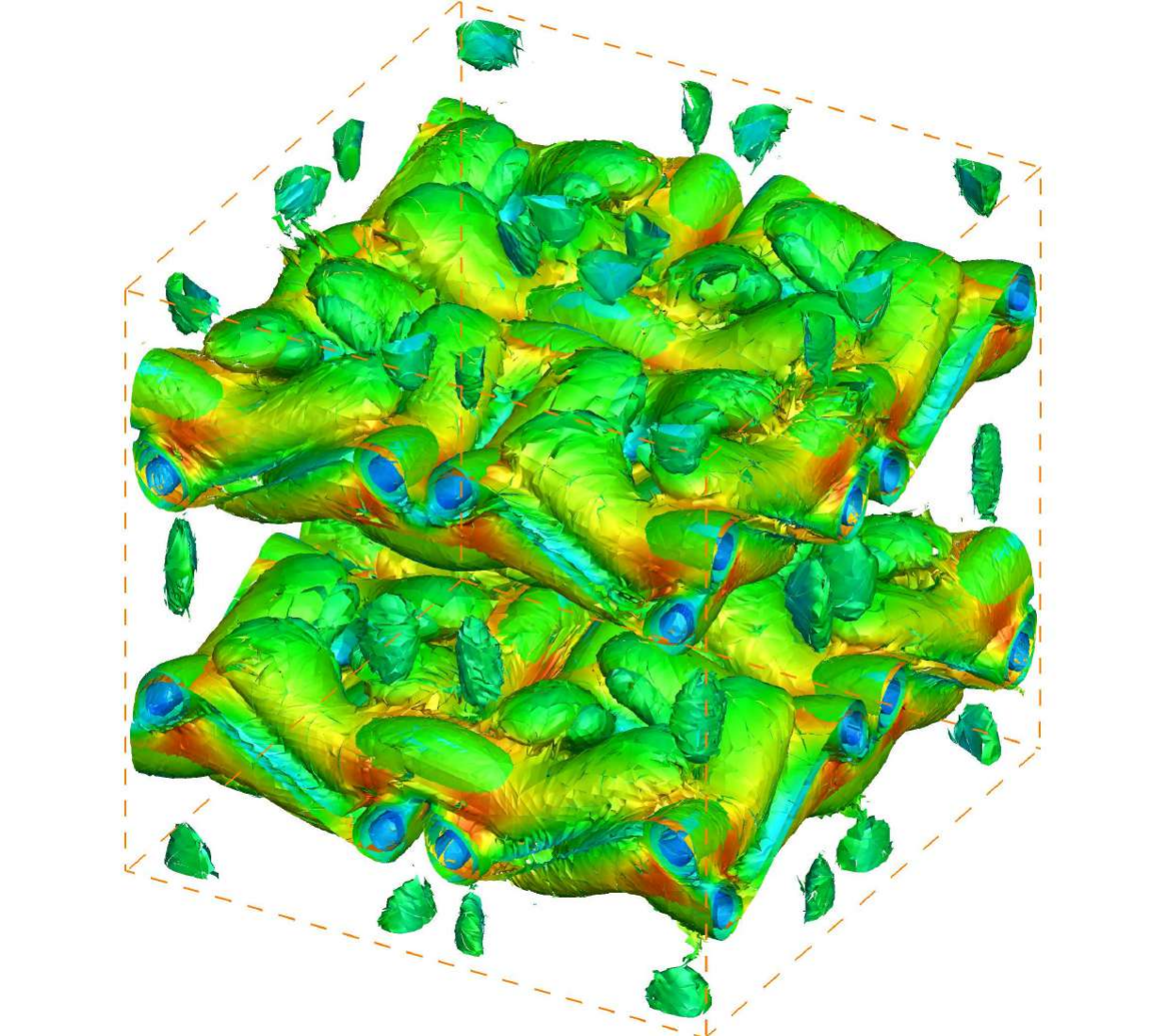} }}%
    \subfloat[\centering $t/t_{c} = 9.0$  ]{{\includegraphics[width=0.4\textwidth]{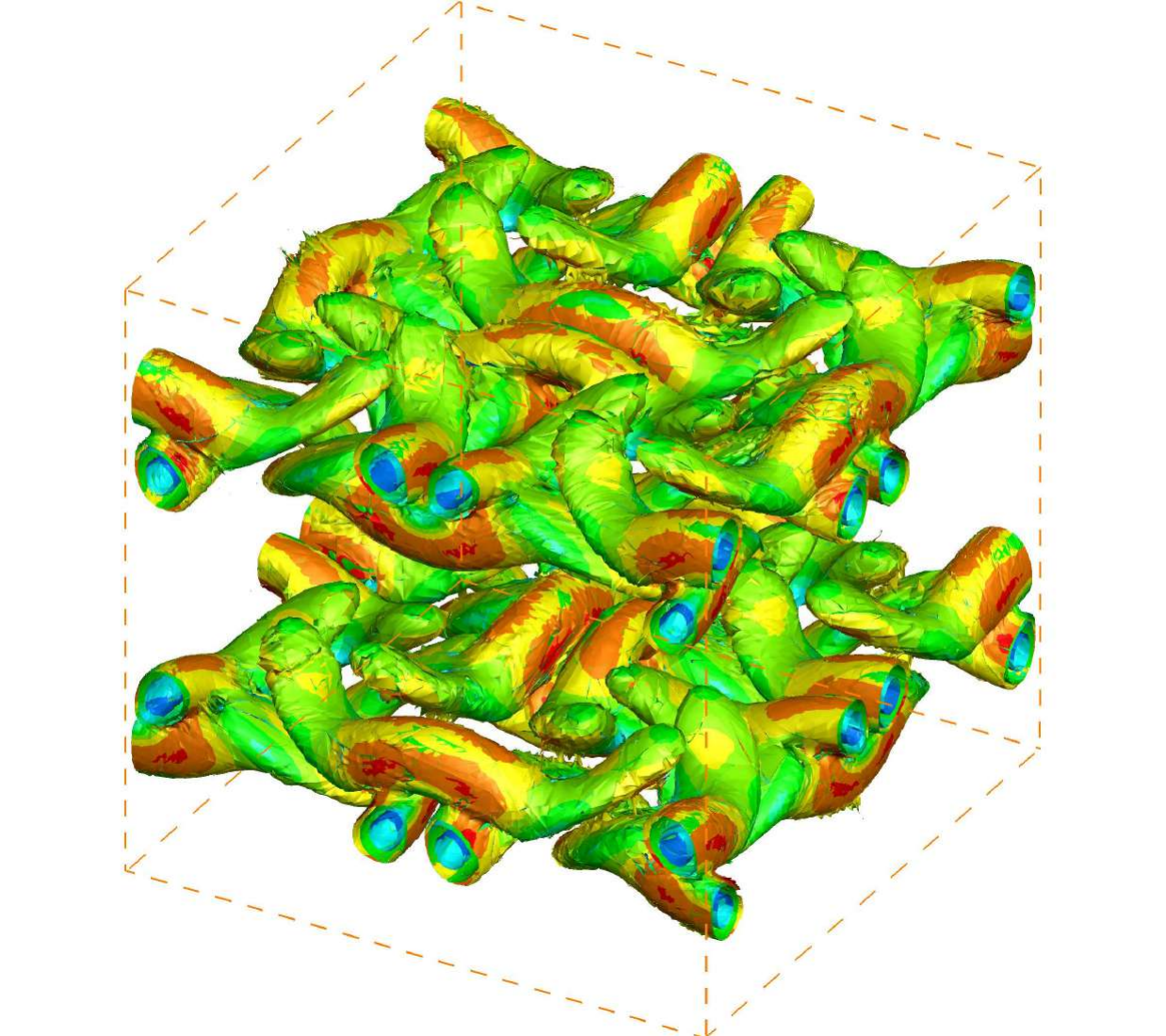} }}\\
    \subfloat[\centering $t/t_{c} = 12.0$ ]{{\includegraphics[width=0.4\textwidth]{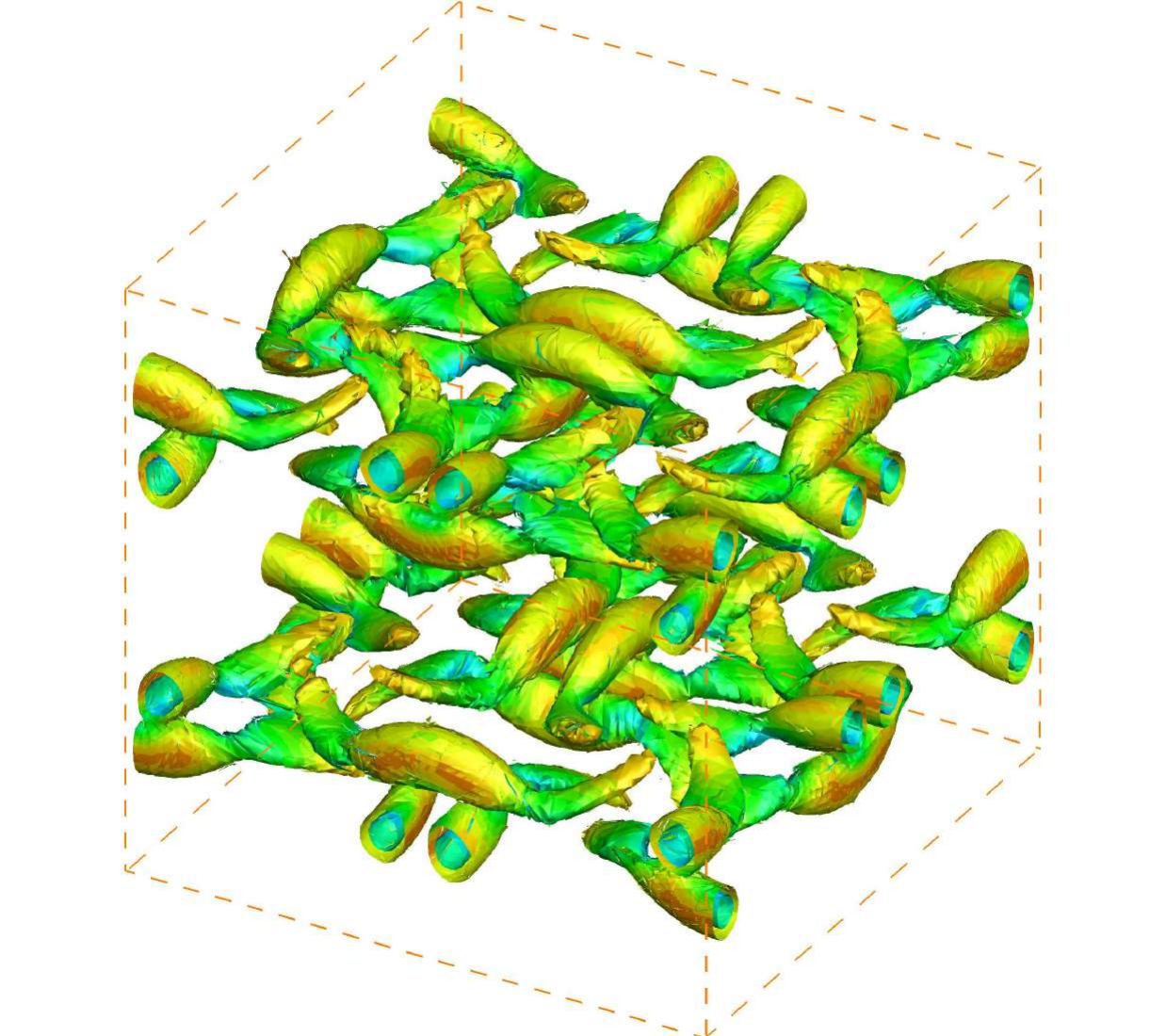} }}%
    \subfloat[\centering $t/t_{c} = 15.0$  ]{{\includegraphics[width=0.4\textwidth]{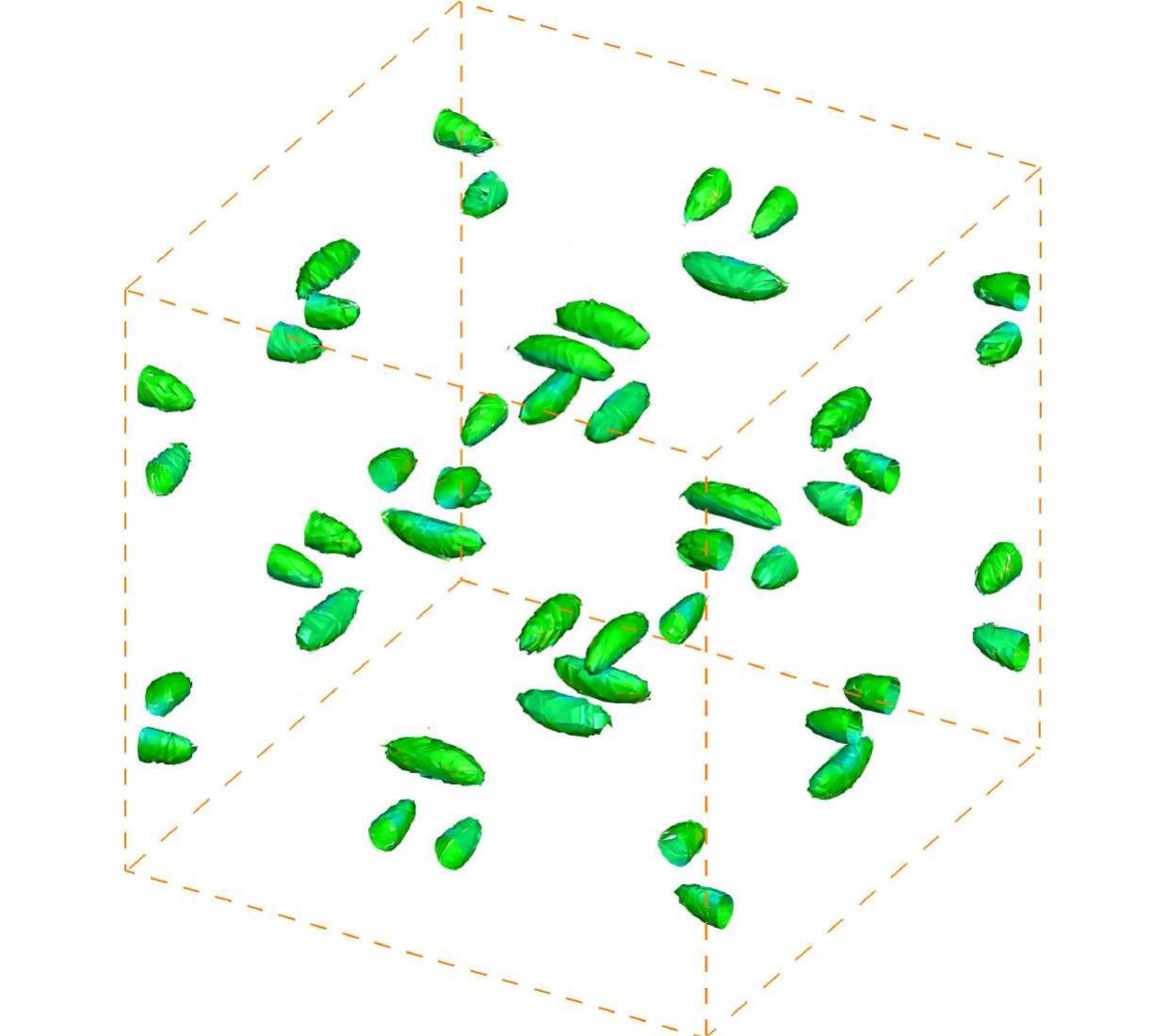} }}
    \caption{Taylor–Green vortex: isocontours of the vorticity magnitude $L / V_{0} \cdot \vert \bm{\omega}\vert = 1.0$ at $Re=100$, computed on the mesh shown in Figure \ref{fig:Couette test case}(a) with $(m,p)=(4,2)$.}%
    \label{fig:TGV RE100}%
\end{figure}

\subsubsection{Verification of accuracy}

We first use the Taylor-Green vortex benchmark to investigate the accuracy that we can obtain with our macro-element HDG method. In terms of discretization accuracy and from a physical perspective, the focus lies on the kinetic energy dissipation rate, illustrated in Figure \ref{fig: TGV Re}. {\color{black} In this context,  the kinetic energy dissipation rate, $\epsilon$, is exactly equal to 
\begin{equation}\label{C5321}
\epsilon = 2\dfrac{\mu}{\Omega} \int_{\Omega} \rho\: \dfrac{\bm{\omega} \cdot \bm{\omega}}{2} \:\mathrm{d}\Omega,
\end{equation} 
for incompressible flow and approximately for compressible flow at low Mach number. The vorticity, $\omega$, is defined as $\omega=\nabla \times V$ in \(\ref{C5321}\).}
We consider the time range $0 \leq t/t_{c}  \leq 15$ and the two Reynolds numbers $Re=100$ and $Re= 400$. The obtained results will be compared against a reference incompressible flow solution \cite{arndt2020exadg}.

The temporal evolution of the flow field is illustrated through isocontours of vorticity magnitude, specifically $L/V_{0} \cdot \vert\omega\vert = 1.0$, as shown in Figure \ref{fig:TGV RE100} for the case of $Re=100$, computed with the macro-element HDG method with the shown mesh and $(m,p)=(4,2)$. In the early stages, corresponding to the initial time, the large-scale vortex structures initiate their evolution and exhibit a rolling-up phenomenon. Around the non-dimensional time instant $t/t_{c} = 6$, the smooth vertical structures give rise to more coherent formations, and by approximately $t/t_{c} = 9$, these coherent structures commence breaking down. Also, a snapshot of the vorticity magnitude $\vert\omega\vert$ on the periodic plane $x = \pi L$, computed with $(m,p)=(4,2)$ at the non-dimensional time instants $t/t_{c} = 3.0, 9.0 , 12.0$ and $Re = 100$ is plotted in Figure \ref{fig:Norm Vorticity}. We conclude from our results that for the benchmark at the chosen parameters, the macro-element HDG method delivers very good accuracy with the chosen mesh resolution and the polynomial degrees.

 \begin{figure}
    \centering
    \subfloat[\centering $t/t_{c} = 3.0$  ]{{\includegraphics[width=0.35\textwidth]{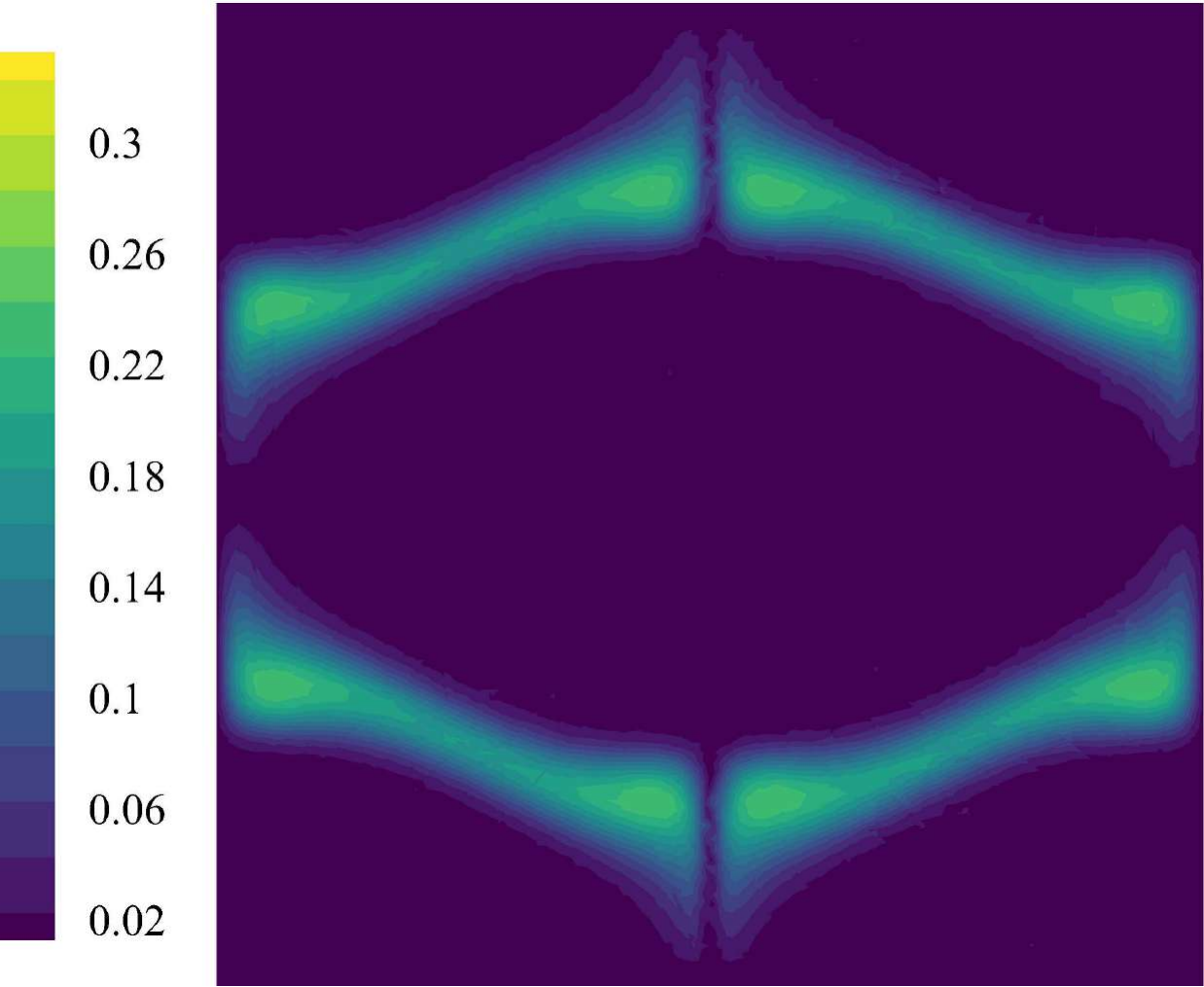} }}%
    \subfloat[\centering $t/t_{c} = 9.0$  ]{{\includegraphics[width=0.325\textwidth]{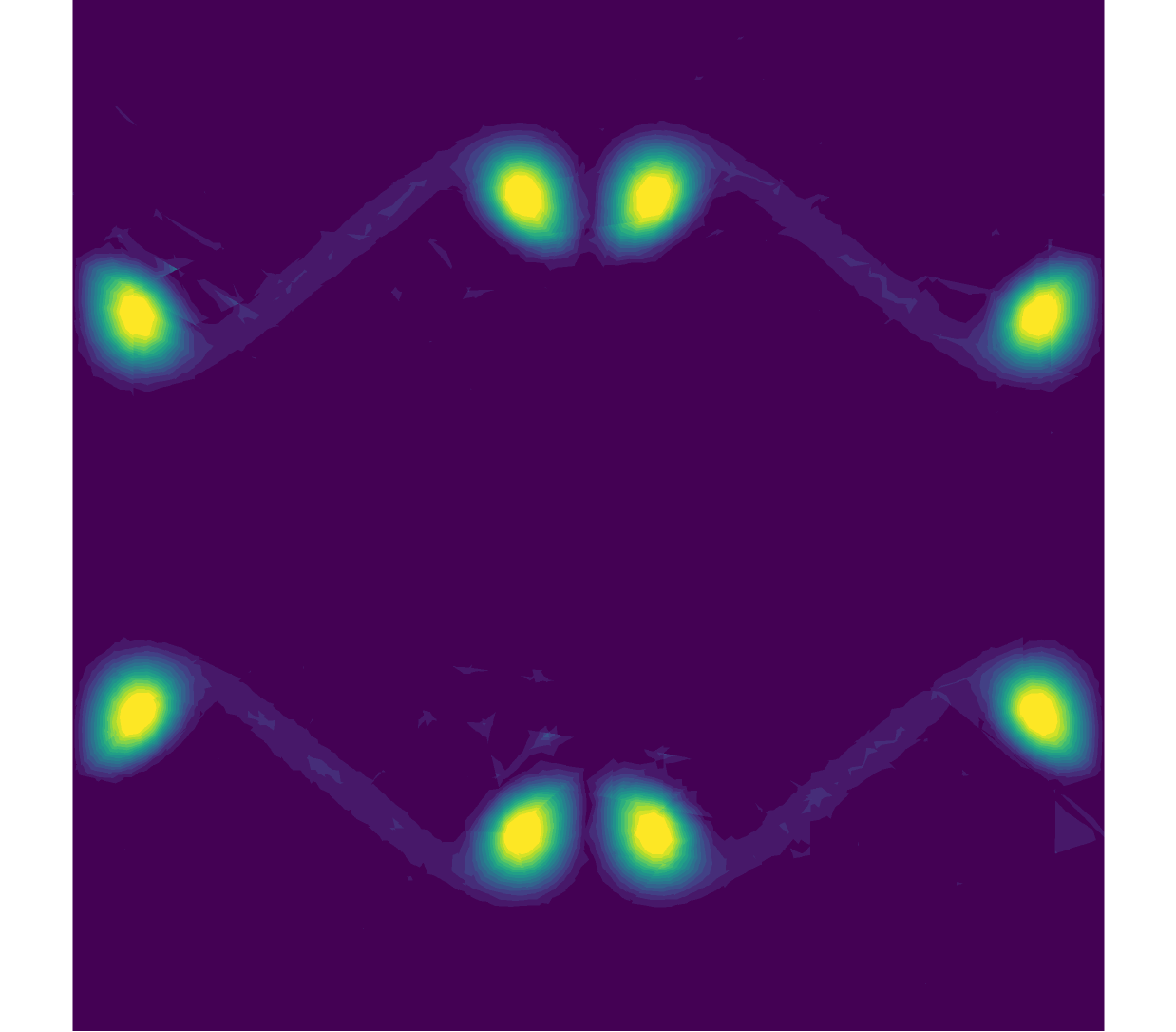} }}%
    \subfloat[\centering $t/t_{c} = 15.0$  ]{{\includegraphics[width=0.325\textwidth]{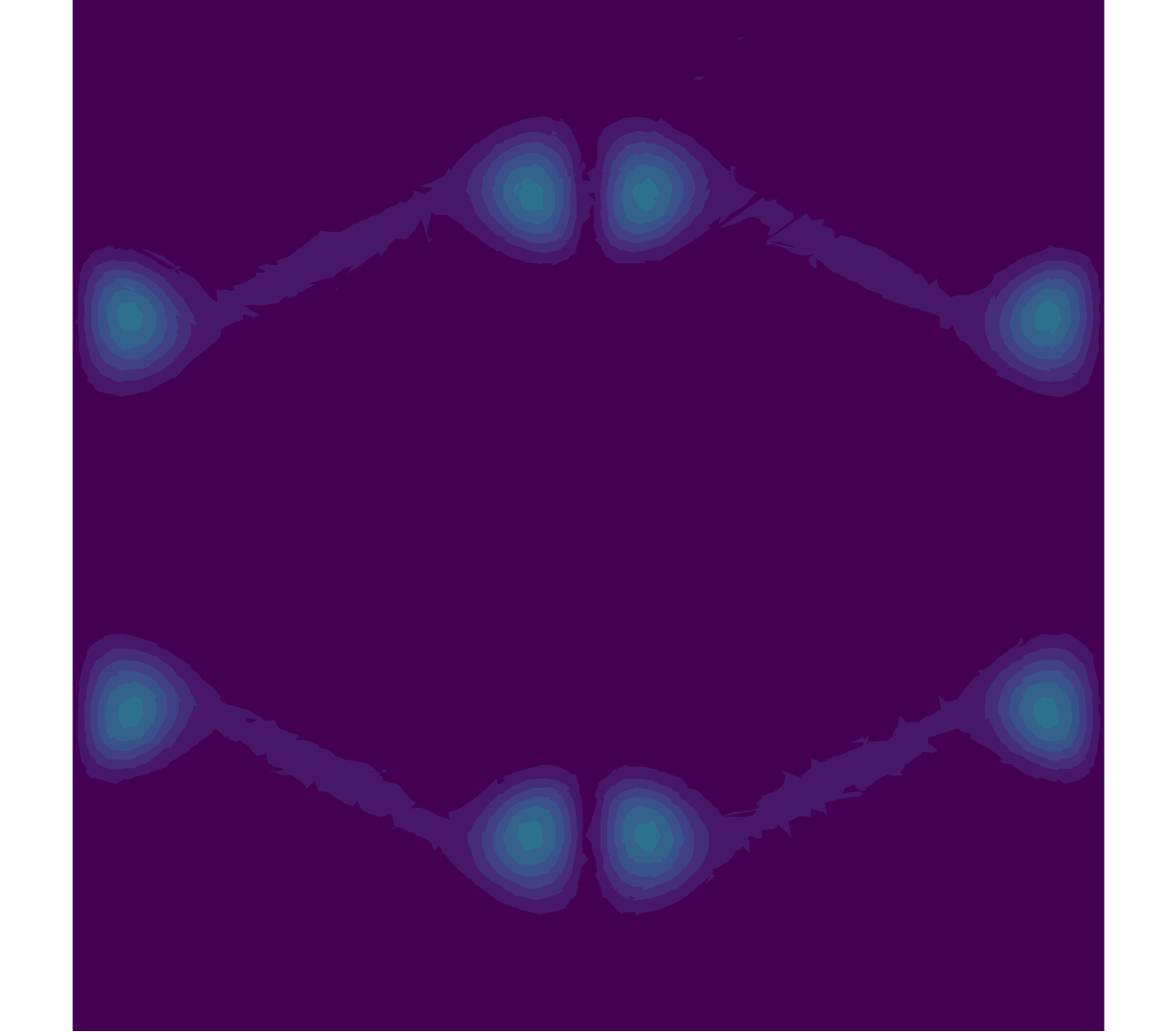} }}%
    \caption{Taylor–Green vortex: snapshot of the vorticity magnitude $\vert\omega\vert$ at $Re=100$, plotted on the periodic plane $x = \pi L$ for three non-dimensional time instants $t/t_{c} = 3.0, 9.0 , 12.0$, computed on the mesh shown in Figure \ref{fig:Couette test case}(a) with $(m,p)=(4,2)$.}
    \label{fig:Norm Vorticity}%
\end{figure}

\subsubsection{Assessment of computational efficiency}

In the next step, we investigate the computational efficiency of the macro-element HDG method for the chosen meshes and polynomial degrees.
In Table \ref{Tab:TGV,TwoRe}, we report the time and number of iterations for the FGMRES linear solver with second-layer static condensation. In particular, we compare timings of the macro-element HDG method with eight elements per macro-element ($m=2$) and the macro-element HDG method with 64 elements per macro-element ($m=4$). We observe that for both mesh resolutions considered here, the computational time required for the global solver decreases as $m$ is increased. This reduction can be expected since the number of FGMRES iterations decreases and the structure of the global matrix changes. We note that this reduction becomes more pronounced with the increase of the mesh resolution for the higher Reynolds number $Re=400$. This improvement can be attributed to an improved performance of our macro-element HDG implementation for larger mesh sizes and the deployment of a larger number of parallel processes.

 \begin{table}
\caption{Taylor-Green vortex: we compare the time for the local solver and the local part of the matrix-free global solver (step 2) vs.\ the time for the remaining parts of the global solver for $Re = 100$ and $Re = 400$.}
\centering
\begin{tabular}{c | c | cc|cc|cc}\toprule
& \multirow{2}{*}{\# Proc's} & \multicolumn{2}{c|}{Time local op's [min]} & \multicolumn{2}{c|}{Time global op's [min]} & \multicolumn{2}{c}{\# iterations} \\
\cmidrule(lr){3-4}\cmidrule(lr){5-6}\cmidrule(lr){7-8}
          				&	& $(m,p) = (2,4)$ 	& $(m,p) = (4,2)$   	    
          					& $(m,p) = (2,4)$		& $(m,p) = (4,2)$    	 	
          					& $(m,p) = (2,4)$  	 & $(m,p) = (4,2)$ 		 \\\midrule         
 $Re = 100$  & 796        	&7.7			&4.1					
 							&18.3			&8.5				
 							&3,863		    &2,314		    \\[5pt]							
 $Re = 400$ & 	5,088		    &29.9			&21.2							
 							&114.1			&66.3							
 				 			&9,604			&8,513	       	    \\\bottomrule	 								
\end{tabular}
\label{Tab:TGV,TwoRe}%
\end{table}


\begin{table}
\caption{Taylor-Green vortex: number of local and global unknowns for a sequence of three meshes, where the last one corresponds to the mesh shown in Figure \ref{fig:Couette test case}(a). These values hold for both $(m,p) = (2,4)$ and $(m,p) = (4,2)$.} 
\centering
\begin{tabular}{l|ccc|cc}\toprule  
 & \multicolumn{3}{c|}{Mesh} & \multicolumn{2}{c}{$\text{\# degrees of freedom}$} \\
\cmidrule(lr){2-4} \cmidrule(lr){5-6} 
	&$N_{ele,1d}$ & $N_{\text{eff}}$  &$N^{\textit{Mcr}}$ 					& $\text{dof}^{local}$   		& $\text{dof}^{global}$  	       		\\\midrule        
Mesh 1 	&4	&$36^{3}$			&495								&1,633,500		&222,750				 \\[5pt] 
Mesh 2 	&5	&$45^{3}$					&955								&3,151,500		&429,750		\\[5pt] 
Mesh 3  	&6	&$54^{3}$						&1592		            			&5,253,600		&716,400				
\\\bottomrule
\end{tabular}
\label{Tab:DOfs Dif Mesh TGV}%
\end{table}  

We then compare the runtime performance of the two macro-element HDG variants with $h$ refinement, {\color{black} see Figure \ref{fig: rate Dif Mesh TGV}}. This comparison is based on computing time for both local and global operations, as well as the number of FGMRES iterations. We compute our example at $Re=100$, and transition to a variable number of processors while maintaining a fixed ratio of the number of macro-elements to the number of processors at two (macro-elements / processors = 2). Table \ref{Tab:DOfs Dif Mesh TGV} reports the number of degrees of freedom for a sequence of three meshes generated by globally increasing the number of macro-elements, $N_{ele,1d}$, in each spatial direction. Both macro-element HDG variants, which use $(m,p)=(2,4)$ and $(m,p)=(4,2)$ utilize the same macro-element meshes, hence the same number of macro-elements $N^{\textit{Mcr}}$, and exhibit the same number of degrees of freedom. Figure \ref{fig: rate Dif Mesh TGV} illustrates that with mesh refinement, the accuracy of both macro-element HDG variants with $(m,p)=(2,4)$ and $(m,p)=(4,2)$ improves and approaches the DNS reference solution. We observe a slight accuracy advantage of the macro-element HDG method that uses $(m,p)=(4,2)$, in particular for the coarsest macro-element mesh (Mesh 1).

\begin{table}
\caption{Taylor-Green vortex at $Re = 100$: we compare the time for the local solver and the local part of the matrix-free global solver (step 2) vs.\ the time for the remaining parts of the global solver. We use the macro-element HDG method with $(m,p)=(2,4)$ and $(m,p)=(4,2)$ (at constant ratio macro-elements / proc's = 2).}
\centering
\begin{tabular}{ l c |cc|cc|cc}\toprule
 & \multirow{2}{*}{\# Proc's} & \multicolumn{2}{c|}{Time local op's [min]} & \multicolumn{2}{c|}{Time global op's [min]} & \multicolumn{2}{c}{\# iterations }\\ 
\cmidrule(lr){3-4}\cmidrule(lr){5-6}\cmidrule(lr){7-8}
          	&				 	& $(m,p) = (2,4)$ 		& $(m,p) = (4,2)$  	    
          						& $(m,p) = (2,4)$			& $(m,p) = (4,2)$    	 	
          						& $(m,p) = (2,4)$  	 	& $(m,p) = (4,2)$		 \\\midrule         
 Mesh 1  	   &	248	 	&5.0					&2.8					
 								&8.3					&4.2				
 								&2,520		    		&1,595		    \\[5pt]
 					
 Mesh 2  		&	478	&6.6					&3.6					
 								&11.9				&5.9				
 								&3,339	    			&1,993		    \\[5pt]	
 									
 Mesh 3  		&	796	&7.7					&4.1					
 								&18.3				&8.5				
 								&3,863	        		&2,314		    \\\bottomrule	 								
\end{tabular}
\label{Tab:Time Dif Mesh TGV}%
\end{table}

Table \ref{Tab:Time Dif Mesh TGV} compares the time for the local parts of the solver and the remaining parts of the global solver, and also reports the number of FGMRES iterations for each case. We observe that at a fixed number of degrees of freedom, using 64 $C^0$-continuous elements within each macro-element at $p=2$ is results in a significant reduction in computing time in comparison to using just 8 $C^0$-continuous elements per macro-element at a higher polynomial degree of $p=4$. An essential factor contributing to this reduction is the improved conditioning of the smaller system, leading to a substantial decrease in the required number of iterations for the FGMRES solver. For the largest mesh with an effective resolution of $N_{\text{eff}}=54^{3}$, the time for the global operations in the macro-element HDG method with $m=4$ is approximately a factor of two smaller than that of the macro-element HDG method with $m=2$, primarily due to the reduction in the number of iterations from 3,863 to just 2,314.

Moreover, we observe that for $m=4$, the ratio between the computing times for local and global operations is closer to the desired optimum of one. Consequently, we conclude that the macro-element HDG method with a larger number of $C^0$-continuous elements per macro-element, owing to its flexibility in adjusting the computational load per macro-element, in general achieves a better balance between local and global operations. This observation indicates that for very large systems, larger macro-elements (at a lower polynomial degree) are preferable to balance local and global operations, leading to the fastest computing times for the overall problem.

 \begin{figure}
    \centering
    \subfloat[\centering Kinetic energy dissipation rates for $(m, p) =(4,2)$]{{\includegraphics[width=0.5\textwidth]{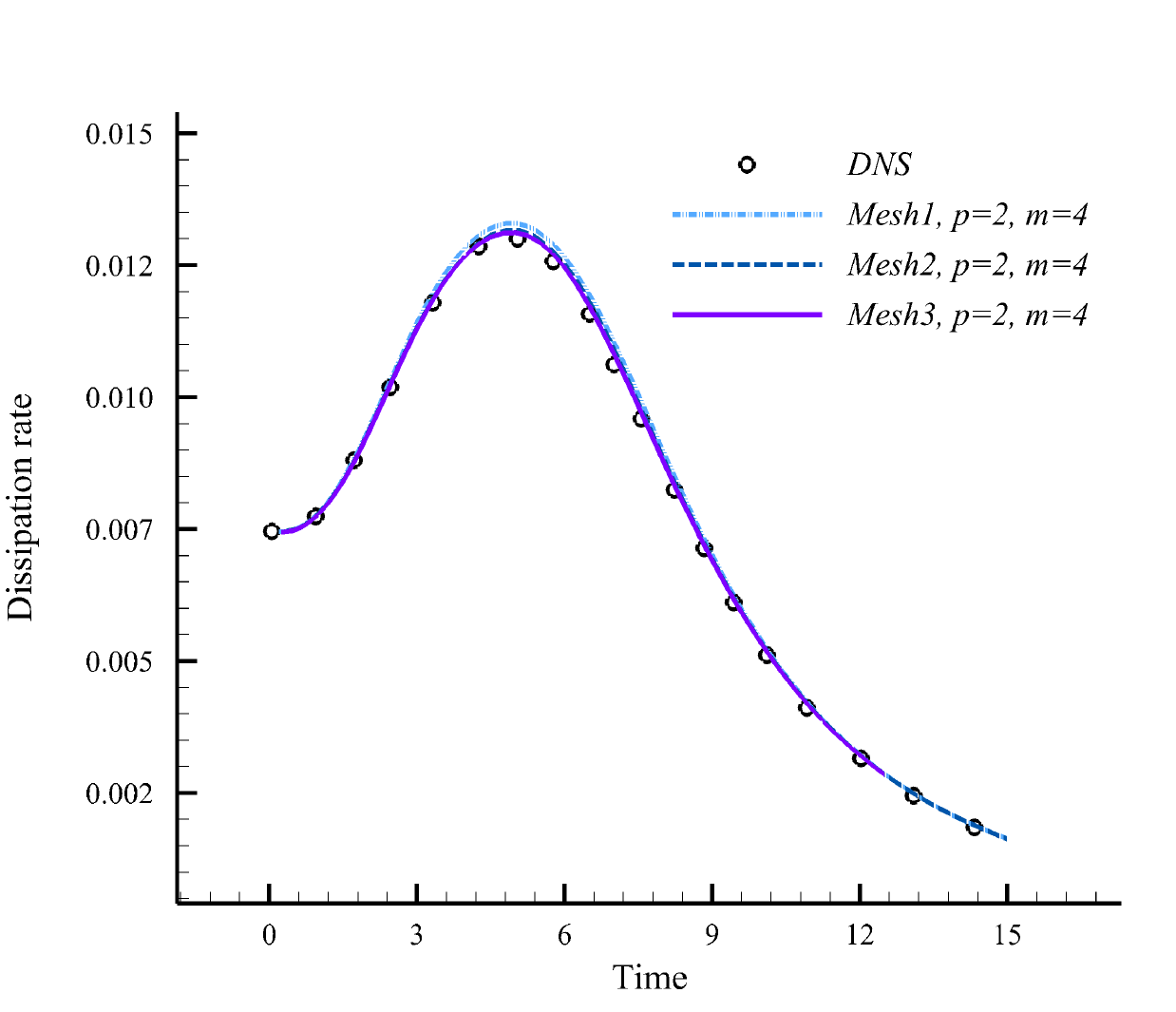} }}%
    \subfloat[\centering Kinetic energy dissipation rates for $(m, p) =(2,4)$]{{\includegraphics[width=0.5\textwidth]{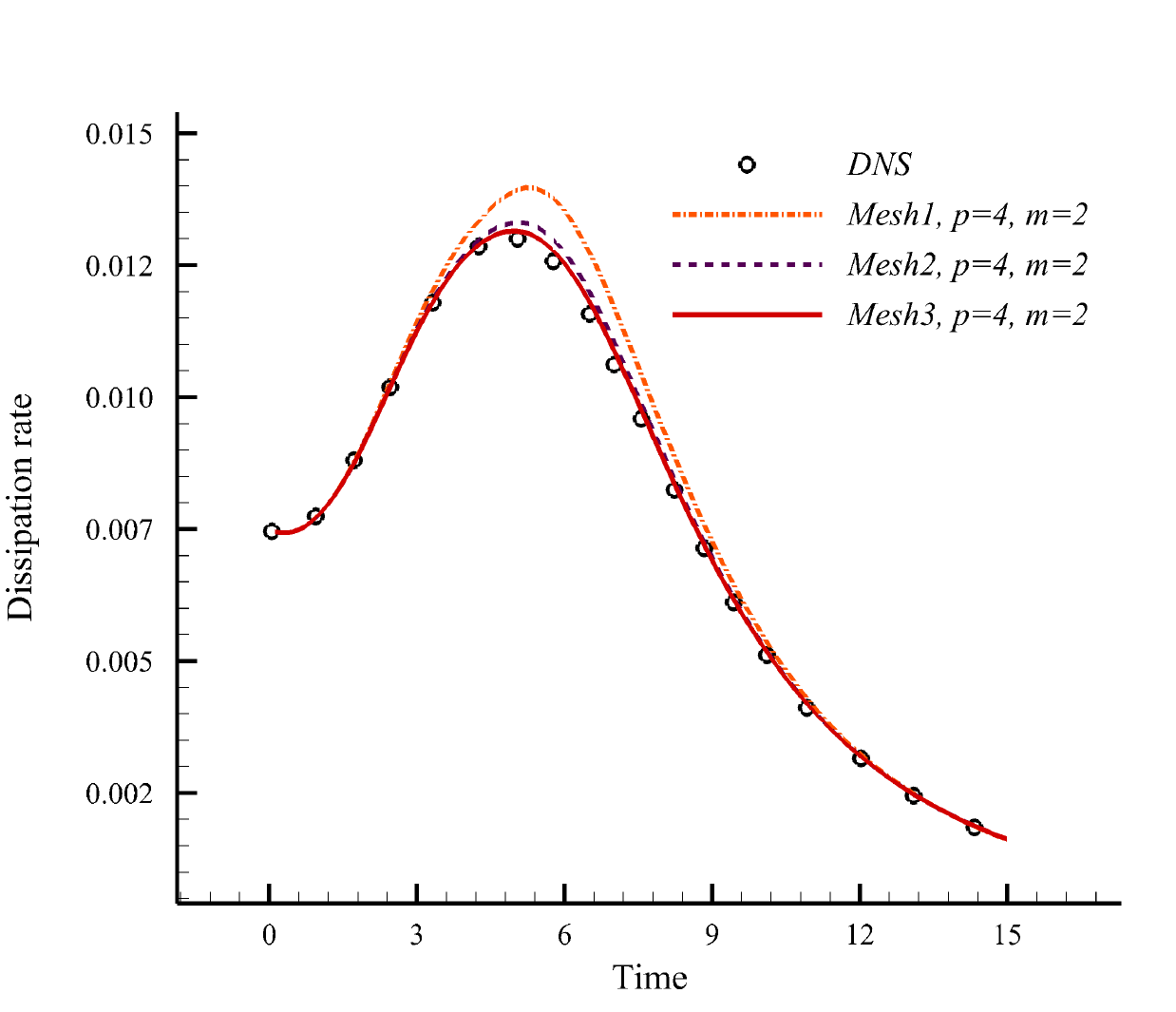} }}%
    \caption{Taylor-Green vortex: time evolution of kinetic energy dissipation rates for $Re = 100$, obtained with the macro-element HDG method and the two different $(m,p)$ pairs on the three different macro-element meshes defined in Table \ref{Tab:DOfs Dif Mesh TGV}.} %
    \label{fig: rate Dif Mesh TGV}%
\end{figure}

\section{Summary and conclusions}

In this paper, we focused on the macro-element variant of the hybridized discontinuous Galerkin (HDG) method and investigated its performance in the analysis of steady and unsteady compressible flow problems at moderate Reynolds numbers. Combining aspects of continuous and hybridized discontinuous finite element discretization, the macro-element HDG variant offers a number of advantages compared to the standard HDG method, such as the mitigation of the proliferation of degrees of freedom, the preservation of the unique domain decomposition mechanism, and automatic load-balancing inherent to the numerical method. In addition, the macro-element HDG method is particularly well suited for a matrix-free solution approach.

In comparison to our earlier work on scalar advection-diffusion problems, compressible flow problems lead to significantly larger systems of equations, due to the requirement of fine meshes with many elements as well as due to the $\left( d+1\right)\times  \left( d+2\right)$  degrees of freedom per basis function. We therefore explored several computational strategies at the level of the local and the global solver, with the aim at enhancing computational efficiency and reducing memory requirements.
For solving the local system per macro-element efficiently, we devised a second-layer static condensation approach that reduces the size of the local system matrix in each macro-element and hence the factorization time of the local solver. For solving the global system efficiently within a matrix-free implementation, we explored the use of a multi-level preconditioner based on the FGMRES method. Based on the multi-level FGMRES implementation available in PETSc, we used the iterative GMRES solver as a preconditioner, and in the second level, we again employed the inverse of the global system matrix computed via a matrix-free Cholesky factorization as a preconditioner for the GMRES solver.


We demonstrated the performance of our developments via parallel implementation of our macro-element HDG variant in Julia and C$^{++}$ that we ported on a modern heterogeneous compute system (Lichtenberg II Phase 1 at the Technical University of Darmstadt, at position 253 in the TOP500 list 11/2023). Our computational results showed that the multi-level FGMRES iterative solver in conjunction with the second-layer static condensation approach indeed enhance the computational efficiency of the macro-element HDG method, benefitting from the reduction in degrees of freedom and communication across compute nodes. Our results also confirmed that unlike standard HDG, the macro-element HDG method is efficient for moderate polynomial degrees such as quadratics, as it is possible to increase the local computational load per macro-element irrespective of the polynomial degree. 
In the context of compressible flow simulations, we observed a shift in computational workload from the global solver to the local solver. This shift helps achieve a balance of local and global operations, enhancing parallelization efficiency as compared to our earlier work on scalar advection-diffusion problems. We demonstrated that - due to this balance, the reduction in degrees of freedom, and the reduction of the global problem size and the number of iterations for its solution - the macro-element HDG method delivers faster computing times than the standard HDG method, particularly for higher Reynolds numbers and mesh resolutions. In particular, we observed that for  large-scale compressible flow computations, our macro-element variant of the HDG method is more efficient, when it employs macro-elements with more $C^0$-continuous elements, but at a lower polynomial degree, rather than using macro-elements with fewer $C^0$-continuous elements, but at a higher polynomial degree.

It remains to be seen how these properties demonstrated here for moderate Reynolds numbers transfer to high-Reynolds-number flow problems that involve turbulence. We plan to investigate this question in the future by applying the macro-element HDG method for the direct
numerical simulation of turbulent flows modeled via the compressible Navier-Stokes equations.

\section*{acknowledgements}
The authors gratefully acknowledge financial support from the German Research Foundation (Deutsche Forschungsgemeinschaft) through the DFG Emmy Noether Grant SCH 1249/2-1. The authors also gratefully acknowledge the computing time provided to them on the high-performance computer Lichtenberg at the NHR Centers NHR4CES at TU Darmstadt. This is funded by the Federal Ministry of Education and Research and the State of Hesse.

\bibliography{sections/references.bib}

\begin{thebibliography}{10}
\providecommand \doibase [0]{http://dx.doi.org/}%

\bibitem{arnold2002unified}
Arnold DN, Brezzi F, Cockburn B, Marini LD. Unified analysis of discontinuous
  Galerkin methods for elliptic problems. {\it SIAM journal on numerical
  analysis} 2002\string; 39(5)\string: 1749--1779.

\bibitem{bassi1997high}
Bassi F, Rebay S. A high-order accurate discontinuous finite element method for
  the numerical solution of the compressible Navier--Stokes equations. {\it
  Journal of computational physics} 1997\string; 131(2)\string: 267--279.

\bibitem{cockburn2018discontinuous}
Cockburn B. Discontinuous Galerkin methods for computational fluid dynamics.
  {\it Encyclopedia of Computational Mechanics Second Edition} 2018\string:
  1--63.

\bibitem{hesthaven2007nodal}
Hesthaven JS, Warburton T. {\it Nodal discontinuous Galerkin methods:
  algorithms, analysis, and applications}.
\newblock Springer Science \& Business Media .
\newblock 2007.

\bibitem{peraire2008compact}
Peraire J, Persson PO. The compact discontinuous Galerkin (CDG) method for
  elliptic problems. {\it SIAM Journal on Scientific Computing} 2008\string;
  30(4)\string: 1806--1824.

\bibitem{cockburn2009unified}
Cockburn B, Gopalakrishnan J, Lazarov R. Unified hybridization of discontinuous
  Galerkin, mixed, and continuous Galerkin methods for second order elliptic
  problems. {\it SIAM Journal on Numerical Analysis} 2009\string; 47(2)\string:
  1319--1365.

\bibitem{nguyen2009implicit}
Nguyen NC, Peraire J, Cockburn B. An implicit high-order hybridizable
  discontinuous Galerkin method for linear convection--diffusion equations.
  {\it Journal of Computational Physics} 2009\string; 228(9)\string:
  3232--3254.

\bibitem{cockburn2009superconvergent}
Cockburn B, Guzm{\'a}n J, Wang H. Superconvergent discontinuous Galerkin
  methods for second-order elliptic problems. {\it Mathematics of Computation}
  2009\string; 78(265)\string: 1--24.

\bibitem{nguyen2012hybridizable}
Nguyen NC, Peraire J. Hybridizable discontinuous Galerkin methods for partial
  differential equations in continuum mechanics. {\it Journal of Computational
  Physics} 2012\string; 231(18)\string: 5955--5988.

\bibitem{peraire2010hybridizable}
Peraire J, Nguyen N, Cockburn B. A hybridizable discontinuous Galerkin method
  for the compressible Euler and Navier-Stokes equations. In:  {\it 48th AIAA
  aerospace sciences meeting including the new horizons forum and aerospace
  exposition}; 2010\string: 363.

\bibitem{cockburn2009hybridizable}
Cockburn B, Dong B, Guzm{\'a}n J, Restelli M, Sacco R. A hybridizable
  discontinuous Galerkin method for steady-state convection-diffusion-reaction
  problems. {\it SIAM Journal on Scientific Computing} 2009\string;
  31(5)\string: 3827--3846.

\bibitem{nguyen2011implicit}
Nguyen NC, Peraire J, Cockburn B. An implicit high-order hybridizable
  discontinuous Galerkin method for the incompressible Navier--Stokes
  equations. {\it Journal of Computational Physics} 2011\string; 230(4)\string:
  1147--1170.

\bibitem{fernandez2017hybridized}
Fernandez P, Nguyen NC, Peraire J. The hybridized discontinuous Galerkin method
  for implicit large-eddy simulation of transitional turbulent flows. {\it
  Journal of Computational Physics} 2017\string; 336\string: 308--329.

\bibitem{vila2021hybridisable}
Vila-P{\'e}rez J, Giacomini M, Sevilla R, Huerta A. Hybridisable discontinuous
  Galerkin formulation of compressible flows. {\it Archives of Computational
  Methods in Engineering} 2021\string; 28(2)\string: 753--784.

\bibitem{nguyen2011high}
Nguyen NC, Peraire J, Cockburn B. High-order implicit hybridizable
  discontinuous Galerkin methods for acoustics and elastodynamics. {\it Journal
  of Computational Physics} 2011\string; 230(10)\string: 3695--3718.

\bibitem{pazner2017stage}
Pazner W, Persson PO. Stage-parallel fully implicit Runge--Kutta solvers for
  discontinuous Galerkin fluid simulations. {\it Journal of Computational
  Physics} 2017\string; 335\string: 700--717.

\bibitem{kronbichler2018performance}
Kronbichler M, Wall WA. A performance comparison of continuous and
  discontinuous Galerkin methods with fast multigrid solvers. {\it SIAM Journal
  on Scientific Computing} 2018\string; 40(5)\string: A3423--A3448.

\bibitem{fabien2019manycore}
Fabien MS, Knepley MG, Mills RT, Rivi{\`e}re BM. Manycore parallel computing
  for a hybridizable discontinuous Galerkin nested multigrid method. {\it SIAM
  Journal on Scientific Computing} 2019\string; 41(2)\string: C73--C96.

\bibitem{roca2013scalable}
Roca X, Nguyen C, Peraire J. Scalable parallelization of the hybridized
  discontinuous Galerkin method for compressible flow. In:  {\it 21st AIAA
  Computational Fluid Dynamics Conference}; 2013\string: 2939.

\bibitem{roca2011gpu}
Roca X, Nguyen NC, Peraire J. GPU-accelerated sparse matrix-vector product for
  a hybridizable discontinuous Galerkin method. In:  {\it 49th AIAA Aerospace
  Sciences Meeting including the New Horizons Forum and Aerospace Exposition};
  2011\string: 687.

\bibitem{hughes2012finite}
Hughes TJ. {\it The finite element method: linear static and dynamic finite
  element analysis}.
\newblock Courier Corporation .
\newblock 2012.

\bibitem{paipuri2019coupling}
Paipuri M, Tiago C, Fern{\'a}ndez-M{\'e}ndez S. Coupling of continuous and
  hybridizable discontinuous Galerkin methods: Application to conjugate heat
  transfer problem. {\it Journal of Scientific Computing} 2019\string;
  78(1)\string: 321--350.

\bibitem{kirby2012cg}
Kirby RM, Sherwin SJ, Cockburn B. To CG or to HDG: a comparative study. {\it
  Journal of Scientific Computing} 2012\string; 51\string: 183--212.

\bibitem{yakovlev2016cg}
Yakovlev S, Moxey D, Kirby RM, Sherwin SJ. To CG or to HDG: a comparative study
  in 3D. {\it Journal of Scientific Computing} 2016\string; 67(1)\string:
  192--220.

\bibitem{badrkhani2023matrix}
Badrkhani V, Hiemstra RR, Mika M, Schillinger D. A matrix-free macro-element
  variant of the hybridized discontinuous Galerkin method. {\it International
  Journal for Numerical Methods in Engineering} 2023\string; 124(20)\string:
  4427--4452.

\bibitem{kronbichler2019fast}
Kronbichler M, Kormann K, Wall WA. Fast matrix-free evaluation of hybridizable
  discontinuous Galerkin operators. In:  {\it Numerical Mathematics and
  Advanced Applications ENUMATH 2017}Springer. ; 2019\string: 581--589.

\bibitem{lermusiaux2007environmental}
Lermusiaux PF, Haley~Jr PJ, Yilmaz NK. Environmental prediction, path planning
  and adaptive sampling-sensing and modeling for efficient ocean monitoring,
  management and pollution control. {\it Sea Technology} 2007\string;
  48(9)\string: 35--38.

\bibitem{HDGLES}
Fernandez P, Nguyen N, Peraire J. The hybridized Discontinuous Galerkin method
  for Implicit Large-Eddy Simulation of transitional turbulent flows. {\it
  Journal of Computational Physics} 2017\string; 336\string: 308-329.
\newblock \href {\doibase 10.1016/j.jcp.2017.02.015} {doi:
  10.1016/j.jcp.2017.02.015}

\bibitem{brooks1982streamline}
Brooks AN, Hughes TJ. Streamline upwind/Petrov-Galerkin formulations for
  convection dominated flows with particular emphasis on the incompressible
  Navier-Stokes equations. {\it Computer methods in applied mechanics and
  engineering} 1982\string; 32(1-3)\string: 199--259.

\bibitem{donea2003finite}
Donea J, Huerta A. {\it Finite element methods for flow problems}.
\newblock John Wiley \& Sons .
\newblock 2003.

\bibitem{xu2017compressible}
Xu F, Moutsanidis G, Kamensky D, et al. Compressible flows on moving domains:
  stabilized methods, weakly enforced essential boundary conditions, sliding
  interfaces, and application to gas-turbine modeling. {\it Computers \&
  Fluids} 2017\string; 158\string: 201--220.

\bibitem{shakib1991new}
Shakib F, Hughes TJ, Johan Z. A new finite element formulation for
  computational fluid dynamics: X. The compressible Euler and Navier-Stokes
  equations. {\it Computer Methods in Applied Mechanics and Engineering}
  1991\string; 89(1-3)\string: 141--219.

\bibitem{tezduyar2006stabilization}
Tezduyar TE, Senga M. Stabilization and shock-capturing parameters in SUPG
  formulation of compressible flows. {\it Computer methods in applied mechanics
  and engineering} 2006\string; 195(13-16)\string: 1621--1632.

\bibitem{tezduyar2006computation}
Tezduyar TE, Senga M, Vicker D. Computation of inviscid supersonic flows around
  cylinders and spheres with the SUPG formulation and YZ $\beta$
  shock-capturing. {\it Computational Mechanics} 2006\string; 38\string:
  469--481.

\bibitem{alexander1977diagonally}
Alexander R. Diagonally implicit Runge--Kutta methods for stiff ODE’s. {\it
  SIAM Journal on Numerical Analysis} 1977\string; 14(6)\string: 1006--1021.

\bibitem{bijl2002implicit}
Bijl H, Carpenter MH, Vatsa VN, Kennedy CA. Implicit time integration schemes
  for the unsteady compressible Navier--Stokes equations: laminar flow. {\it
  Journal of Computational Physics} 2002\string; 179(1)\string: 313--329.

\bibitem{jameson1991time}
Jameson A. Time dependent calculations using multigrid, with applications to
  unsteady flows past airfoils and wings. In:  {\it 10th Computational fluid
  dynamics conference}; 1991\string: 1596.

\bibitem{mulder1985experiments}
Mulder WA, Van~Leer B. Experiments with implicit upwind methods for the Euler
  equations. {\it Journal of Computational Physics} 1985\string; 59(2)\string:
  232--246.

\bibitem{cockburn2016static}
Cockburn B. Static condensation, hybridization, and the devising of the HDG
  methods. In:  {\it Building bridges: connections and challenges in modern
  approaches to numerical partial differential equations}Springer.  2016 (pp.
  129--177).

\bibitem{nguyen2015class}
Nguyen NC, Peraire J, Cockburn B. A class of embedded discontinuous Galerkin
  methods for computational fluid dynamics. {\it Journal of Computational
  Physics} 2015\string; 302\string: 674--692.

\bibitem{saad1986gmres}
Saad Y, Schultz MH. GMRES: A generalized minimal residual algorithm for solving
  nonsymmetric linear systems. {\it SIAM Journal on scientific and statistical
  computing} 1986\string; 7(3)\string: 856--869.

\bibitem{saad1993flexible}
Saad Y. A flexible inner-outer preconditioned GMRES algorithm. {\it SIAM
  Journal on Scientific Computing} 1993\string; 14(2)\string: 461--469.

\bibitem{wang2007implicit}
Wang L, Mavriplis DJ. Implicit solution of the unsteady Euler equations for
  high-order accurate discontinuous Galerkin discretizations. {\it Journal of
  Computational Physics} 2007\string; 225(2)\string: 1994--2015.

\bibitem{diosady2009preconditioning}
Diosady LT, Darmofal DL. Preconditioning methods for discontinuous Galerkin
  solutions of the Navier--Stokes equations. {\it Journal of Computational
  Physics} 2009\string; 228(11)\string: 3917--3935.

\bibitem{fidkowski2005p}
Fidkowski KJ, Oliver TA, Lu J, Darmofal DL. p-Multigrid solution of high-order
  discontinuous Galerkin discretizations of the compressible Navier--Stokes
  equations. {\it Journal of Computational Physics} 2005\string; 207(1)\string:
  92--113.

\bibitem{badrkhani2017development}
Badrkhani V, Nejat A, Tahani M. Development of implicit Algorithm for
  High-order discontinuous Galerkin methods to solve compressible flows using
  Newton-Krylov methods. {\it Modares Mechanical Engineering} 2017\string;
  17(3)\string: 281--292.

\bibitem{botti2017h}
Botti L, Colombo A, Bassi F. h-multigrid agglomeration based solution
  strategies for discontinuous Galerkin discretizations of incompressible flow
  problems. {\it Journal of Computational Physics} 2017\string; 347\string:
  382--415.

\bibitem{franciolini2020p}
Franciolini M, Botti L, Colombo A, Crivellini A. p-Multigrid matrix-free
  discontinuous Galerkin solution strategies for the under-resolved simulation
  of incompressible turbulent flows. {\it Computers \& Fluids} 2020\string;
  206\string: 104558.

\bibitem{vakilipour2019developing}
Vakilipour S, Mohammadi M, Badrkhani V, Ormiston S. Developing a physical
  influence upwind scheme for pressure-based cell-centered finite volume
  methods. {\it International Journal for Numerical Methods in Fluids}
  2019\string; 89(1-2)\string: 43--70.

\bibitem{davis2004algorithm}
Davis TA. Algorithm 832: UMFPACK V4. 3---an unsymmetric-pattern multifrontal
  method. {\it ACM Transactions on Mathematical Software (TOMS)} 2004\string;
  30(2)\string: 196--199.

\bibitem{schutz2012hybridized}
Sch{\"u}tz J, Woopen M, May G. A hybridized DG/mixed scheme for nonlinear
  advection-diffusion systems, including the compressible Navier-Stokes
  equations. In:  {\it 50th AIAA Aerospace Sciences Meeting including the New
  Horizons Forum and Aerospace Exposition}; 2012\string: 729.

\bibitem{franciolini2020efficient}
Franciolini M, Fidkowski KJ, Crivellini A. Efficient discontinuous Galerkin
  implementations and preconditioners for implicit unsteady compressible flow
  simulations. {\it Computers \& Fluids} 2020\string; 203\string: 104542.

\bibitem{johnson1999flow}
Johnson T, Patel V. Flow past a sphere up to a Reynolds number of 300. {\it
  Journal of Fluid Mechanics} 1999\string; 378\string: 19--70.

\bibitem{taneda1956experimental}
Taneda S. Experimental investigation of the wake behind a sphere at low
  Reynolds numbers. {\it Journal of the physical society of Japan} 1956\string;
  11(10)\string: 1104--1108.

\bibitem{taylor1937mechanism}
Taylor GI, Green AE. Mechanism of the production of small eddies from large
  ones. {\it Proceedings of the Royal Society of London. Series A-Mathematical
  and Physical Sciences} 1937\string; 158(895)\string: 499--521.

\bibitem{brachet1991direct}
Brachet M. Direct simulation of three-dimensional turbulence in the
  Taylor--Green vortex. {\it Fluid dynamics research} 1991\string;
  8(1-4)\string: 1.

\bibitem{fehn2018efficiency}
Fehn N, Wall WA, Kronbichler M. Efficiency of high-performance discontinuous
  Galerkin spectral element methods for under-resolved turbulent incompressible
  flows. {\it International Journal for Numerical Methods in Fluids}
  2018\string; 88(1)\string: 32--54.

\bibitem{arndt2020exadg}
Arndt D, Fehn N, Kanschat G, et al. ExaDG: High-order discontinuous Galerkin
  for the exa-scale. In:  {\it Software for exascale computing-SPPEXA
  2016-2019}Springer International Publishing. ; 2020\string: 189--224.

\end{thebibliography}

\end{document}